\PassOptionsToPackage{unicode}{hyperref}
\PassOptionsToPackage{hyphens}{url}
\PassOptionsToPackage{dvipsnames,svgnames,x11names}{xcolor}
\documentclass[
  11pt,
  abstract]{scrartcl}

\usepackage{adjustbox}
\usepackage{amsmath,amssymb,amsthm}
\usepackage{iftex}
\ifPDFTeX
  \usepackage[T1]{fontenc}
  \usepackage[utf8]{inputenc}
  \usepackage{textcomp}
\else
  \usepackage{unicode-math}
  \defaultfontfeatures{Scale=MatchLowercase}
  \defaultfontfeatures[\rmfamily]{Ligatures=TeX,Scale=1}
\fi
\usepackage{lmodern}
\usepackage[backend=biber, style=apa]{biblatex}
\ifPDFTeX\else  
\fi
\IfFileExists{upquote.sty}{\usepackage{upquote}}{}
\IfFileExists{microtype.sty}{%
  \usepackage[nopatch=footnote]{microtype}
  \UseMicrotypeSet[protrusion]{basicmath}
}{}
\makeatletter
\@ifundefined{KOMAClassName}{%
  \IfFileExists{parskip.sty}{%
    \usepackage{parskip}
  }{%
    \setlength{\parindent}{0pt}
    \setlength{\parskip}{6pt plus 2pt minus 1pt}}
}{%
  \KOMAoptions{parskip=off}}
\makeatother
\usepackage{xcolor}
\usepackage[bottom=3cm,left=1in,right=1in]{geometry}
\ifx\paragraph\undefined\else
  \let\oldparagraph\paragraph
  \renewcommand{\paragraph}[1]{\oldparagraph{#1}\mbox{}}
\fi
\ifx\subparagraph\undefined\else
  \let\oldsubparagraph\subparagraph
  \renewcommand{\subparagraph}[1]{\oldsubparagraph{#1}\mbox{}}
\fi

\providecommand{\tightlist}{\setlength{\itemsep}{0pt}\setlength{\parskip}{0pt}}
\usepackage{longtable,booktabs,array}
\usepackage{calc}
\usepackage{etoolbox}
\makeatletter
\patchcmd\longtable{\par}{\if@noskipsec\mbox{}\fi\par}{}{}
\makeatother
\IfFileExists{footnotehyper.sty}{\usepackage{footnotehyper}}{\usepackage{footnote}}
\makesavenoteenv{longtable}
\usepackage{graphicx}
\makeatletter
\def\maxwidth{\ifdim\Gin@nat@width>\linewidth\linewidth\else\Gin@nat@width\fi}
\def\maxheight{\ifdim\Gin@nat@height>\textheight\textheight\else\Gin@nat@height\fi}
\makeatother
\setkeys{Gin}{width=\maxwidth,height=\maxheight,keepaspectratio}
\newlength{\cslhangindent}
\newlength{\csllabelwidth}
\newlength{\cslentryspacingunit}
 {}
\usepackage{calc}

\usepackage{orcidlink}
\usepackage{fvextra}
\usepackage{authblk}
\usepackage{placeins}
\usepackage[ruled,vlined,linesnumbered]{algorithm2e}
\usepackage{listings}
\usepackage{setspace}
\usepackage[all]{nowidow}
\usepackage{tabularx}
\usepackage{makecell}
\usepackage{csquotes}
\usepackage{enumitem}

\usepackage[
font=small,
labelfont={bf,sf},
labelsep=quad,
format=plain
]{caption}

\newcommand{\bsa}{\boldsymbol{a}}
\newcommand{\bsb}{\boldsymbol{b}}

\newcommand{\bsx}{\boldsymbol{x}}
\newcommand{\bsy}{\boldsymbol{y}}

\newcommand{\bfm}{\mathbf{m}}

\newcommand{\bfB}{\mathbf{B}}

\newcommand{\bfD}{\mathbf{D}}

\newcommand{\bfF}{\mathbf{F}}

\newcommand{\bfI}{\mathbf{I}}

\newcommand{\bfK}{\mathbf{K}}

\newcommand{\bfM}{\mathbf{M}}

\newcommand{\bfW}{\mathbf{W}}

\newcommand{\sfT}{\mathsf{T}}

\newcommand{\bsbeta}{\boldsymbol{\beta}}
\newcommand{\bsgamma}{\boldsymbol{\gamma}}
\newcommand{\bsdelta}{\boldsymbol{\delta}}

\newcommand{\bstheta}{\boldsymbol{\theta}}

\newcommand{\bsmu}{\boldsymbol{\mu}}

\newcommand{\bssigma}{\boldsymbol{\sigma}}
\newcommand{\bstau}{\boldsymbol{\tau}}

\newcommand{\bsphi}{\boldsymbol{\phi}}

\newcommand{\bsGamma}{\boldsymbol{\Gamma}}

\newcommand{\bsOmega}{\boldsymbol{\Omega}}

\newcommand{\bfzero}{\mathbf{0}}

\newcommand{\calN}{\mathcal{N}}

\newcommand{\calU}{\mathcal{U}}
\newcommand{\calI}{\mathcal{I}}
\newcommand{\calG}{\mathcal{G}}
\newcommand{\calL}{\mathcal{L}}
\newcommand{\calP}{\mathcal{P}}

\providecommand{\bbR}{\mathbb{R}}
\providecommand{\bbP}{\mathbb{P}}
\providecommand{\bbE}{\mathbb{E}}
\providecommand{\bbV}{\mathbb{V}}
\providecommand{\bbI}{\mathbb{I}}
\providecommand{\bbVar}{\bbV\text{ar}}

\DeclareMathOperator*{\argmax}{arg\,max}

\theoremstyle{plain}
\newtheorem{lemma}{Lemma}[section]
\theoremstyle{definition}
\newtheorem{definition}{Definition}[section]
\theoremstyle{plain}
\newtheorem{theorem}{Theorem}[section]
\theoremstyle{plain}
\newtheorem{proposition}{Proposition}[section]
\theoremstyle{remark}
\AtBeginDocument{}

\usepackage{booktabs}
\usepackage{longtable}
\usepackage{array}
\usepackage{multirow}
\usepackage{float}
\usepackage{colortbl}
\usepackage{pdflscape}
\usepackage{tabu}
\usepackage{threeparttable}
\usepackage{threeparttablex}
\usepackage[normalem]{ulem}
\usepackage{makecell}
\usepackage{xcolor}
\makeatletter
\AtBeginDocument{%
\ifdefined\contentsname
  \renewcommand*\contentsname{Table of contents}
\else
  \newcommand\contentsname{Table of contents}
\fi
\ifdefined\listfigurename
  \renewcommand*\listfigurename{List of Figures}
\else
  \newcommand\listfigurename{List of Figures}
\fi
\ifdefined\listtablename
  \renewcommand*\listtablename{List of Tables}
\else
  \newcommand\listtablename{List of Tables}
\fi
\ifdefined\figurename
  \renewcommand*\figurename{Figure}
\else
  \newcommand\figurename{Figure}
\fi
\ifdefined\tablename
  \renewcommand*\tablename{Table}
\else
  \newcommand\tablename{Table}
\fi
}
\@ifpackageloaded{float}{}{\usepackage{float}}
\@ifundefined{c@chapter}{\newfloat{codelisting}{h}{lop}}{\newfloat{codelisting}{h}{lop}[chapter]}
\floatname{codelisting}{Listing}

\makeatother
\makeatletter
\@ifpackageloaded{subcaption}{}{\usepackage{subcaption}}
\makeatother
\makeatletter
\@ifpackageloaded{tcolorbox}{}{\usepackage[skins,breakable]{tcolorbox}}
\makeatother
\makeatletter
\@ifundefined{shadecolor}{\definecolor{shadecolor}{rgb}{.97, .97, .97}}
\makeatother
\ifLuaTeX
  \usepackage{selnolig}
\fi
\IfFileExists{bookmark.sty}{\usepackage{bookmark}}{\usepackage{hyperref}}
\IfFileExists{xurl.sty}{\usepackage{xurl}}{}
\usepackage{threeparttable}

\usepackage{titletoc}

\usepackage{xstring}
\usepackage{refcount}

\newcommand{\appautoref}[1]{Appendix~\autoref{#1}}
\newcommand{\appautorefsec}[1]{\autoref{#1}}

\newcommand{\mainref}[1]{\ref{#1}}
\newcommand{\mainautoref}[1]{\autoref{#1}}

\hypersetup{
  pdftitle={Bayesian Penalized Transformation Models: Structured Additive Location-Scale Regression for Arbitrary Conditional Distributions},
  pdfauthor={Johannes Brachem; Paul F. V. Wiemann; Thomas Kneib},
  colorlinks=true,
  linkcolor={blue},
  filecolor={teal},
  citecolor={blue},
  urlcolor={Blue},
  pdfcreator={LaTeX via pandoc}}

\author[1]{
\large
Johannes Brachem
\orcidlink{0000-0001-7884-4631}
\thanks{Corresponding author: \texttt{brachem@uni-goettingen.de}}
}
\author[2]{
\large
Paul F. V. Wiemann
\orcidlink{0000-0003-1901-0295}
}
\author[1]{
\large
Thomas Kneib
\orcidlink{0000-0003-3390-0972}
}

\affil[1]{Chair of Statistics, University of Göttingen, Germany
}
\affil[2]{Department of Statistics, The Ohio State University
}

\date{}

\title{
Bayesian Penalized Transformation Models
}
\subtitle{Structured Additive Location-Scale Regression for Arbitrary Conditional Distributions
\rule{\textwidth}{1pt}
}

\begin{document}
\maketitle
\begin{refsection}
\begin{abstract}
    Penalized transformation models (PTMs) are a semiparametric location-scale regression family that estimate a response’s conditional distribution directly from the data,
    and model the location and scale through structured additive predictors.
    The core of the model is a monotonically increasing transformation
    function that relates the response distribution to a reference
    distribution. The transformation function is equipped with a smoothness
    prior that regularizes how much the estimated distribution diverges from
    the reference. PTMs can be seen as a bridge between
    conditional transformation models and generalized additive models for
    location, scale and shape. Markov chain Monte Carlo inference for PTMs
    offers straightforward
    uncertainty quantification for the conditional distribution as well as
    for the covariate effects. A simulation study demonstrates the effectiveness
    of the approach and includes comparisons to many alternative methods.
    Applications to the Fourth Dutch Growth Study and the Framingham Heart
    Study illustrate the usage and practical utility. A full-featured
    implementation is available as a Python library. Supplementary material for this article is available online.

\end{abstract}
\ifdefined\Shaded\renewenvironment{Shaded}{\begin{tcolorbox}[enhanced, boxrule=0pt, interior hidden, borderline west={3pt}{0pt}{shadecolor}, sharp corners, frame hidden, breakable]}{\end{tcolorbox}}\fi

\clearpage

\section{Introduction}\label{introduction}

\doublespacing

We consider the location-scale regression model $Y = \mu(\bsx) + \sigma(\bsx)R$
for a univariate, absolutely continuous random variable $R$ and
response \(Y\). The location \(\mu(\bsx) \in \bbR\) and scale
\(\sigma(\bsx) \in \bbR_{>0}\) depend on a vector of covariate
observations \(\bsx\). We write both terms as depending on the same
vector \(\bsx\) only for notational convenience; in practice, different
covariates can be used in the two terms. The variable $R$
is assumed to have expectation zero and variance one. It can be
interpreted as a standardized error,
and its distribution can be thought of as a standardized version of the
conditional distribution of $Y$.
This model can account
for heteroscedasticity and opens up
possibilities for insight not only into how covariates relate to the
mean of a response, but also its scale. For both frequentist and
Bayesian inference, the model is generally completed by assuming a
parametric distribution for $R$; most commonly the
standard Gaussian distribution. In this paper, we introduce penalized
transformation models (PTMs) as a way to relax this assumption. In PTMs, the
cumulative distribution function (CDF) of $R$ is specified as
\(F_R(r) = F_Z\bigl(h(r)\bigr)\), where
\(h: \bbR \rightarrow \bbR\) is a monotonically increasing and at least
once differentiable function that transforms $R$ to follow a
reference distribution with continuous CDF
\(F_Z: \bbR \rightarrow [0, 1]\). This construction directly implies a
definition of the conditional probability density function of the response via
the change of variables theorem as
\begin{equation}\protect\hypertarget{eq-response-density}{}{
        \begin{aligned}
            f_Y(y | \bsx) = \frac{\partial}{\partial y} F_Z\left( h\left(\frac{y - \mu(\bsx)}{\sigma(\bsx)} \right) \right) = f_Z\bigl( h\left(r \right) \bigr) \frac{\partial}{\partial r} h(r) \frac{1}{\sigma(\bsx)},
        \end{aligned}
    }\label{eq-response-density}\end{equation} where \(f_Z\) denotes the
density of the selected reference distribution, and we note that
$r = (y - \mu(\bsx)) / \sigma(\bsx)$.
If we choose the standard Gaussian distribution as
the reference distribution and the identity function for \(h\), so that
\(F_R(r) = \Phi(r)\), the model falls back
to the Gaussian location-scale model
\(Y \sim \calN\bigl(\mu(\bsx), \sigma(\bsx)^2\bigr)\).
The specification and estimation
of the transformation
function \(h\) is a major focus in penalized transformation models, because it
allows researchers to estimate the conditional response distribution directly from
their data, while regularizing towards the reference distribution.
We construct the transformation function
around a monotonically increasing penalized spline \parencite[see][]{Lang2004-BayesianPsplines,Pya2015-ShapeConstrainedAdditive} with modifications for increased interpretability
and smooth extrapolation. The transformation function is penalized towards the
identity function, providing regularization towards the reference distribution $F_Z$.

\paragraph{Related literature.} There are a number of flexible distributional regression
approaches present in the recent statistical literature \parencite[see][for a review]{Kneib2021-RageMeanReview}. Generalized additive models
for location, scale and shape \parencite[GAMLSS,][]{Rigby2005-GeneralizedAdditiveModels,Umlauf2018-BamlssBayesianAdditive}
allow all parameters of a wide range of parametric distributions to be
modeled additively using smooth functions of covariates in structured additive predictors.
The limiting factor of GAMLSS is the necessity to choose a fixed parametric form for
the conditional distribution of the response.

In conditional transformation models \parencite[CTMs,][]{Hothorn2014-ConditionalTransformationModels,Hothorn2018-MostLikelyTransformations,Carlan2024-BayesianConditionalTransformation}, the response's
conditional CDF is specified nonparametrically as
\(F_{Y}(y | \bsx ) = F_Z\bigl(g(y | \bsx)\bigr)\), with inference and
modeling focused on a conditional transformation function
\(g(y|\bsx)\). This function is additively decomposed as
\(g(y|\bsx) = \sum_{j=1}^Jg_j(y|\bsx)\), and the partial transformation
functions \(g_j\) are parameterized as tensor products
\(g_j(y | \bsx) = (\bsa_j(y)^\sfT \otimes \bsb_j(\bsx)^\sfT)^\sfT \bsgamma_j\),
where \(\otimes\) denotes the Kronecker product. Depending on the
researcher's choice for the bases \(\bsa_j(y)\) and \(\bsb_j(\bsx)\),
CTMs can capture all moments of the conditional response distribution
dependent on covariates. They can provide great flexibility, but the interpretation
of covariate effects is challenging.

\textcite{Siegfried2023-DistributionfreeLocationscaleRegression}
recently presented transformation additive models for location and scale
(TAMLS), where the response's conditional distribution is set up as
\(F_{Y}(y | \bsx ) = F_Z\bigl(z\bigr)\),
with $z = \sigma(\bsx)^{-1}g(y) - \mu(\bsx)$,
and the transformation function \(g\) is parameterized with Bernstein
polynomials. When rearranging to
\(g(y) =\sigma(\bsx) \mu(\bsx) + \sigma(\bsx) z\), where
\(z \sim F_Z\), it becomes evident that $g$ plays the role of a pre-transformation of
the response. The additive predictors \(\mu(\bsx)\) and
\(\sigma(\bsx)\) reflect the location and scale of the transformed
response \(g(Y)\). In a similar spirit, \textcite{Kowal2024-MonteCarloInference}
present a Bayesian model of the form $g(Y) = f(\bsx) + \sigma \varepsilon$, where the
transformation function $g$ is inferred nonparametrically via Bayesian
bootstrap.
The authors consider linear and Gaussian process
specifications for $f(\bsx)$ and assume a parametric distribution with fixed location
and scale for $\epsilon$; in most cases standard Gaussian. The conceptual model
formulation is similar to TAMLS, but does not include covariate models for the
scale of the response. Both are distinctly different from penalized transformation
models in the fact that they model the location --- and scale, in the case of TAMLS ---
of a non- or semiparametrically \textit{transformed response}, which makes
interpretation challenging. In penalized transformation models, the
location and scale are instead modeled directly on the level of the response, providing a more
straightforward interpretation.

Dirichlet process mixture models \parencite[DPMMs,][]{Antoniak1974-MixturesDirichletProcesses,Ferguson1973-BayesianAnalysisNonparametric} are a classic Bayesian nonparametric approach to modeling unknown distributions. They express the response distribution as an infinite mixture of simpler parametric kernels, most commonly Gaussian ones. \textcite{Chib2010-AdditiveCubicSpline} and \textcite{Wiesenfarth2014-BayesianNonparametricInstrumental} modeled the error distribution in a smooth mean regression and simultaneous equation model, respectively, with a DPMM. \textcite{Leslie2007-GeneralApproachHeteroscedastic} presented a linear location-scale regression model with a DPMM error distribution. \textcite{Villani2009-RegressionDensityEstimationa} use a DPMM in which the location and scale of the mixture components and the mixture weights were modeled as smooth functions on covariates;
an approach that maximizes flexibility. \textcite{Rodriguez-Alvarez2024-DensityRegressionDirichleta} present a DPMM with full structured additive predictors on the locations of the mixture components, keeping the mixture variances and weights constant. Of these approaches, only
\textcite{Rodriguez-Alvarez2024-DensityRegressionDirichleta} provide an easily accessible
implementation in an R package.

Quantile regression \parencite{Koenker2001-QuantileRegression,Waldmann2018-QuantileRegressionShorta} constitutes yet another popular approach to distributional regression without strong theoretical assumptions by directly and separately modeling the quantiles of interest. A common issue in quantile regression is quantile crossing, which is a consequence of the independence of models for different quantiles; but solutions for this issue exist \parencite[see, for instance,][]{Schnabel2013-SimultaneousEstimationQuantile}.

\paragraph{Contributions.} From an applied perspective, penalized transformation models are particularly notable for the following features.
First, they provide interpretability in terms of location and scale directly
on the level of the response.
Second, they flexibly estimate the conditional response distribution semiparametrically
from the data. Third, they offer regularization towards a reference
distribution through penalization of the transformation function towards identity.
Fourth, they provide the full flexibility of structured additive predictors for
modeling the location and scale.
Fifth, we present Bayesian inference via Markov chain Monte Carlo (MCMC) methods, which provides
natural uncertainty quantification on model components, including the
structured additive regression predictors and the features of the conditional response
distribution.
Sixth, we provide an open-source Python implementation
built on the probabilistic programming framework Liesel \parencite{Riebl2023-LieselProbabilisticProgramming}, with automatic differentiation and
just-in-time compilation provided by the machine learning library JAX \parencite{deepmind2020jax}.

From a more technical perspective, penalized transformation models are attractive
because they provide a closed-form, coherent response density that in principle
allows for both Bayesian and frequentist inference. The approach of modeling a flexible
distribution by transformation to a known reference with a single, monotonically
increasing function additionally strikes us as a useful one, since it combines
flexibility with conceptual simplicity and can readily be regularized towards a reference distribution.
In our view, this simplicity is a notable
secondary benefit that is valuable for both adoption of the
method in applications and further development.

While we acknowledge that strong alternative approaches exist in the statistical literature,
and none of the mentioned features is entirely unique on its own,
we argue that the combination of features makes penalized transformation models (PTMs) an
attractive option. Put succinctly, PTMs are more flexible than GAMLSS, easier to interpret than CTMs, TAMLS and many DPMMs, more accessible than the most similar DPMMs, and provide a more coherent model than quantile regression.

The remainder of this paper is structured as follows. In  \autoref{sec-ptm}, we
provide the technical details of our model setup, including our parameterization of the transformation function $h$, our regularizing priors, a brief review of structured
additive predictors, and modifications for count data and censored responses.
In \autoref{sec-posterior-inference}, we provide details on our Markov chain Monte Carlo
inference. In \autoref{sec-simulation}, we present a simulation study to assess the performance of PTMs for recovering a wide range of different residual distributions and obtaining unbiased and efficient estimates for nonlinear location and scale effects. We compare PTMs to many of the previously mentioned alternative models. In \autoref{sec-applications}, we apply the model to data from the
Fourth Dutch Growth Study and the Framingham Heart Study, showcasing its usage and providing additional comparisons to the alternative models, before we conclude the paper in \autoref{sec-conclusion}.

\section{Penalized Transformation Models}\label{sec-ptm}

The basic model formulation is given by $Y = \mu(\bsx) + \sigma(\bsx)R$, with
\(\bbE(R) = 0\), \(\bbVar(R) = 1\), and the
cumulative distribution function (CDF) of $R$ given by
\(F_R(r) = F_Z(h(r))\), where $h$ is the strictly monotonically increasing
transformation function and \(F_Z\) is the CDF of a fully specified continuous reference
distribution with expectation zero and variance one.

\hypertarget{sec-transformation}{%
    \subsection{Transformation function}\label{sec-transformation}}

We set up the transformation function \(h: \bbR \rightarrow \bbR\) around a monotonically
increasing B-spline based on a commonly used approach, described for example by \textcite{Pya2015-ShapeConstrainedAdditive}, with some modifications of our own. Outside the subdomain $[a,b] \subset \bbR$,
where little to no information is available from the data, the function smoothly transitions to a
linear function with slope one. The core of the function is a B-spline segment given by
\begin{equation}\protect\hypertarget{eq-transformation-core}{}{
        h(r) = \alpha + \frac{1}{s(\bsdelta)}
        \sum_{j=2}^J B_j(r) \sum_{\ell = 2}^j \exp\bigl(\delta_{\ell-1}\bigr) \quad \text{if} \quad r \in [a, b],
    }\label{eq-transformation-core}\end{equation}
with real-valued
parameters \(\bsdelta^\sfT = [\delta_1, \dots, \delta_{J-1}]\), constituting log-increments in the spline control points.
\(B_j(r)\)
is a third-order (cubic) B-spline basis evaluation,
and $J$ is the number of bases, which is manually chosen by the researcher. The number
of bases determines the flexibility of the function, typical values range from
$J=15$ to $J=40$. Due to a regularizing random walk prior that we place on the
log-increment parameters $\bsdelta$ (see \autoref{sec-ptm-prior}), the exact choice for $J$ is not critical in
practice, as long as it is high enough to allow for sufficient flexibility.
The summation starts at $j=2$, because \eqref{eq-transformation-core} is a re-written
version of a B-spline $h(r) = \sum_{j=1}^J B_j(r) \omega_j$, where $\omega_1 = \alpha$
and $\omega_j = \alpha + \sum_{\ell=2}^j \exp(\delta_{\ell-1})$ for $j=2, \dots, J$;
using the unity decomposition property of B-spline bases, $\sum_{j=1}^J B_j(r) = 1$.
The bases are set up using
\(J + 4\) equidistant knots
\(k_{-2} < k_{-1} < k_{0} < \dots < k_{J} < k_{J+1}\), where
\(k_{1}, \dots, k_m\) with $m = J-2$ and $k_1=a$, $k_m=b$ lie inside the core domain
and are called interior knots, and the
remaining knots lie outside $[a,b]$ and are called outer knots. The outer knots are
required by the recursive definition of B-spline bases.
The distance between adjacent knots is $d = k_{j+1} - k_j$.
The boundaries of the function's core domain are $a=k_1$ and $b=k_m$, and they are manually chosen by the researcher. In conjunction with a choice for $J$, this determines the full knot sequence. Theorem~\ref{thm-core-probability} will provide background for informing these choices of $a$ and $b$.
The intercept parameter $\alpha$ is fixed to $\alpha = a - h(a | \alpha = 0)$, such that $h(a)=a$ for any choice of $\bsdelta$.
The function $s(\bsdelta)$ normalizes the average slope over the core domain $[a, b]$ to one and is given by
\begin{equation} \label{eq:sfn}
    s(\bsdelta) = \frac{1}{b-a}
    \sum_{j=1}^{J-3}
    \left(
    \frac{\exp(\delta_{j}) + \exp(\delta_{j+2})}{6} + \frac{2\exp(\delta_{j+1})}{3}  \right).
\end{equation}
In the core segment, the first derivative of $h$ with respect to $r$ is given by
\begin{equation} \label{eq:hderiv}
    \frac{\partial}{\partial r} h(r) = \frac{1}{s(\bsdelta)\cdot d} \sum_{j=2}^J B_j^{(2)}(r) \exp(\delta_{j-1}),
\end{equation}
where $d = k_{j+1} - k_j$ is the (constant) distance between adjacent knots and $B_j^{(2)}(r)$ is a quadratic B-spline basis function evaluation. This derivative is required to evaluate the density of $Y$ in \eqref{eq-response-density}. Its form follows from a well-known property of B-splines: the derivative of a B-spline is another B-spline of lower order, with the coefficients given by differences in the original coefficients, scaled by the distance between knots \parencite[see, for example,][p. 20]{Eilers2021-PracticalSmoothingJoys}. See also \appautoref{app-prop:trafo-derivative}.

\begin{figure}[tb]
    {\centering \includegraphics[width=0.97\textwidth]{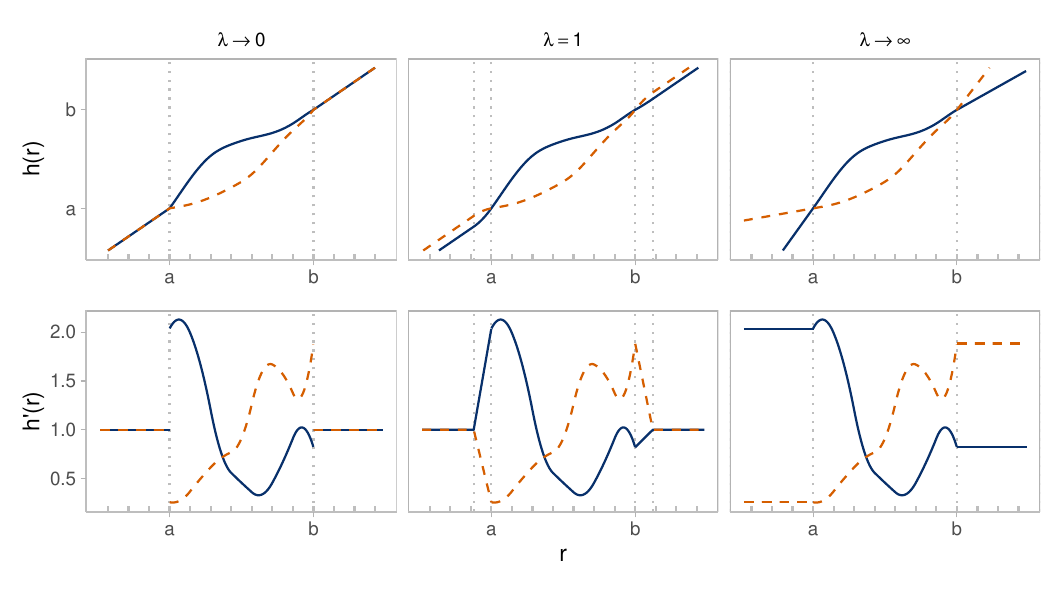}
    }
    \caption{\label{fig-trafo}Two examples for possible transformation
        functions (top row) and their derivative (bottom row) given the same fixed knot sequence with different choices for $\lambda$. The
        knots are shown by gray upticks on the x-axis. Dotted vertical lines
        mark the boundaries of the function segments. Note that, for $\lambda\rightarrow0$ and $\lambda\rightarrow \infty$, there is no transition segment.}
\end{figure}

Outside the core domain, the transformation function smoothly transitions to a linear function with unit slope. The transition segments on subdomains $r \in [a-\lambda, a)$ and $r \in (b, b+\lambda]$ are derived from linearly interpolating between $h'(a)$ and $h'(b)$, and $h'(a-\lambda) = 1$ and $h'(b + \lambda)=1$ on the left and right side of the core segment, respectively; $h'$ denotes the derivative of $h$ with respect to $r$.
The parameter $\lambda \in \bbR_{>0}$ is a manually chosen hyperparameter determining the width of the transition segment, such that $\lambda \rightarrow 0$ implies an abrupt change to the identity function, leading to discontinuous derivatives of $h$ at $a-\lambda$ and $b+\lambda$ and $\lambda \rightarrow \infty$ implies simple linear extrapolation using the slope of $h$ at $a$ and $b$ for left and right extrapolation, respectively. \autoref{fig-trafo} shows illustrations of the transformation
function and its derivative using arbitrarily chosen
parameters in the spline segments and three different choices for $\lambda$.
The exact parameterization of the transition and extrapolation segments are included in \appautorefsec{app-sec-trafo}.
Theoretical considerations about the meaning of $\lambda$ for the tail behavior
of the implied distribution of $R$ are included in the appendix in Proposition~\ref{app-prop-tails}. \appautorefsec{app-sec:sim1} includes simulation results in an unconditional setting, suggesting that if $\lambda$ is chosen very small, sampling can become less efficient.
The first application illustration includes models with moderate $\lambda$ and $\lambda \to \infty$, showing differences in tail densities that arise due to the different extrapolation behaviors. In the shown application, no difference in predictive performance remains between different choices for $\lambda$ when the boundaries $a$ and $b$ are chosen wide enough.

The use of log-increment parameters is a common technique for ensuring strict monotonicity in a B-spline, see for instance \textcite{Pya2015-ShapeConstrainedAdditive}. The normalization of the average slope and the smooth transition to linear extrapolation are our modifications.

We now state some relevant properties of the transformation function more formally. For all the following statements, let the function \(h: \bbR \rightarrow \bbR\) be given by
\eqref{eq-transformation-core} on $[a,b]$, and with extrapolation on $(-\infty, a) \cup (b, \infty)$ as described in detail in \appautorefsec{app-sec-trafo}. The full proofs are given in \appautorefsec{app-proofs}.

\begin{proposition}[Monotonicity of the transformation
        function]\protect\hypertarget{thm-monotonicity}{}\label{thm-monotonicity}
    \(h(r)\) is strictly monotonically
    increasing in \(r\).
\end{proposition}
Strict monotonicity of $h$ is necessary for us to be able to define the cumulative distribution function of $R$ as $F_R(r) = F_Z(h(r))$. It follows from defining the spline coefficients in terms of log-increments.

\begin{theorem}[Core probability]\protect\hypertarget{thm-core-probability}{}\label{thm-core-probability}
    Let $R$ be an absolutely continuous random variable with cumulative distribution
    function $F_R(r) = F_Z(h(r))$, where $F_Z$ is
    a continuous cumulative distribution function without unknown parameters.
    Then $\bbP(R \leq a) = F_Z(a)$ and $\bbP(R \leq b) = F_Z(b)$.
\end{theorem}
This result is important, because it can be used to define the knots of the spline part of $h$ by choosing $a$ and $b$ as appropriate quantiles of the reference distribution.
Specifically, using a standard Gaussian reference distribution, we can set up a default knot basis by letting $a=-4$ and $b = 4$, such that we impose the assumption $\bbP(R \in [-4, 4]) \approx 0.99994$. Since we assume $R$ to be standardized to expectation zero and variance one, we consider the additional assumption implied by letting $a=-4$ and $b = 4$ to be a useful default. In the first application illustration, we show how a quantile-quantile plot can be used to diagnose the adequacy of this choice, such that researchers
can choose $a$ and $b$ to meet the requirements of their particular applications.

\begin{theorem}[Reduction to identity]\label{thm-identity}
    If $\delta_j = c$ for all $j = 1, \dots, J-1$ and some constant $c \in \bbR$, then $h(r)=r$ for all $r \in \bbR$.
\end{theorem}
Reduction to identity is an important special case, because if $h$ is the identity function, we can write $F_R(r) = F_Z(h(r)) = F_Z(r)$, i.e., the distribution of $R$ becomes the reference distribution. We will use this property in \autoref{sec-ptm-prior} to develop a prior for $\bsdelta$ that penalizes deviations of $F_R$ from the CDF of the reference distribution. Note that the theorem holds for any value of $c$; the only requirement is that all $\delta_j$ are equal.

\subsection{Prior for \texorpdfstring{$\bsdelta$}{transformation parameter}}\label{sec-ptm-prior}

\begin{figure}[tb]
    \includegraphics[width=\textwidth]{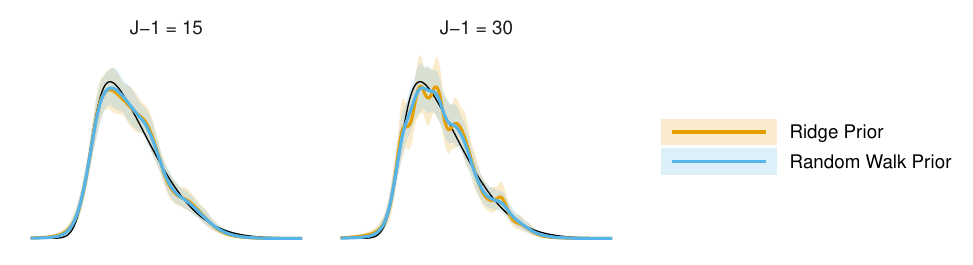}
    \caption{\label{fig-ridge-vs-rw-illustration}Comparison of average posterior densities fitted on $500$ observations under a random walk prior or a ridge prior for $\bsdelta$, and $J-1 = 15$ or $J-1=30$ parameters in $\bsdelta$. The ridge prior is an i.i.d. Gaussian prior $\bsdelta \sim \calN(\bfzero, \tau^2_\delta \bfI)$. The true data-generating density is given as a solid black line. The shaded areas are pointwise 90\%  credible bands.}
\end{figure}

When specifying a prior for the parameters \(\bsdelta = [\delta_1, \dots, \delta_{J-1}]^\sfT\), we
focus on two main considerations. First, we penalize deviations of $F_R$ from the reference CDF $F_Z$, i.e., we regularize the distribution of $R$ towards the reference distribution. Second, we generally penalize overly wiggly estimates for $h$. Note that the number of parameters
(\(J-1\)) determines the flexibility of the transformation function.
Choosing a high value like \(J-1=30\) leads to a very flexible
transformation function, but without penalization it can also make the model prone to
overfitting. Overfitting may show itself in an overly wiggly
transformation function that results in a likewise wiggly estimate of
the probability density function \(f_Y(y | \bsx)\). We therefore choose
a regularizing prior that moderates the trade-off between flexibility
and overfitting. \autoref{fig-ridge-vs-rw-illustration} illustrates the effect of this prior in
comparison to an i.i.d. Gaussian (ridge) prior.
Note how, in the ridge prior setting with
$J-1 = 30$ parameters, the estimated density is more wiggly, indicating overfitting,
while in the random walk prior setting, the estimated density and credible bands remain smooth.

By Theorem~\ref{thm-identity}, the transformation function becomes the identity function, and consequently the distribution of $R$ becomes the reference distribution, if all elements of $\bsdelta$ take the same (arbitrary) real value $c$. We can thus penalize deviations from the reference distribution by penalizing differences between the elements of $\bsdelta$, which can be achieved with a first-order random walk prior
\(\delta_j | \delta_{j-1} \sim \mathcal{N}(\delta_{j-1}, \tau^2_\delta)\)
with \(j = 2, \dots, {J-1}\) and a constant prior for \(\delta_1\).
It resembles the
general Bayesian P-spline priors introduced by \textcite{Lang2004-BayesianPsplines} and was used
in a slightly modified version by \textcite{Carlan2024-BayesianConditionalTransformation}
in the construction of Bayesian conditional transformation models. The
variance of the random walk, \(\tau^2_\delta\), regulates the overall
smoothness of the transformation function, with $\tau^2 \rightarrow 0$ implying that $h$ becomes the identity function. The joint representation of this prior for \(\bsdelta\) is a rank-deficient multivariate normal prior
\begin{equation}\protect\hypertarget{eq-delta-prior}{}{
        \pi(\bsdelta | \tau_\delta) \propto \left(\frac{1}{\tau^2_\delta}\right)^{\text{rk}(\bfK)/2}  \exp\left( -\frac{1}{2\tau^2_\delta} \bsdelta^\sfT \bfK \bsdelta \right),
    }\label{eq-delta-prior}\end{equation} with the \((J-1) \times (J-1)\)
penalty matrix \(\bfK = \bfD^\sfT\bfD\) of rank \(J-2\) constructed from
a \((J-2) \times (J-1)\) first-difference matrix \(\bfD\) that is zero
everywhere except that \(\bfD_{[d,d+1]} = -\bfD_{[d,d]} = 1\) for
\(d = 1, \dots, J-2\), where $\bfD_{[d,d+1]}$ refers to the element of $\bfD$ at row $d$ and
column $d+1$.
The rank-deficiency of this prior arises by construction,
since we include the starting condition $\delta_1$ with a constant prior in \eqref{eq-delta-prior},
as illustrated with an example in \autoref{app-sec-rwprior-illustration}. This starting condition can be regarded as an additive constant which is applied to the remaining parameters, and which is
unpenalized by the prior. In fact, the starting condition $\delta_1$ is unidentified in the model, since any added
constant to $\bsdelta$ cancels out in the transformation function \eqref{eq-transformation-core} by
$s(\bsdelta)$, see \appautoref{app-sec:cancellation-of-constants}.
We thus remove the constant part by applying a reparameterization through a mixed model decomposition, following \textcite{Kneib2019-ModularRegressionLego}. The reparameterization yields a transformed parameter $\tilde \bsdelta$ with full-rank prior $\tilde \bsdelta \sim \calN(\bfzero_{J-2}, \bfI_{J-2})$. More details are provided in \appautoref{app-sec:sum-to-zero}. We conduct inference on the level of $\tilde \bsdelta$.
An equivalent prior for $\bsdelta$ could be set up by setting $\delta_1 \equiv 0$; then
the prior for $[\delta_2, \dots, \delta_{J-1}]^\sfT$ is a multivariate normal prior
$\calN(\bfzero_{J-2}, \tau^2_\delta \tilde \bfK)$, where $\tilde \bfK$ can be obtained
by dropping the first row and column from $\bfK$.

In combination with a suitable hyperprior for the
variance parameter \(\tau^2_\delta\), this regularizing prior allows us
to equip the model with a generous number of parameters, for example
\(J-1=30\), and let the inference algorithm arrive at a sensible amount
of smoothness. \appautorefsec{app-sec:sim1} includes an empirical comparison of models
with $J-1=15$ and $J-1=30$ parameters on data generated from different distributions,
showing that for challenging scenarios the higher number of parameters is beneficial,
while it has no substantial downsides in easier scenarios.

\subsection{Hyperprior for \texorpdfstring{$\tau^2_\delta$}{variance parameter}}
By controlling the
deviation of the transformation function from the identity function, the
variance parameter \(\tau^2_\delta\) controls how much the
distribution of $R$ can deviate from the reference distribution.
For \(\tau^2_\delta \rightarrow 0\), the model reduces to the base
model, where $R$ follows the reference distribution and $Y$ follows a location-scale version of the reference distribution.
We set up a penalized complexity prior, following the principles outlined
by \textcite{Simpson2017-PenalisingModelComponent}.
First, the prior should favor the base model unless there is clear
information to support the more complex model. Second, additional
complexity is measured as $d(\pi \| \pi_b) = \sqrt{2 \text{KLD}(\pi \| \pi_b)}$, where $\pi$ is \eqref{eq-delta-prior} and $\pi_b$ is the prior corresponding to the base model, i.e. \eqref{eq-delta-prior} with $\tau^2_\delta \rightarrow 0$ and $\text{KLD}(\pi \| \pi_b) = \int \pi(\bsdelta) (\log(\pi(\bsdelta)) - \log(\pi_b(\bsdelta))) \mathrm d \bsdelta$ is the Kullback-Leibler divergence between $\pi$ and $\pi_b$.
Third, we place an exponential prior with density $\pi_d(d) = \tilde \lambda \exp(-\tilde \lambda d)$ on the distance measure $d$, such that the density for stronger deviations from the base model decays at a constant rate. We write $\tilde \lambda$ for the rate here to avoid confusion with the hyperparameter $\lambda$ of the transformation function.
As shown in detail by \textcite[Theorem 1]{Klein2016-ScaledependentPriorsVariance},
for a multivariate normal prior like \eqref{eq-delta-prior}, this setup is equivalent to using a Weibull prior $\tau^2_\delta \sim \operatorname{Weibull}(0.5, \psi)$
with the shape fixed to \(0.5\) and scale
\(\psi\). This scale determines the rate of decay in the prior for $d$.

By principle four in \textcite{Simpson2017-PenalisingModelComponent} and
\textcite{Klein2016-ScaledependentPriorsVariance}, $\psi$ is chosen such that the condition $\bbP(q(\psi) \leq \tilde c) = 1-\tilde \alpha$ is fulfilled for some meaningfully interpretable function $q$ of $\psi$ and user-specified values $\tilde c > 0$ and $\tilde \alpha \in (0, 1)$. We choose $q$ as the total variation distance
$\text{TV}(f_R, f_Z) = \frac{1}{2} \int | f_R(r | \bsdelta) - f_Z(r) | \mathrm d r$
between the transformation distribution with density $f_R(r | \bsdelta) = f_Z(h(r | \bsdelta))\frac{\partial}{\partial r}h(r | \bsdelta)$ and the reference density $f_Z$, averaged over $\bsdelta$ and $\tau^2_\delta$:
$$
    q(\psi) = \int \pi(\tau^2_\delta | \psi) \int \pi(\bsdelta | \tau^2_\delta) \left( \frac{1}{2} \int | f_R(r | \bsdelta) - f_Z(r) | \mathrm d r \right) \mathrm d \bsdelta\, \mathrm d \tau^2_\delta.
$$
The total variation distance is an attractive choice here because it takes values in $[0, 1]$ and can be interpreted as the relative overlap between the two densities, with $\text{TV}(f_R, f_Z) = 0$ implying that $f_R$ and $f_Z$ overlap completely, and $\text{TV}(f_R, f_Z) = 1$ implying that there is no overlap at all between $f_R$ and $f_Z$. \appautoref{app-fig-sdprior-references} includes an illustration of the total variation distance for four example densities in comparison to a standard Gaussian density.

\begin{figure}[tb]
    \begin{subfigure}{0.43\linewidth}
        \centering \includegraphics[width=\textwidth]{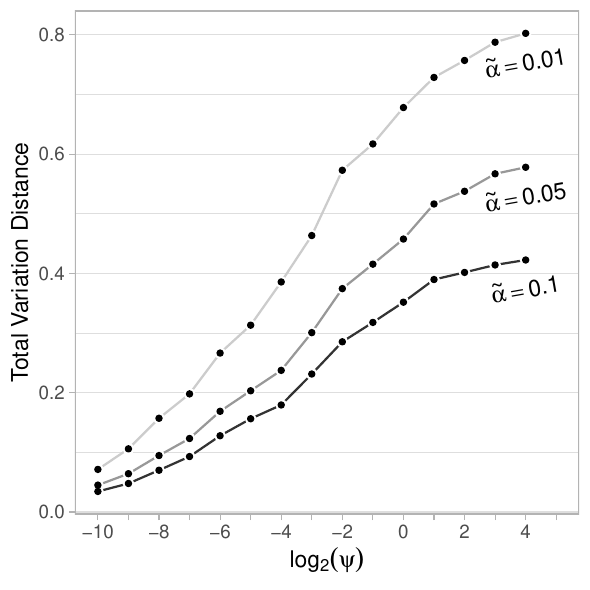}
    \end{subfigure}
    \begin{subfigure}{0.56\linewidth}
        \centering \includegraphics[width=\textwidth]{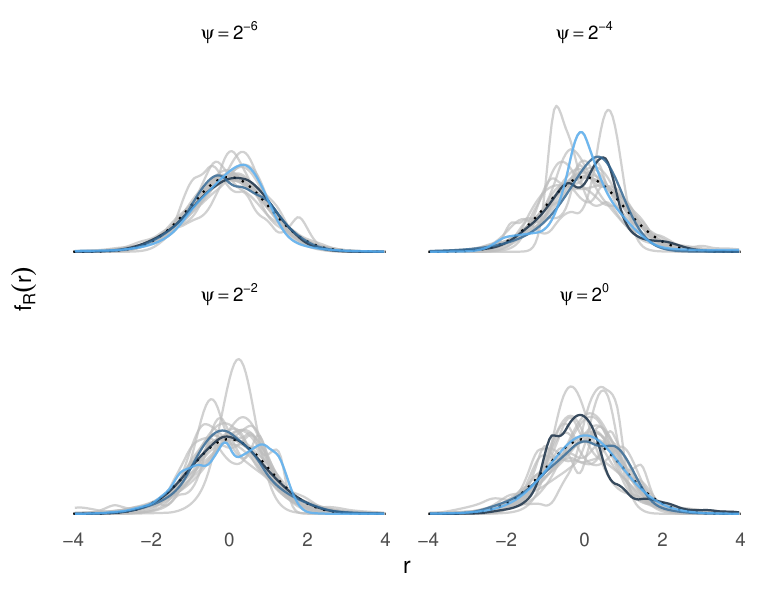}
    \end{subfigure}

    \caption{\label{fig-prior-predictive}
        Left panel: Quantile curves of total variation distance as a function of $\log_2(\psi)$ for three quantile levels $\tilde \alpha$, using $J-1=30$ parameters in $\bsdelta$.
        Right panel: Twenty densities
        obtained from prior predictive simulations using different scale
        parameters \(\psi\) in the hyperprior for \(\tau^2_\delta\). The gray
        lines are individual samples, the dotted black line is the standard
        Gaussian distribution for reference. The colored lines are three
        arbitrarily chosen densities for each setting, intended as
        highlighted examples. Note that we fix the upper limit of the y-axis for
        better general visibility, which causes some clipping for individual
        samples.
    }

\end{figure}

We conducted prior predictive simulations to inform the selection of a
value for \(\psi\). The left panel of \autoref{fig-prior-predictive} shows the results for 15 values $\psi = 2^k$ for $k = -10, -9, \dots, 4$. For each $\psi$, we drew $100$ values of $\tau^2_\delta \sim \operatorname{Weibull}(0.5, \psi)$, followed by another $100$ draws of $\bsdelta$ from its prior for each value of $\tau^2_\delta$, resulting in $10\,000$ draws for each $\psi$. We then numerically evaluated the total variation distance of the densities $f_R(r|\bsdelta)$ and the standard Gaussian density, resulting in a sample of $10\,000$ total variation distances for each $\psi$. The figure displays the $.1$, the $.05$ and the $.01$ quantiles of these samples. As our model setup requires, all
densities are standardized to mean zero and standard deviation one, such that
the total variation distance captures only differences in shape, not in location or scale.
We conducted the simulation for three different values of the hyperparameter $\lambda$, which controls the extrapolation behavior of the transformation function ($\lambda \rightarrow 0$, $\lambda = 1$, and $\lambda \rightarrow \infty$). Since we did not observe meaningful differences between different values for $\lambda$, we pooled the individual samples. While \autoref{fig-prior-predictive} displays the results for $J-1 = 30$ parameters in $\bsdelta$, \appautoref{app-fig-sdprior-15vs30} includes a comparison to $J-1 = 15$ parameters, where the pattern is similar, but the total variation difference to the standard Gaussian distribution tends to be smaller compared to $J-1=30$. \autoref{fig-prior-predictive} (right) shows twenty
densities obtained from hierarchical draws from the priors of
\(\tau^2_\delta\) and \(\bsdelta\) using \(J-1 = 30\) for four different values of $\psi$.

From the results, we judge that a value of $\psi = 2^{-1} = 0.5$, which corresponds to $\tilde \alpha = 0.01$ and $\tilde c \approx 0.62$, can be viewed as a reasonable default, allocating about 90\% prior probability to distributions that have at least a 70\% overlap with the standard Gaussian distribution, and about 99\% prior probability to distributions that have at least a 40\% overlap with the standard Gaussian. This provides some scepticism with respect to extreme deviations from Gaussianity while allowing substantial deviations if warranted by the data. Researchers can use the results displayed in \autoref{fig-prior-predictive} to inform their choice of $\psi$ according to their problem-specific scepticism about deviations from the standard Gaussian density.
For the remainder of this paper, we use $\psi = 0.5$ unless stated otherwise.

\begin{figure}
    \includegraphics[width=\textwidth]{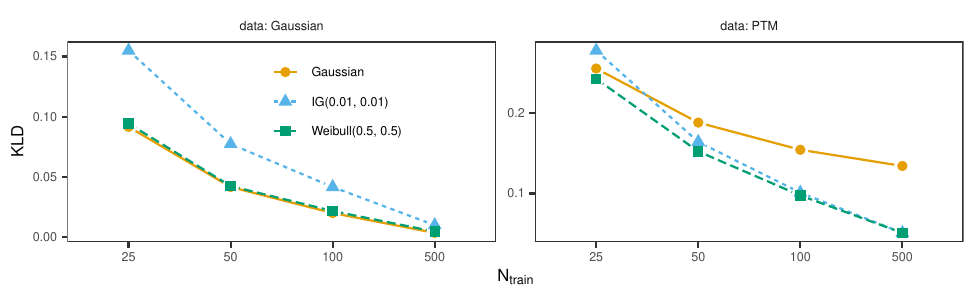}
    \caption{Performance of PTMs with different priors for $\tau^2_\delta$ for a univariate model $\bbP(R \leq r_i) = \Phi(h(r_i | \bsdelta))$. Results for a simple Gaussian model $r_i \sim \calN(\mu, \sigma^2)$ are included for reference.
        Observations $r_i$, $i = 1, \dots, N_{\text{train}}$ were drawn from a standard Gaussian distribution (left) or from varying PTM distributions obtained by drawing
        $\bsdelta$ from its prior (right); each simulation run includes a different true distribution in the PTM scenarios. For each sample size, the plot shows the Kullback-Leibler divergence $\operatorname{KLD}(f \| \hat{f})$ between the true  and estimated distribution, averaged over 100 replications.
        \label{fig:sim1-kld}}
\end{figure}

\autoref{fig:sim1-kld} shows results from a simulation study on unconditional
data, described in detail in \appautorefsec{app-sec:sim1}. The figure shows that the scale-dependent prior improves the PTM's ability to recover a Gaussian distribution to a point where its performance is indistinguishable from the true Gaussian model. The advantage compared to a model with an inverse gamma prior is particularly large when the sample size is small. We even see a small advantage of the scale-dependent prior with small sample sizes for non-Gaussian data generated from the PTM. The simulation study also compares a case with a highly dispersed prior $\tau^2_\delta \sim \text{Weibull}(0.5, 25)$, which appears to lead to less efficient sampling and worse convergence compared to the scale-dependent configuration.

\subsection{Structured additive predictors for location and
    scale}\label{sec-covariates}

We use structured additive predictors to model \(\mu(\bsx)\) and
\(\sigma(\bsx)\). Structured additive predictors can represent a wide
array of linear, non-linear, random, and spatial functions in a unified
framework using basis expansions. Specifically, we define
\(\mu(\bsx) = \beta_0 + \sum_{\ell=1}^L f_\ell(\bsx)\), where
\(\beta_0\) is an intercept and each \(f_\ell(\bsx)\) is given by
\(f_\ell(\bsx) = \sum_{d=1}^{D} A_{d,\ell}(\bsx) \beta_{d,\ell} = \bsa_\ell^\sfT\bsbeta_\ell\).
Here, \(A_{1,\ell}(\bsx), \dots, A_{D,\ell}(\bsx)\) are the evaluations
of basis functions of possibly different types, chosen based on which
kind of function should be represented by \(f_\ell(\bsx)\). The
coefficient vectors \(\bsbeta_1, \dots, \bsbeta_L\) are each equipped
with a suitable smoothness prior, which is a potentially rank-deficient
multivariate Gaussian prior of the form
\begin{equation}\protect\hypertarget{eq-stap-prior}{}{
        \pi(\bsbeta_\ell | \tau^2_\ell) \propto
        \left( \frac{1}{\tau^2_\ell} \right)^{\text{rk}(\bfK_\ell) / 2}
        \exp\left(-\frac{1}{2\tau^2_\ell} \bsbeta_\ell^\sfT\bfK_\ell\bsbeta_\ell \right),
    }\label{eq-stap-prior}\end{equation} where \(\ell = 1, \dots, L\)
indicates the function, \(\bfK_\ell\) denotes a \(D \times D\) penalty
matrix for a function represented by \(D\) bases and \(\tau^2_\ell\)
denotes an inverse smoothing parameter. Collecting
\(\bsa = [\bsa_1^\sfT, \dots, \bsa^\sfT_{L}]^\sfT\) and
\(\bsbeta = [\bsbeta_1^\sfT, \dots, \bsbeta_{L}^\sfT]^\sfT\), the
predictor can be written as \(\mu(\bsx) = \beta_0 + \bsa^\sfT \bsbeta\).
The predictor \(\sigma(\bsx)\) is defined analogously, with intercept
\(\gamma_0\), basis vector \(\bsb\) and coefficient vector \(\bsgamma\).
However, since \(\sigma(\bsx)\) is defined only on the positive real
line, we use a log link, so that we have
\(\ln \sigma(\bsx) = \gamma_0 + \bsb^\sfT \bsgamma\).
The choice of basis functions \(A_{d,\ell}(\bsx)\) and penalty matrix
\(\bfK_\ell\) jointly determines the form of the individual terms. \textcite[p. 554]{Fahrmeir2013-RegressionModelsMethods} provide a comprehensive overview of useful specifications. The
setup is completed with hyperpriors on the inverse smoothing parameters
\(\tau^2_\ell\). As a default, we use inverse gamma priors
\(\tau^2_\ell \sim \mathcal{IG}(a_\ell, b_\ell)\) with \(a_\ell=1\) and \(b_\ell=0.001\). The
inverse gamma prior is conditionally conjugate to the multivariate Gaussian prior,
enabling us to use Gibbs updates for $\tau^2_\ell$ during MCMC sampling.

Although our notation suggests that the same covariates \(\bsx\) are used in both the location and scale predictors, the two predictors can in fact be specified independently and tailored to the research question at hand. Possible specifications include, but are not limited to, using the same predictor structure for both components; including only an intercept in one or both predictors; employing different functional forms of the same covariate (e.g., a P-spline in the location predictor and a linear effect in the scale predictor); or including a given covariate in one predictor but not the other. The use of \(\bsx\) in both predictors is intended solely to achieve notational compactness.

\subsection{Extensions}\label{sec-ptm-extensions}

While this paper is focused on continuous responses, the general model setup
can be applied to other scenarios by adapting the likelihood.
Below, we provide outlines of such adaptations
to censored and count responses that we deem promising for further exploration.

\paragraph{Censored responses.}
Censoring of response observations
can be taken into account by adjusting \eqref{eq-response-density}. The
likelihood contributions for right-, left-, and interval-censored
continuous or discrete observations are given by \[
    \begin{aligned}
        F_Z\left( h\left(\frac{\overline{y} - \mu(\bsx)}{\sigma(\bsx)} \right) \right)                                                                                   & \quad  \text{for } y \in (-\infty, \overline{y})      & \quad \text{(left censored)}      \\
        1 - F_Z\left( h\left(\frac{\underline{y} - \mu(\bsx)}{\sigma(\bsx)} \right) \right)                                                                              & \quad \text{for } y \in (\underline{y}, \infty)       & \quad \text{(right censored)}     \\
        F_Z\left( h\left(\frac{\overline{y} - \mu(\bsx)}{\sigma(\bsx)} \right) \right) - F_Z\left( h\left(\frac{\underline{y} - \mu(\bsx)}{\sigma(\bsx)} \right) \right) & \quad \text{for } y \in (\underline{y}, \overline{y}) & \quad \text{(interval censored)}. \\
    \end{aligned}
\]

\paragraph{Count responses.}
PTMs can be applied to count
responses, explicitly taking zero-inflation into account with a
two-component model. A first component covers the probability of an
observation \(y = 0\) as a function of covariates using
\(\bbP(Y = 0 | \bsx) = F_Z\bigl(\eta_0(\bsx)\bigr)\), where
\(\eta_0(\bsx) = \alpha_0 + \sum_{k=1}^K f_{0,k}(\bsx)\) is a structured
additive predictor. A second component covers the nonzero observations
using
\(\bbP(Y \leq y | Y > 0, \bsx) = F_Z\bigl( h(r) \bigr)\),
where \(r = \bigl(\lfloor y \rfloor - \mu(\bsx)\bigr) / \sigma(\bsx)\) and $\lfloor \cdot \rfloor$ is the floor function. In
this case, \(\mu(\bsx)\) and \(\sigma(\bsx)\) represent the location and
scale of the nonzero component, respectively. Combining the two parts, a
two-component count-PTM can be set up as \[
    \bbP(Y \leq y | \bsx) = F_Z\bigl(\eta_0(\bsx)\bigr) + \mathbb{I}_{y > 0} \bigl[1 - F_Z\bigl(\eta_0(\bsx)\bigr)\bigr] F_Z\left( h\left(\frac{\lfloor y \rfloor - \mu(\bsx)}{\sigma(\bsx)}\right) \right),
\] where the indicator function \(\mathbb{I}_{y > 0} = 1\) if the
condition \(y > 0\) is true and \(\mathbb{I}_{y > 0} = 0\) otherwise.
Note that the first component does not require a transformation
function, and instead of $F_Z$, which implies a probit model if $F_Z$ is the
standard Gaussian CDF, other maps from $\bbR \to [0, 1]$ like the
logistic function can be chosen in this component, too.

\section{Posterior inference}\label{sec-posterior-inference}

\subsection{Posterior sampling}\label{posterior-sampling}

Assuming conditional independence between the model parameters, the
joint unnormalized posterior is given by
\begin{align}
    p(\bsbeta, \bsgamma, \tilde \bsdelta, \bstau^2 | \bsy) \propto
    \calL(\bsy | \bsbeta, \bsgamma, \tilde \bsdelta)
    \left[\prod_{\ell=1}^{L} \pi(\bsbeta_\ell| \tau^2_{\ell,\mu}) \pi(\tau^2_{\ell,\mu}) \right]
    \left[\prod_{s=1}^{S} \pi(\bsgamma_{s} | \tau^2_{s,\sigma}) \pi(\tau^2_{s,\sigma}) \right] \pi(\tilde \bsdelta|\tau^2_\delta) \pi(\tau^2_\delta),
\end{align}
where $\calL(\bsy | \bsbeta, \bsgamma, \tilde \bsdelta) = \prod_{i=1}^N f(y_i | \bsx_i, \bsbeta, \bsgamma, \tilde \bsdelta)$ denotes the likelihood with individual observations identified by the index \(i = 1, \dots, N\). The indices $\ell = 1, \dots, L$ and $s = 1, \dots, S$ identify the individual terms of the structured additive predictors in the location and scale model parts, respectively.
For MCMC sampling, we transform $\tau^2_\delta$ to the log-level and
evaluate its prior using the change of variables theorem, such that
$\ln \tau^2_\delta$ can be sampled on the whole real number line. We use a
Metropolis-within-Gibbs sampling scheme, where the full parameter vector is partitioned
into blocks, and each block is updated by drawing from its full conditional.
In the location and scale model parts, each term's parameter vector
$\bsbeta_\ell$ and $\bsgamma_s$ is treated as a block. Linear terms within each
parameter's predictor are grouped into a single block.
We use Metropolis-Hastings steps with iteratively weighted least squares (IWLS) proposals for
parameter blocks in $\bsbeta$, and $\bsgamma$,
Gibbs updates for $\tau^2_{\ell, \mu}$ and $\tau^2_{s, \sigma}$,
and the No-U-turn Sampler (NUTS), a variant of Hamiltonian Monte Carlo,
for $\tilde \bsdelta$ and $\ln \tau^2_\delta$.
Our implementation is done in Python,
building on the probabilistic programming framework Liesel \parencite{Riebl2023-LieselProbabilisticProgramming}.
Liesel provides a general NUTS interface via
BlackJAX \parencite{blackjax2020github}, as well as IWLS and Gibbs interfaces,
combined with automatic differentiation and just-in-time
compilation via JAX \parencite{deepmind2020jax}.
We treat the intercepts
$\beta_0$ and $\gamma_0$ in the covariate models as constants that are identified by the model
assumptions $\bbE(R)=0$ and $\bbVar(R)=1$ when the remaining parameters are held fixed.
They are updated deterministically in each iteration; details are provided in
\appautoref{app-seq:intercepts}.
We use a grid approximation to evaluate the basis matrix required in
the transformation function, see \appautoref{app-sec:basis-grid}.
We initialize the model parameters by finding approximations to their
posterior modes via gradient ascent, see \appautoref{app-sec:init} for details.
This initialization is essential for successful sampling. See \appautoref{app-sec:sim2-diagnostics} for a comparison of diagnostic criteria of a model initialized at the posterior modes and a model that included initialization with random jittering from a Gaussian distribution.
Algorithm~\ref{alg-sampling} gives an overview of the sampling algorithm's
overall structure.

We monitor standard MCMC diagnostics and have found that the target acceptance rates were
reached reliably, $\hat{R}$ indicated successful convergence, and effective sample sizes
were sufficient for posterior inference; \appautoref{app-sec:sim2-diagnostics} includes
the corresponding details for our main simulation study.
Below, we provide additional details on the individual
sampling steps.

\begin{algorithm}[bth]
    \DontPrintSemicolon
    \caption{Markov chain Monte Carlo inference for Bayesian penalized transformation models.\label{alg-sampling}}
    \vspace{0.5em}
    \begin{enumerate}[leftmargin=*,itemsep=0pt]
        \small
        \item Initialize the model parameters according to Algorithm \ref{app-alg-init} in \appautoref{app-sec:init}.
        \item Draw warmup samples; updating the step sizes for all samplers and the matrix $\bfM$ of the No-U-Turn Sampler as described in the text.
        \item Draw posterior samples via Markov chain Monte Carlo as follows. \\[0.5em] \For{MCMC iterations $t = 1, \dots, T$}{
                  \For{location terms $\ell = 1, \dots, L$}{
                      $\bsbeta_\ell^{[t]} \gets \bsbeta_\ell^{[t]*}$ drawn from $\calN(\bfm_{\ell}^{[t-1]}, \epsilon^2_\ell\bfF(\bsbeta_\ell^{[t-1]})^{-1})$ with acceptance probability $\alpha_{\ell}^{[t]}$, else $\bsbeta_\ell^{[t-1]}$.\;
                      $\tau_\ell^{2[t]} \gets \tau^{2*}_{\ell}$, drawn from $\calI \calG(\tilde a_{\ell}, \tilde b_{\ell})$\;
                  }
                  \For{scale terms $s = 1, \dots, S$}{
                      $\bsgamma_s^{[t]} \gets \bsgamma_s^{[t]*}$ drawn from $\calN(\bfm_{s}^{[t-1]}, \epsilon^2_s\bfF(\bsgamma_s^{[t-1]})^{-1})$ with acceptance probability $\alpha_{s}^{[t]}$, else $\bsgamma_s^{[t-1]}$.\;
                      $\tau_s^{2[t]} \gets \tau^{2*}_{s}$, drawn from $\calI \calG(\tilde a_{s}, \tilde b_{s})$\;
                  }
                  $\beta_0, \gamma_0 \gets$ updated using~\eqref{app-eq-intercepts} in \appautoref{app-seq:intercepts}.\;
                  $\tilde \bsdelta^{[t]}, \ln(\tau_\delta^{2})^{[t]} \gets$ $\tilde \bsdelta^{[t]*}, \ln(\tau_\delta^{2})^{[t]*}$ proposed via \text{NUTS} with acceptance probability $\tilde \alpha^{[t]}$, else $\tilde \bsdelta^{[t-1]}, \ln(\tau_\delta^{2})^{[t-1]}$\;
              }
    \end{enumerate}
\end{algorithm}

\paragraph{Iteratively Weighted Least Squares Proposals.}
Iteratively weighted least squares proposals are a common choice for Bayesian structured
additive distributional regression models; they were developed by \textcite{Gamerman1997-SamplingPosteriorDistributiona} and \textcite{Brezger2006-GeneralizedStructuredAdditivea}, and adapted for distributional regression by
\textcite{Klein2015-BayesianStructuredAdditiveb}. For a generic parameter block
$\bstheta$, they can be motivated by constructing a second-order Taylor approximation
to the parameter's log full conditional distribution around the
current value $\bstheta^{[t-1]}$.
In addition, we include a step size $\epsilon$ \parencite{Riebl2023-LieselProbabilisticProgramming}, inspired by the closely related Metropolis-adjusted Langevin algorithm \parencite{Girolami2011-RiemannManifoldLangevin}.
The proposal distribution is a Gaussian distribution $\calN(\bfm, \epsilon^2\bfF(\bstheta^{[t-1]})^{-1})$
with expectation $\bfm = \bstheta^{[t-1]} + \frac{\epsilon^2}{2}\bfF(\bstheta^{[t-1]})^{-1} s(\bstheta^{[t-1]})$
and covariance $\epsilon^2\bfF(\bstheta^{[t-1]})^{-1}$,
where $\bfF(\bstheta)$ and $s(\bstheta)$ denote the negative Hessian
(i.e., the information matrix)
and the gradient of the log full conditional evaluated at $\bstheta$, respectively.
A proposal $\bstheta^*$ is accepted in a Metropolis-Hastings acceptance step
with probability
\begin{equation}
    \alpha = \min\left\{1,
    \frac{p(\bstheta^* | \cdot)}{p(\bstheta^{[t-1]} | \cdot)}
    \frac{q(\bstheta^{[t-1]} | \bstheta^*)}{q(\bstheta^* | \bstheta^{[t-1]})}
    \right\},
\end{equation}
where $p(\bstheta | \cdot)$ denotes the full conditional density and
$q(\bstheta^* | \bstheta^{[t-1]})$ denotes the density of the Gaussian proposal
distribution given the current $\bstheta^{[t-1]}$, evaluated at the proposed value
$\bstheta^*$.
We evaluate the derivatives using automatic differentiation via JAX \parencite{jax2018github}.

\textcite{Klein2015-BayesianStructuredAdditiveb} found that using the expectation of
the negative Hessian $\bbE(\bfF(\bstheta^{[t-1]}))$
in $\bfF(\bstheta^{[t-1]})^{-1}$ improved sampling efficiency
in distributional regression models.
For $\bstheta = \bsbeta_\ell$ and $\bstheta = \bsgamma_s$, we can approximate
$\bbE(\bfF(\cdot))$ by the expected information of the log full
conditionals derived under the assumption of a Gaussian working model; see
\appautorefsec{app-sec:inference} for the details.
This choice is motivated as follows.
Because the terms in $\mu(\bsx)$ and $\sigma(\bsx)$ primarily shift and rescale the response, their posterior geometry is close to Gaussian regression. Using the Gaussian expected information hence gives a stable, computationally cheap approximation to the local curvature.
Crucially, we still accept/reject using the actual non-Gaussian likelihood,
so the target posterior is exact; the Gaussian working model is only used to
generate the proposal.

Similar to the procedure in the No-U-Turn Sampler described below,
we find suitable values for the step size $\epsilon$ in a warmup phase,
where $\epsilon$ is updated to reach a user-defined target acceptance probability
using the dual averaging algorithm
\parencite{Nesterov2009-PrimaldualSubgradientMethods,Hoffman2014-NoUTurnSamplerAdaptively}.
We found that a target acceptance probability of $.5$ yielded satisfying results.

\paragraph{Gibbs updates.} For each $\tau^2_{\ell}$, the full conditional distribution is available in closed form as
$\tau^2_{\ell} \sim \calI \calG(\tilde a_{\ell}, \tilde b_{\ell})$, where $\tilde a_{\ell} = a + \frac{1}{2} \operatorname{rk}(\bfK_\ell)$ and $\tilde b_{\ell} = b + \frac{1}{2}(\bsbeta_\ell^\sfT \bfK_\ell \bsbeta_\ell)$, a common result in Bayesian additive models \parencite[see, for example,][]{Umlauf2018-BamlssBayesianAdditive}. Analogous Gibbs updates are available for $\tau^2_{s}$. If different priors are assigned for $\tau^2_{\ell}$ or $\tau^2_{s}$ --- for example, scale-dependent penalized complexity priors as defined for $\tau^2_\delta$ --- they can be log-transformed and sampled by Hamiltonian Monte Carlo.

\paragraph{Hamiltonian Monte Carlo.}
For parameters in $\tilde \bsdelta$, the posterior geometry can include
tight curvature, flat regions where the
likelihood is weakly identified, and strong correlations.
We found observed-information based IWLS
proposals to be numerically unstable, including near-singular Hessians and
frequent step-rejections, and we found that using a fixed observed information
(e.g., at a mode), is fragile and can be inefficient away from that point.
We therefore update $\tilde \bsdelta$ using Hamiltonian Monte Carlo (HMC),
which uses gradients but does not require repeated Hessian evaluations.
Hamiltonian Monte Carlo is known to explore curved posteriors effectively.
It treats the gradient
of the negative log posterior as a high-dimensional surface and simulates the movement
of a mass-free particle on this surface via Hamiltonian dynamics \parencite{Neal2011-MCMCUsingHamiltonian,Betancourt2018-ConceptualIntroductionHamiltonian}.
An introduction tailored towards statisticians is given by \textcite{Thomas2020-LearningHamiltonianMonte}.
HMC augments a generic parameter $\bstheta$ with an auxiliary momentum
variable $\bsphi$ of matching dimension and constructs the Hamiltonian $H(\bstheta, \bsphi) = U(\bstheta) + K(\bsphi)$,
where the \textit{potential energy} is given by the negative log posterior,
$U(\bstheta) = - \ln p(\bstheta|\bsy)$. For MCMC sampling, the momentum is
randomly drawn from a multivariate Gaussian distribution $\bsphi \sim \calN(\bfzero, \bfM)$,
such that the \textit{kinetic energy} is $K(\bsphi) = \frac{1}{2}\bsphi^\sfT \bfM^{-1} \bsphi$.
The momentum $\bsphi$ can be understood as a random initial push applied to $\bstheta$ in the physical simulation.
The Hamiltonian equations then describe the movement of $\bstheta$ over time:
\begin{align}
    \frac{d \bsphi}{dt}
    = & - \frac{\partial H(\bstheta, \bsphi)}{\partial \bstheta}
    =  - \frac{\partial U(\bstheta)}{\partial \bstheta}
    =  \nabla_{\bstheta} \ln p(\bstheta),                        \\
    \frac{d \bstheta}{dt}
    = & \frac{\partial H(\bstheta, \bsphi)}{\partial \bsphi}
    =  \frac{\partial K(\bsphi)}{\partial \bsphi}
    =  \bfM^{-1}\bsphi.
\end{align}
These differential equations are approximated using the leapfrog algorithm
\parencite{Ruth1983-CanonicalIntegrationTechnique}, which breaks the solution down into
$L$ discrete steps of size $\epsilon$. The leapfrog algorithm requires repeated
evaluations of $\nabla_{\bstheta} \ln p(\bstheta)$, for which we employ automatic
differentiation via JAX \parencite{deepmind2020jax}.
It generates a proposal $\bstheta^*$, which is accepted with probability
\begin{equation}
    \tilde \alpha = \min\left\{1,
    \frac{p(\bstheta^*)q(\bsphi^*)}{p(\bstheta^{[t-1]})q(\bsphi^{[t-1]})},
    \right\},
\end{equation}
where $q$ denotes the momentum density and $t = 1, \dots, T$ identifies the current
iteration.
The
No-U-Turn Sampler \parencite[NUTS,][]{Hoffman2014-NoUTurnSamplerAdaptively} is a
variant of HMC that automatically determines an appropriate number of steps $L$, which
can significantly improve the efficiency of the sampler by producing samples with
minimal autocorrelation. We find suitable values for the hyperparameters $\epsilon$ and $\bfM$ in a warmup phase, where $\epsilon$ is updated to reach a user-defined target acceptance probability using the dual averaging algorithm \parencite{Nesterov2009-PrimaldualSubgradientMethods,Hoffman2014-NoUTurnSamplerAdaptively}, and the empirical covariance of the warmup samples for $\bstheta$ is used for $\bfM$, following the recommendation in \textcite{Betancourt2018-ConceptualIntroductionHamiltonian}. The warmup phase is automatically controlled by Liesel \parencite{Riebl2023-LieselProbabilisticProgramming}. We found that a target acceptance probability of $.9$ yielded satisfying results. We initialize $\bfM$ to the identity matrix and find an initial value for $\epsilon$ algorithmically according to Algorithm 4 in \textcite{Hoffman2014-NoUTurnSamplerAdaptively} as implemented in Liesel. For our Metropolis-within-Gibbs scheme, we use the full conditional
distribution $p(\bstheta|\cdot)$ of the current parameter block $\bstheta$ given all
remaining parameters instead of the full joint posterior $p(\bstheta)$.

\subsection{Posterior estimates}\label{posterior-estimates}

Bayesian inference gives us access not only to the posterior
distribution of the directly sampled parameters, but also of derived
quantities. Specifically, by propagating the posterior samples through
the model, we can obtain posterior predictive distributions of the
response's conditional CDF, PDF, and quantiles. For each posterior sample indexed by \(t = 1, \dots, T\), an estimate
of the conditional cumulative distribution given a vector of covariate
observations \(\bsx^*\) can be obtained as
\[
    \hat{F}^{[t]}_Y(y | \bsx^*) = F_Z\left( h\left(\frac{y - \mu\bigl(\bsx^* | \bsbeta^{[t]}\bigr)}{\sigma\bigl(\bsx^* | \bsgamma^{[t]}\bigr)}\ \middle| \bsdelta^{[t]} \right) \right).
\]
These samples
\(\hat{F}_Y^{[1]}(y | \bsx^*), \dots, \hat{F}_Y^{[T]}(y | \bsx^*)\) can
be used to derive summary statistics like posterior quantiles or highest
posterior density intervals of the posterior CDF, or the posterior mean
\(\hat{F}_Y(y | \bsx^*) = \frac{1}{T}\sum_{t=1}^T \hat{F}_Y^{[t]}(y | \bsx^*)\)
as a point estimate. Similarly, we can obtain the posterior
samples of the predictive conditional density
\(\hat{f}_Y^{[1]}(y | \bsx^*), \dots, \hat{f}_Y^{[T]}(y | \bsx^*)\), and samples of conditional quantiles of the response distribution can be obtained by numerical inversion of the transformation function, i.e.
$$
    y^{*[t]} = \sigma\bigl(\bsx^* | \bsgamma^{[t]}\bigr) \cdot h^{-1}\bigl( F_Z^{-1}(u) |  \bsdelta^{[t]}\bigr) + \mu\bigl(\bsx^* | \bsbeta^{[t]}\bigr),
$$
where $u \in (0, 1)$ is the desired probability level.

\section{Simulation study}\label{sec-simulation}

We conduct simulations to compare the performance of penalized
transformation models with alternative models.

\paragraph{Data generation and scenarios.}
We generate data from the model $y_i = \mu(\bsx_i) + \sigma(\bsx_i) R_i$ for
\(i = 1, \dots, N_{\text{obs}}\). The covariate effects are defined as
$\mu(\bsx) = s_1(x_1) + s_2(x_2) + s_3(x_3) + s_4(x_4)$, and
$\ln \sigma(\bsx) = 0.1\mu(\bsx)$.
All covariates are generated as i.i.d. realizations from
\(\calU(-2, 2)\). The four functions consist of a linear function \(s_1(x) = x\), a u-shaped
function with a linear trend \(s_2(x) = x + \frac{(2x)^2}{5.5}\), an
oscillating function with a negative trend
\(s_3(x) = -x + \pi \sin(\pi x)\), and a bell-shaped function with a
small positive trend \(s_4(x) = 0.5x + 15 \phi(2(x-0.2)) - \phi(x + 0.4)\).
All four functions lie on similar scales. The functions are included in \autoref{fig:sim2-fitted-cov}.
$R_i$ is drawn from a Gaussian, a skewed Gaussian, a bimodal mixture
of Gaussians, and from
random PTM distributions via $R_i = h^{-1}(Z_i | \bsdelta)$
with $Z_i \sim \calN(0, 1)$ and $\bsdelta$ drawn from its prior. \appautorefsec{app-sec:data}
gives details on all data-generating mechanisms.

We used sample sizes
\(N_{\text{obs}} \in \{250, 500, 1\,000\}\) and
\(N_{\text{sim}} = 100\) replications for all data-generating mechanisms, leading to a
total of $1\,200$ scenarios. Note that, in the PTM data generation,
a new sample of \(\bsdelta\) is drawn in every replication, such that each
replication uses a different true density.

\paragraph{Tested models.}
We fit the simulated data using the following models:
\begin{itemize}
    \tightlist
    \item PTM: A penalized transformation model with P-splines for all four covariates
          in both the location and scale predictors. We use $J-1=30$ and $\tau^2 \sim \operatorname{Weibull}(0.5, \psi)$ with $\psi=0.5$, and set $a=-4$ and $b=4$.
    \item Gaussian: A corresponding Gaussian location-scale model with P-splines for all four covariates
          in both the location and scale predictors.
    \item TAMLS: A transformation additive model with P-splines for all four covariates in both the location and scale predictors \parencite{Siegfried2023-DistributionfreeLocationscaleRegression}.
    \item SBGP: A semiparametric Bayesian Gaussian process model as proposed by \textcite{Kowal2024-MonteCarloInference}. The authors also describe a linear and a linear quantile regression version of their model. We select the Gaussian process model, since the simulation study is targeted at nonlinear covariate effects.
    \item DDPstar: A density regression via Dirichlet process mixtures of normal structured additive regression models \parencite{Rodriguez-Alvarez2024-DensityRegressionDirichleta}, as available in the R package \texttt{DDPstar}.
    \item BCTM-LS: A Bayesian conditional transformation model \parencite{Carlan2024-BayesianConditionalTransformation} in location-shift form, as available in the Python library \texttt{liesel-bctm} \parencite{Brachem2023-LieselbctmBayesianConditional}.
    \item BCTM-TE: A full Bayesian conditional transformation model with tensor product terms for all four covariates.
    \item QGAM: A fully specified additive quantile regression model with P-splines for all four covariates, as available in the R package \texttt{qgam} \parencite{Fasiolo2021-FastCalibratedAdditive,Fasiolo2021-QgamBayesianNonparametric}. We fit QGAM for a grid of 25 evenly spaced probability levels in $\{0.005, \dots, 0.995\}$.
\end{itemize}
Additional details on model specifications are included in
\appautoref{app-sec:sim2-models}.

\paragraph{Performance measures.}
We evaluate predictive performance using the Kullback-Leibler
divergence $\mathrm{KL}(F_{\text{true}} \| \hat F)$ from the true to the estimated distribution,
the mean absolute difference (MAD) between the true and the estimated CDF, and the continuous ranked probability score \parencite[CRPS,][]{Gneiting2007-StrictlyProperScoring}.
Additionally, we include the Watanabe-Akaike Information Criterion \parencite[WAIC,][]{Watanabe2010-AsymptoticEquivalenceBayes,Vehtari2017-PracticalBayesianModel} where applicable to validate its function as a model selection tool.
We also record the coverage rate of a 90\% credible or confidence
interval for the CDF, and the width of the interval.
For the covariate functions, we use the mean squared error (MSE),
the coverage rate of a 90\% credible or confidence
interval, and the width of the interval. The performance criteria
are estimated using $N_{\text{test}}=5\,000$ test observations in each scenario.
We provide details on the computations carried out in \appautoref{app-sec:sim2-performance}.
Note that not all models allow for the computation of all performance criteria.

\subsection{Results}

\begin{figure}
    \includegraphics[width=\textwidth]{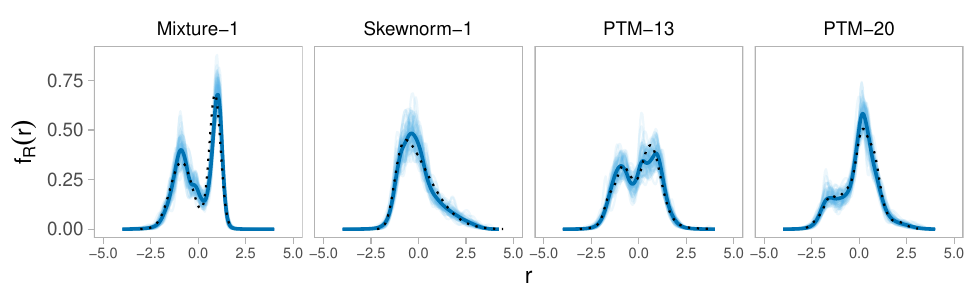}
    \caption{Examples for fitted standardized conditional densities $f_R(r)$ via PTM in the simulation study. The dotted black line shows the true data-generating density for reference. The shaded area shows a $90\%$ pointwise posterior credible band. The thick blue line shows the posterior mean. Thin, transparent blue lines show 50 random posterior samples. The data show estimates based on $500$ training observations. The panel titles indicate the data type and the data generation seed.\label{fig:sim2-fitted-dens}}
\end{figure}

\autoref{tab:sim-dist} shows results for distributional predictive
performance.
For a concise presentation, the results are averaged over all data types, seeds, and sample sizes. We include disaggregated results
in \appautoref{app-sec:sim2-results}; they do not lead to different conclusions. We can see that the PTM obtains the lowest
Kullback-Leibler divergence, the lowest MAD and the second-lowest CRPS. The WAIC is consistent with these results, confirming its adequacy for model selection. At 91\%, the
coverage rate for the PTM's credible intervals for the conditional CDF is very close to
the nominal level, and the average width of the interval is the second-lowest. The uncompetitive results for SBGP in all performance criteria are striking, 
but this is most likely due to a mismatch between the model formulation and the data-generating mechanism used in our simulation study. In the application examples, SBGP performs much more similar to the other included models. \autoref{fig:sim2-fitted-dens} shows four examples for fitted densities. \appautoref{app-tab:sim2-gauss} adds a detailed comparison of the PTM and the Gaussian model for the Gaussian data scenario: the results of both models are equivalent, confirming that the PTM can operate as a drop-in replacement for a Gaussian location-scale model.

\begin{table}
    \caption{Predictive performance and CDF coverage for the tested models. The results are averaged over all seeds, data types, and sample sizes. For the KLD, MAD, and CDF, smaller values indicate a better predictive performance. Coverages are reported for 90\% credible intervals.\label{tab:sim-dist}}
    \small
\begin{tabu} to \linewidth {>{\raggedleft}X>{\centering}X>{\centering}X>{\centering}X>{\centering}X>{\centering}X>{\raggedleft}X}
\toprule
\multicolumn{5}{c}{ } & \multicolumn{2}{c}{CDF CI} \\
\cmidrule(l{3pt}r{3pt}){6-7}
Model & KLD $\downarrow$ & CRPS $\downarrow$ & WAIC $\downarrow$ & MAD $\downarrow$ & Coverage & Width\\
\midrule
PTM & 0.075 & 0.819 & 1897 & 0.081 & 0.912 & 0.238\\
Gaussian & 0.157 & 0.833 & 2002 & 0.093 & 0.848 & 0.254\\
DDPstar & 0.193 & 0.839 & 2024 & 0.094 & 0.791 & 0.243\\
BCTM-TE & 0.221 & 0.864 & 2078 & 0.109 & 0.650 & 0.225\\
BCTM-LS & 0.228 & 0.862 & 2097 & 0.108 & 0.681 & 0.236\\
TAMLS & 0.236 & 0.841 & -$^a$ & 0.095 & 0.671 & 0.249\\
SBGP$^b$ & 1.556 & 1.392 & 3365 & 0.234 & 0.290 & 0.210\\
QGAM & -$^c$ & 0.801 & -$^c$ & -$^c$ & -$^c$ & -$^c$\\
\bottomrule
\multicolumn{7}{p{\linewidth}}{%
    \scriptsize 
    $a$: Bayesian criterion; not available for the frequentist TAMLS.\newline
    $b$: For SBGP, 197 seeds (roughly 20 seeds per condition) produced outliers that highly skewed the performance criteria towards worse performance. Here, we show the results including only the 1003 non-outlier seeds. \newline
    $c$: These measures are not available for quantile regression.
    } \\
\end{tabu}

\end{table}

\begin{figure}[tb]
    \includegraphics[width=\textwidth]{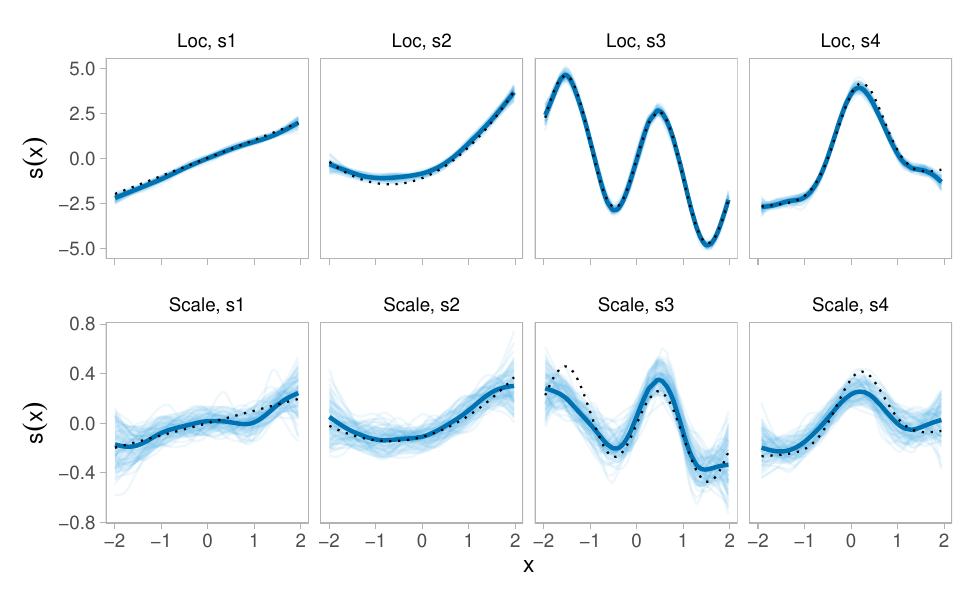}
    \caption{Fitted covariate functions via PTM in the simulation study. The dotted black line shows the true functions for reference. The shaded area shows a $90\%$ pointwise posterior credible band. The thick blue line shows the posterior mean. Thin, transparent blue lines show 50 random posterior samples. The data show estimates based on $500$ training observations from the PTM data generation scenario.\label{fig:sim2-fitted-cov}}
\end{figure}

\autoref{tab:sim-cov} shows results for the evaluation of covariate function recovery, similarly aggregated for a concise presentation. The table includes only the PTM and Gaussian model, since they allow for an evaluation of covariate functions that corresponds to the data-generating process.
We include disaggregated results and the other sample sizes
in \appautoref{app-sec:sim2-results}; they do not lead to different conclusions.
The PTM achieves the smallest MSE for both model parts. Coverage for the location and scale terms is close to nominal. For the scale terms, the average interval width is smaller than for the
Gaussian model.
For the location terms, the interval width yielded by the PTM is the smallest. \autoref{fig:sim2-fitted-cov} shows fitted covariate functions for one dataset of the PTM data generation scenario based on $500$ training observations. The plots suggest that the location functions are easily recovered, while the estimation of the scale functions is more challenging; a plausible result given the moderately small sample size. \appautoref{app-sec:sim2-diagnostics} includes diagnostic information for the PTM fits.

Overall, the results confirm that PTMs are effective for both recovering diverse
conditional distributions and for recovering nonlinear
covariate effects on the location and scale of such distributions. The results also
confirm that PTMs yield well-calibrated uncertainty quantification for both the
conditional CDF and for the covariate effects. The comparison to the Gaussian baseline
is of particular interest: As the results in \autoref{tab:sim-cov} show, the PTMs
match the performance of the Gaussian model for the covariate effects, while the
results in \autoref{tab:sim-dist} show that the PTMs additionally capture the shape of
the response's conditional distribution, even including well-calibrated uncertainty
quantification for the CDF. PTMs can thus be thought of, and used as, a
drop-in replacement for structured additive Gaussian location-scale models
that retains the Gaussian model's interpretability and predictor structure
while easily including additional information about potential non-Gaussianity of the response's conditional distribution. Note also that, in the case of Gaussian
data, the regularizing prior of PTMs will allow the model to reduce to the Gaussian,
as illustrated in \autoref{fig:sim1-kld} and further substantiated in the additional
results included in \appautoref{app-sec:sim2-results}.

\begin{table}
    \caption{Results for covariate effects. The results are averaged over all seeds, data types, sample sizes, and over all four functions for the location and scale model parts, respectively. Coverages are reported for 90\% credible intervals.\label{tab:sim-cov}}
    \small
    
\begin{tabu} to \linewidth {>{\raggedleft}X>{\centering}X>{\centering}X>{\centering}X>{\centering}X>{\centering}X>{\centering}X}
\toprule
\multicolumn{1}{c}{ } & \multicolumn{3}{c}{Location Terms} & \multicolumn{3}{c}{Scale Terms} \\
\cmidrule(l{3pt}r{3pt}){2-4} \cmidrule(l{3pt}r{3pt}){5-7}
Model & MSE $\downarrow$ & Coverage & Width & MSE $\downarrow$ & Coverage & Width\\
\midrule
PTM & 0.042 & 0.917 & 0.429 & 0.014 & 0.905 & 0.251\\
Gaussian & 0.055 & 0.924 & 0.504 & 0.016 & 0.914 & 0.276\\
\bottomrule
\end{tabu}
\end{table}

\subsection{Scaling to higher sample sizes}

\begin{figure}[bth]
    \includegraphics[width=\textwidth]{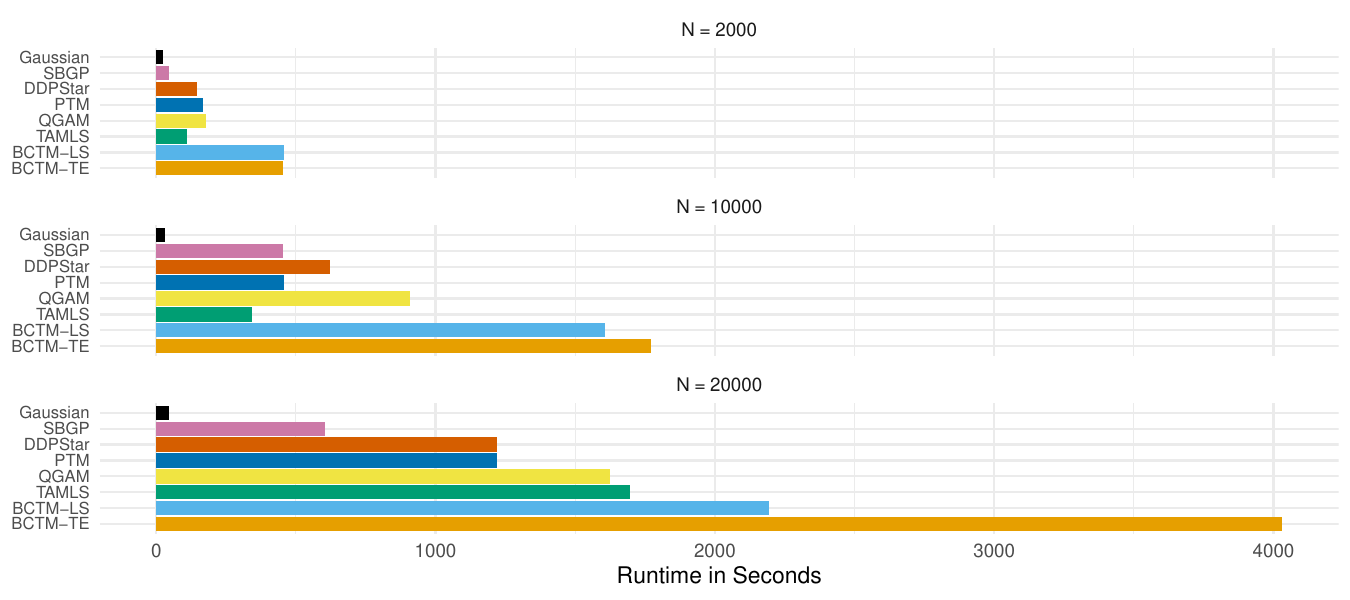}
    \caption{Runtime in seconds for different sample sizes for the eight model configurations considered in the simulation study. All runtimes are reported from one model run on data from the bimodal mixture of Gaussians scenario.
        The MCMC methods Gaussian, PTM, BCTM-LS, and BCTM-TE use $1\,000$ warmup samples and $2\,000$ posterior samples in four chains each, totaling $8\,000$ posterior samples. DDPstar does not offer sampling in parallel chains, and so simply uses $1\,000$ burnin samples and $2\,000$ posterior samples. SBGP uses $2\,000$ Monte Carlo samples. QGAM was run until convergence for the set of 25 probability levels considered in the simulation study. TAMLS was run until convergence.\label{fig:scaling}}
\end{figure}

We conducted individual model runs on data from the bimodal mixture of Gaussians scenario with sample sizes $2\,000$, $10\,000$ and $20\,000$ on a 2021 M1 MacBook Pro with 32 GB of RAM and 10 CPUs. The results are displayed in \autoref{fig:scaling}.
The Gaussian model scales best out of all tested models by a large margin, finishing four chains of $2\,000$ posterior iterations in about 45 seconds. Among the other models, PTMs take a middle place with runtimes of 167, 457, and $1\,220$ seconds, respectively (roughly 3, 10, and 20 minutes). The global minimum bulk effective sample sizes for the PTM over all parameters are 161, 242 and 251 samples; the average minimum effective sample sizes (i.e., taking the minimum per parameter block, then averaging over the minima) are 610, $1\,195$ and $1\,344$. Predictive performance of the PTM improves with rising sample size as expected. For example, the Kullback-Leibler divergences to the true distribution recorded in the three scenarios are $0.0225$, $0.0063$, and $0.0037$.

These results suggest that scaling the PTM to moderate sample sizes is generally feasible, but scaling to high sample sizes beyond $20\,000$ observations is likely to be challenging. Based on our observations,
the bottleneck for computation time lies in potentially high numbers of
gradient evaluations in the NUTS proposal for $\tilde \bsdelta$. In some cases,
speed can be improved by instead using ordinary Hamiltonian Monte Carlo with
fixed number of integration steps $L < 100$.

\section{Applications}\label{sec-applications}

\subsection{Fourth Dutch Growth Study}\label{sec-db}

For a first illustration, we use a dataset consisting of cross-sectional measurements
of the body mass index (BMI) and age of Dutch children and adolescents aged 0
to 21 years. The data come from the Fourth Dutch Growth Study
\parencite{Fredriks2000-BodyIndexMeasurements,Fredriks2000-ContinuingPositiveSecular}. We use a subset of $7\,294$
observations provided in the R package \texttt{gamlss.data} \parencite{Stasinopoulos2021-GamlssdataDataGeneralised}.
We set up location-scale penalized transformation models (PTM)
$\text{bmi}_i = \mu(\text{age}_i) + \sigma(\text{age}_i) R_i$,
where \(i = 1, \dots, 7\,294\) with a standard Gaussian reference
distribution. The location and scale functions \(\mu(\text{age}_i)\) and
\(\sigma(\text{age}_i)\) are modelled using Bayesian P-splines with
\(20\) parameters each and include intercepts. For \(\sigma(\text{age}_i)\), we use a
logarithmic link function to ensure positivity.
We vary the knot boundaries $a$ and $b$
and the number of parameters in $\bsdelta$ to correspond to changes in the width of the interval $[a,b]$; a wider interval warrants more parameters. We consider two basic model
variants: 1) $a = -b = 4$ and $J-1 = 20$, 2) $a = -4$, $b = 7$, $J-1 = 30$. Each
of these variants is fitted with
$\lambda \rightarrow \infty$ for direct linear extrapolation of the transformation
function and with $\lambda = 0.1(b-a)$ for a smooth transition to unit slope, where a wider
core interval is associated with a wider transition interval. This leads to a total of
four PTM variants. All variants use $\tau^2_\delta \sim \operatorname{Weibull}(0.5, \psi)$ with $\psi=0.5$.

\begin{figure}[bt]
    \includegraphics[width=\textwidth]{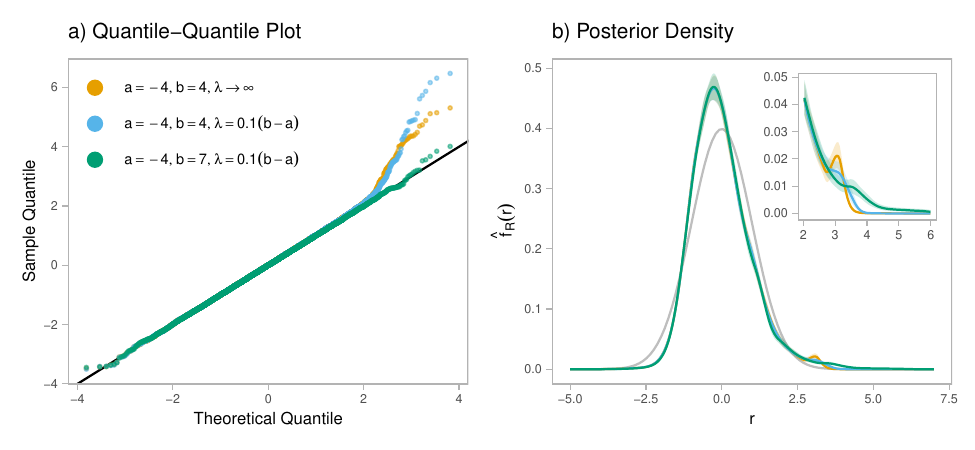}
    \includegraphics[width=\textwidth]{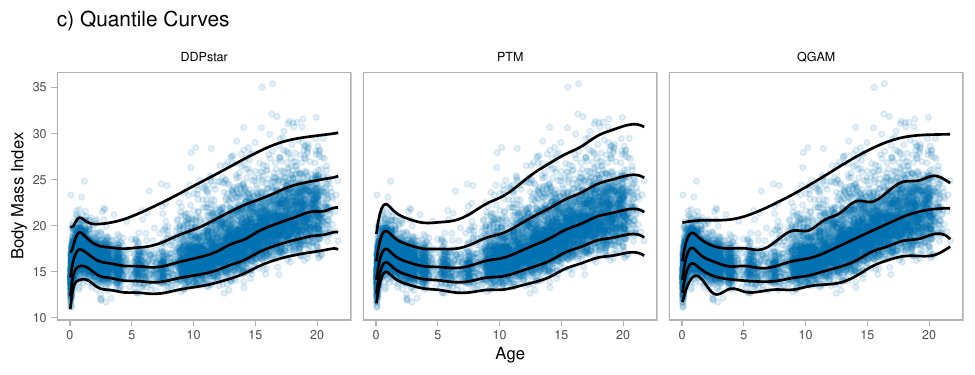}
    \caption{
        Panel a): Quantile-quantile plots for three PTM variants. A black line shows the identity function for reference. Panel b): Mean posterior density of $R$ for the same three PTM variants. The shaded areas indicate 90\% posterior credible bands. The inset provides a closer look at the right tails.
        The gray line is a standard Gaussian density for reference. Panel c): Quantile curves for DDPstar, PTM, and QGAM at probability levels $\{0.01, 0.1, 0.5, 0.9, 0.99\}$.\label{fig:db-summary}
    }
\end{figure}

\autoref{fig:db-summary} shows quantile-quantile plots (panel a)
and the mean posterior estimate for the density of $R$ (panel b) for three of these
models. The quantile-quantile plots compare the distribution of the posterior mean
transformed responses
$\hat{z}_i =T^{-1} \sum_{t=1}^T h(\sigma^{[t]}(\text{age}_i)^{-1} (\text{bmi}_i - \mu^{[t]}(\text{age}_i)))$
to the theoretical quantiles of a standard Gaussian distribution. The model construction
implies that the distribution of the transformed values should correspond to the
standard Gaussian, so the QQ-plots can be used for model diagnosis. Here, the plots
indicate a misfit in the right tail for both model variants that use $b=4$, while the
fit using $a = -4$, $b=7$ and $\lambda = 0.1(b-a)$ appears to be reasonable; the
fit of the model with $a = -4$, $b=7$ and $\lambda \to \infty$ is omitted from
the plot because it looks identical. Panel b) shows the differences in the right tail
of the $R$ distributions estimated by the three models. Consistent with the diagnostic information in the QQ-plots, we found that the PTM with $a = -4$, $b=7$ and $\lambda = 0.1(b-a)$ shows
better predictive performance in terms of log score and continuous ranked probability score than the variants using $b=4$.

\autoref{tab:app-scores} shows a comparison of the best PTM's predictive performance to
all competing methods included in the simulation study using log scores and the
continuous ranked probability score based on 10-fold cross-validation. In terms of log
score, the PTM achieves a second place, closely behind DDPstar, while the CRPS favors
most other models, except for the Gaussian and SBGP, over the PTM. This discrepancy
could be due to different emphasis placed on distributional characteristics by log scores
and CRPS.

\autoref{fig:db-summary},
panel c) shows quantile curves for probability levels $\{0.01, 0.1, 0.5, 0.9, 0.99\}$
using the full dataset for the PTM and the top contenders in terms of log score (DDPstar) and CRPS (QGAM).
\appautorefsec{app-application-dutch-boys} includes similar quantile curve plots for all
methods, as well as details on their model specifications.
Compared to the predicted $0.01$ and $0.9$ quantiles obtained with QGAM, the estimates yielded by DDPstar and PTM appear to be more smooth. All models appear to reasonably capture the
overall pattern of the data, which involves a quick rise of BMI in the first 1-2 years
of life, followed by a dip lasting to about age seven, and then a rise and increased
spread until the maximum age in the sample. A peak at about age one seems to be
attenuated in the top quantile in DDPstar, lost in QGAM, but preserved in PTM,
while a dip in the lowest quantile at about age three is lost in DDPstar and PTM but
captured in QGAM.

\begin{table}[bt]
    \small
    \caption{Comparison of predictive performance in terms of log score and continuous ranked probability score (CRPS) based on 10-fold cross validation. Smaller values indicate a better performance. The best performance for each measure and dataset is marked in bold print. \label{tab:app-scores}}

\begin{tabu} to \linewidth {>{\raggedleft\arraybackslash}p{9em}>{\centering}X>{\centering}X>{\centering}X>{\centering}X>{\centering}X}
\toprule
\multicolumn{1}{c}{ } & \multicolumn{2}{c}{4th Dutch Growth Study} & \multicolumn{3}{c}{Framingham Heart Study} \\
\cmidrule(l{3pt}r{3pt}){2-3} \cmidrule(l{3pt}r{3pt}){4-6}
\multicolumn{3}{c}{ } & \multicolumn{2}{c}{Without RI} & \multicolumn{1}{c}{With RI} \\
\cmidrule(l{3pt}r{3pt}){4-5} \cmidrule(l{3pt}r{3pt}){6-6}
Model & Log Score & CRPS & Log Score & CRPS & CRPS\\
\midrule
PTM (nonlin) & 688.1 & 0.372 & \textbf{138.6} & 0.527 & 0.534\\
PTM (lin) & -$^a$ & -$^a$ & 139.0 & 0.528 & 0.537\\
\addlinespace
Gaussian (nonlin) & 731.9 & 0.375 & 142.7 & 0.529 & 0.534\\
Gaussian (lin) & -$^a$ & -$^a$ & 142.5 & 0.530 & 0.536\\
\addlinespace
BCTM & 699.1 & 0.365 & 138.7 & 0.528 & 0.531\\
DDPstar & \textbf{687.0} & 0.361 & 140.3 & 0.515 & -$^c$\\
SBGP & 723.7 & 0.378 & 146.5 & 0.563 & -$^c$\\
QGAM & -$^b$ & \textbf{0.360} & -$^b$ & 0.511 & -$^c$\\
TAMLS & 702.2 & 0.362 & 139.4 & \textbf{0.510} & 0.530\\
\bottomrule
\multicolumn{6}{p{\linewidth}}{%
    \scriptsize 
    $a$: Linear models were not run for the 4$^{th}$ Dutch Growth Study data.\newline
    $b$: Log score is not available for quantile regression.\newline
    $c$: Random intercept models were not run for DDPstar, SBGP, and QGAM.
    } \\
\end{tabu}
\end{table}

An additional feature of the data is summarized by \autoref{fig:db-summary} panel b), which
shows the standardized conditional density of BMI obtained in the PTM. It reveals a light left and heavy right tail. While the quantile curves and performance criteria indicate
that both DDPstar and QGAM capture similar features of the conditional distribution,
we view it as a benefit of PTMs that the standardized conditional density can be
easily investigated to aid interpretation.

\subsection{Framingham Heart Study}\label{sec-framingham}

As a second application illustration, we now turn to the Framingham Heart Study \parencite{Zhang2001-LinearMixedModels}. The data comprise a total of
$1\,044$ cholesterol level measurements taken of 200 study participants over
the course of up to ten years. For 133 participants, there are six data
points, while for the remaining 67 participants we have between one and
five observations each. The data further contains participants'
\emph{sex} (binary, with \(0= \text{female}\) and \(1= \text{male}\)),
their \emph{age} in years at study onset and the \emph{years} between
the start of the study and the current measurement. For modeling, we add
the age and year variables to obtain a single \textit{age} variable that indicates
the participant's actual age at the time of each measurement. We fit
penalized transformation models (PTMs) as variants of the model
\begin{align}
    \text{cholst}_{ij}       & = \mu_{ij} + \sigma_{ij} R_{ij}                                     \\
    \mu_{ij}                 & = \beta_0 + s(\text{age}_{ij}) + \beta_1\text{sex}_{i}  + \zeta_{i} \\
    \ln \sigma_{ij}          & = \gamma_0 + g(\text{age}_{ij}) + \gamma_1\text{sex}_{i}            \\
    \bbP(R_{ij} \leq r_{ij}) & = \Phi(h(r_{ij} | \bsdelta)),
\end{align}
where the index \(i = 1, \dots, 200\) identifies the participant and
\(j = 1, \dots, n_i\) indicates the individual observations available
for participant \(i\). We use $a = -7$,  $b = 7$ and $J-1 = 40$ parameters
in $\bsdelta$, providing a wide knot base with a moderate number of parameters. As hyperprior for $\tau^2_\delta$, we use  $\tau^2_\delta \sim \operatorname{Weibull}(0.5, \psi_\delta)$ with $\psi_\delta=0.5$. For the functions $s$ and $g$, we use Bayesian P-splines with 20 parameters. On the location,
we use an i.i.d. random intercept $\zeta_i \sim \calN(0, \psi^2_\zeta)$. The model variants
include models that omit the random intercept and models that use linear effects of age
in place of $s$ and $g$. While it would be possible to also include a random intercept
in the scale predictor $\ln \sigma_{ij}$, exploratory model runs indicated that this
added complexity provides no practical benefit.

\autoref{tab:app-scores} shows a comparison of the best PTM's predictive performance to
all competing methods included in the simulation study using log scores and the
continuous ranked probability score based on 10-fold cross-validation, leaving out
10\% of participants in each fold, such that we evaluate out-of-sample prediction.
In the random intercept models, the predictive CRPS was obtained for marginal
predictions, drawing samples of $\zeta_i \sim \calN(0, \psi^2)$ using the estimated
random effect variance $\psi^2$ or draws from its posterior distribution, if applicable,
for the clusters in the test sets. The CRPS indicates that all random intercept models
do slightly worse for marginal predictions than their respective counterparts without
random intercepts. The nonlinear PTM without random intercept yields the best predictive
log score, albeit closely followed by the BCTM. The
linear PTM takes the third place. The CRPS again favors different models, in this case
TAMLS, QGAM and DDPstar, in this order. \appautorefsec{app-application-framingham-heart-study}
includes details on the model specifications for the competitors and plots of quantile
curves for all models.

\begin{figure}[tb]
    \includegraphics[width=\textwidth]{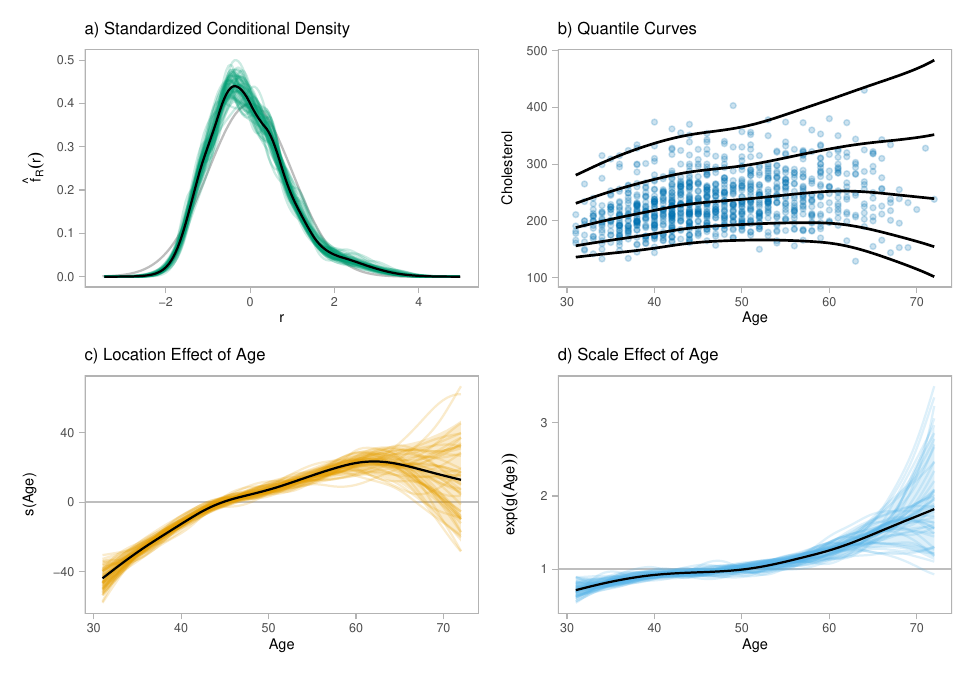}
    \caption{Panel a): Mean posterior density of $R$ for nonlinear PTM without random intercept. The gray line is a standard Gaussian density for reference. Panel b): Quantile curves at probability levels $\{0.01, 0.1, 0.5, 0.9, 0.99\}$. Panel c): Posterior mean change in average Cholesterol level as a function of age. Panel d): Posterior mean multiplicative change in the scale of Cholesterol level as a function of age. In panels a, b, c: The shaded areas indicate 90\% posterior credible bands, colored lines are 50 random draws from the respective collection of posterior functions.\label{fig:fh-summary}}
\end{figure}

\autoref{fig:fh-summary} shows a visual summary of the nonlinear PTM without random
intercept. The standardized conditional density estimate (panel a) indicates a deviation from
Gaussianity in terms of a lighter left tail and a slightly heavier right tail. The
location effect (panel c) suggests that Cholesterol levels are rising with participants'
age up to about age 60. After age 60, Cholesterol levels among patients in the dataset start to fall slightly, but uncertainty grows large due to the
small number of observations associated with participants over the age of 60. The
scale effect (panel d) shows a multiplicative effect, i.e., if $\exp(g(\text{age}))=1$, the
scale is at baseline level, while $\exp(g(\text{age}))=2$ indicates a scale at twice the baseline level. We see that the scale increases with participants' age, while
uncertainty grows large past the age of 60. The picture is completed by quantile curves
for probability levels $\{0.01, 0.1, 0.5, 0.9, 0.99\}$ in panel b). In spite of the
downward slope of the location effect at large age, the predictions for the
two top quantiles do not level off like the median and the lower quantiles do, because
of the increase in variance induced by the scale effect and the heavier right tail of
the conditional distribution.

\section{Conclusion}\label{sec-conclusion}

With penalized transformation models, we contribute a well-interpretable
model for the location and scale of arbitrary distributions. The model can be conveniently
summarized by plotting the estimated standardized conditional
density $f_R(r)$ alongside the
location and scale effects. The Bayesian approach to PTMs is especially
attractive due to its straightforward quantification of uncertainty
about derived estimates for the conditional distribution.
Through the random walk prior on the
transformation function's parameters \(\bsdelta\), the Bayesian view also provides a natural
interpretation of regularization: it moderates the trade-off between a
more complex model and the base model defined by the reference
distribution.

PTMs provide a fruitful basis for future developments. On one hand, an
attractive road for development lies in tailoring PTMs to specific
applications like count data and censored data, as sketched in \autoref{sec-ptm-extensions}, or heavy-tailed data. On
the other hand, we can envision general extensions like introducing a
binary switch between the base model and the more general model through
the use of spike and slab priors, or extensions to modeling higher
moments of the response distribution as functions of covariates by letting
$\bsdelta$ depend on covariates. Currently, scaling to large sample sizes
with tens of thousands is challenging for PTMs. Thus, another logical
step would be the development of faster, approximative inference alternatives to MCMC,
such as variational inference.

In the application examples, the predictive performance of the PTM is competitive with,
but not uniformly superior to, that of other sophisticated
distributional models. This observation underscores that no general dominance of the PTM
should be expected: in several respects, it is more restrictive than highly flexible
alternatives such as Dirichlet process mixture models (e.g., DDPstar) or conditional
transformation models (e.g., BCTM).
Instead, the appeal of the PTM lies in its structured formulation, which allows applied
researchers to reason transparently about location and scale effects while still
incorporating
additional information about the shape of the response's conditional distribution
that would be lost in a purely Gaussian model.

In conclusion, we view PTMs as a highly promising option in
statisticians' toolkit with a broad range of possibilities for application and extension
due to its combination of flexibility and interpretability.

\subsubsection*{Supplemental materials and reproducibility}

Supplemental materials are available from \url{https://github.com/liesel-devs/ptm-supplement}.
The supplemental materials include an online appendix as well as code
and data for the application examples.
Our open-source implementation in a Python library is available from
\url{https://github.com/liesel-devs/liesel-ptm}.

\subsubsection*{Funding}

We are grateful to the German Research Foundation (DFG) for financial support through grant 443179956.

\subsubsection*{Conflicts of Interest}

There are no competing interests to declare.

{\footnotesize
    \onehalfspacing

    \printbibliography
}
\end{refsection}

\clearpage
\section*{Appendix}
\addcontentsline{toc}{section}{Appendix}
\appendix
\counterwithin{figure}{section}
\counterwithin{table}{section}
\setcounter{figure}{0}
\setcounter{table}{0}
\startcontents[appendix]
\startlist[appendix]{lof}
\startlist[appendix]{lot}

\printcontents[appendix]{}{1}{}
\clearpage

\begin{refsection}
\section{Transformation function details}
\label{app-sec-trafo}

The full transformation function
is given by

\small

\begin{equation}\protect\hypertarget{app-eq-transformation}{}{
        h(r) =
        \left\{
        \begin{array}{l @{\quad} l r l}
            r + \tilde{A}                  & \text{if} \quad
            r \in (- \infty, a - \lambda ) & \text{(linear)},     \\[0.5em]
            \text{left}(r) + A
                                           & \text{if} \quad
            r \in [a - \lambda, a )        & \text{(transition)}, \\[0.5em]
            \text{spline}(r)               & \text{if} \quad
            r \in [a, b]                   & \text{(core)},       \\[0.5em]
            \text{right}(r) + B
                                           & \text{if} \quad
            r \in(b, b + \lambda ]         & \text{(transition)}, \\[0.5em]
            r  + \tilde{B}                 & \text{if} \quad
            r \in (b + \lambda,  \infty )  & \text{(linear)},
        \end{array}
        \right.
    }\label{app-eq-transformation}\end{equation} \normalsize
where \(\lambda \in \bbR_{>0}\), \(a < b\), and \(a, b, \tilde{A}, A, \tilde{B}, B \in \bbR\).
The core segment is a modified increasing B-spline with real-valued
parameters \(\bsdelta^\sfT = [\delta_1, \dots, \delta_{J-1}]\) given by
\begin{equation}\protect\hypertarget{app-eq-transformation-core}{}{
        \text{spline}(r) = \alpha + \underbrace{\frac{1}{s(\bsdelta)}
            \sum_{j=2}^J B_j(r) \left( \sum_{\ell = 2}^j \exp\bigl(\delta_{\ell-1}\bigr) \right)}_{:=g(r)},
    }\label{app-eq-transformation-core}\end{equation}
where \(B_j(r)\)
is a third-order (cubic) B-spline basis evaluation. The bases are set up using
\(J + 4\) equidistant knots
\(k_{-2} < k_{-1} < k_{0} < \dots < k_{J} < k_{J+1}\), where
\(k_{1}, \dots, k_m\) with $m = J-2$ are called the interior knots and the distance between adjacent knots is $d = k_{j+1} - k_j$.
The boundary knots are $k_1=a$ and $k_m=b$.
The intercept parameter $\alpha$ is given by $\alpha = a - g(a)$.
The function $s(\bsdelta)$ is given by
\begin{equation} \label{app-eq:sfn}
    s(\bsdelta) = \frac{1}{b-a}
    \sum_{j=1}^{J-3}
    \left(
    \frac{\exp(\delta_{j}) + \exp(\delta_{j+2})}{6} + \frac{2\exp(\delta_{j+1})}{3}  \right).
\end{equation}
The left and right transition segments provide smooth transitions to the linear extrapolation segments. They are given by
\begin{align}
    \label{app-eq-trafo-left}
    \text{left}(r)  & = r\left[
        \frac{a - r/2}{\lambda}
        \right]
    + r \left[
        1 - \frac{a - r/2}{\lambda}
    \right] \frac{\partial}{\partial a} \text{spline}(a),\text{ and} \\[0.5em]
    \text{right}(r) & = r \left[
        \frac{r/2 - b}{\lambda}
        \right]
    + r \left[
        1 - \frac{r/2 - b}{\lambda}
        \right]
    \frac{\partial}{\partial b} \text{spline}(b).
\end{align}
The constants \(A = \text{spline}(a) - \text{left}(a)\) and \(B = \text{spline}(b) - \text{right}(b)\) in the transition segments induce vertical shifts, ensuring that the
function values at the segment boundaries align. The constant shifts in the linear function segments are $\tilde{A} = \text{left}(a-\lambda) - (a-\lambda) + A$ and
$\tilde{B} = \text{right}(b + \lambda) - (b + \lambda) + B$.

\hypertarget{app-proofs}{%
    \section{Proofs}\label{app-proofs}}

We first use
Definition~\ref{app-def-inc-bspline} to define a monotonically increasing B-spline.

\begin{definition}[B-spline]\protect\hypertarget{app-def-inc-bspline}{}\label{app-def-inc-bspline}

    Let \(f: \bbR \rightarrow \bbR\) be given by
    \begin{equation}\protect\hypertarget{app-eq-mispline}{}{
            f(r) = \sum_{j=1}^J B_j^{(3)}(r) \omega_j,
        }\label{app-eq-mispline}\end{equation} with spline coefficients $\omega_1, \dots, \omega_J \in \bbR$ and
    B-spline bases \(B_j^{(3)}\) of order \(3\) using an increasing knot
    base \(k_{-2} < k_{-1} < k_{0} < \dots < k_{J} < k_{J+1}\). Then we call
    \eqref{app-eq-mispline} a B-spline.

\end{definition}

We use the symbol $f$ to denote the function here instead of $h$ as used in the main text to distinguish between a pure spline and the full transformation function $h$. The knots \(k_{1}, \dots, k_m\) with $m = J-2$ are called the interior knots. We are working exclusively with equidistant knots, and denote the distance between adjacent knots as $d = k_{j+1} - k_j$. The boundary knots are $k_1=a$ and $k_m=b$.

We use the recursive definition of a B-spline basis as given by \textcite{Fahrmeir2013-RegressionModelsMethods}:
\begin{equation}\protect\hypertarget{app-eq-bspline-bases}{}{
        B_j^{(p)}(r) = \frac{r-k_{j-p}}{k_j - k_{j-p}}B_{j-1}^{(p-1)}(r) + \frac{k_{j+1}-r}{k_{j+1} - k_{j+1-p}}B_j^{(p-1)}(r).
    }\label{app-eq-bspline-bases}\end{equation} The order of the basis is given
by \(p\), with \(p=3\) for cubic splines. The basis of order \(p=0\) is
given by \[
    B_j^{(0)}(r) = I(k_j \leq r < k_{j+1}), \qquad j = 1, \dots, J-1,
\] where \(I(\cdot)\) is the indicator function for the enclosed
condition. For equidistant knots \(k_{j+1} - k_j = d\), the basis
\eqref{app-eq-bspline-bases} simplifies to
\begin{equation}\protect\hypertarget{app-eq-bspline-bases-equidist}{}{
        B_j^{(p)}(r) = \frac{r-k_{j-p}}{p d}B_{j-1}^{(p-1)}(r) + \frac{k_{j+1}-r}{p d}B_j^{(p-1)}(r).
    }\label{app-eq-bspline-bases-equidist}\end{equation}

\subsection{Proposition~\ref{app-prop-trafo-derivative} (Derivative)}
\label{app-prop:trafo-derivative}

\begin{proposition}
    \label{app-prop-trafo-derivative}
    The first derivative of $h$ with respect to $r$ is given by
    $$
        \frac{\partial}{\partial r} h(r) = \frac{1}{s(\bsdelta)\cdot d} \sum_{j=2}^J B_j^{(2)}(r) \exp(\delta_{j-1}),
    $$
    where $d = k_{j+1} - k_j$ is the (constant) distance between adjacent knots and $B_j^{(2)}(r)$ is a quadratic B-spline basis function evaluation.
\end{proposition}

\begin{proof}
    The proposition is concerned with the derivative of the transformation function's core segment, which is given by
    $$
        h(r) = \alpha + \frac{1}{s(\bsdelta)}
        \sum_{j=1}^J B_j(r) \sum_{\ell = 2}^j \exp\bigl(\delta_{\ell-1}\bigr) \quad \text{if} \quad r \in [a, b],
    $$
    where $B_j$ are cubic B-spline bases. The function can be written as a B-spline $h(r) = s(\bsdelta)^{-1} \tilde{h}(r)$, where $\tilde{h}(r) = \sum_{j=1}^J B_j(r) \omega_j$ is a cubic B-spline with spline coefficients defined as $\omega_j = \alpha + \sum_{\ell = 2}^j \exp(\delta_{\ell-1})$. It is a well-known property of B-splines that the derivative of a B-spline of order $p$ is another B-spline of order $p-1$, with the spline coefficients given by differences in the original coefficients divided by the distance between the corresponding knots \parencite[see, for example,][p. 20]{Eilers2021-PracticalSmoothingJoys}. Since we use equidistant knots, this distance is the constant $d$ in our case. Since we use a B-spline of order $p=3$, the derivative of $\tilde h$ is a B-spline of order $p=2$, given by
    $$
        \frac{\partial}{\partial r} \tilde h(r) = \sum_{j=2}^J B_j^{(2)}(r) \frac{\omega_j - \omega_{j-1}}{d} = \frac{1}{d} \sum_{j=2}^J B_j^{(2)}(r) \exp(\delta_{-1}).
    $$
    Combining this partial result with the scaling by $s(\bsdelta)^{-1}$, we arrive at the result given in Proposition~\ref{app-prop-trafo-derivative}.

\end{proof}

\subsection{Proof of Proposition~\mainref{thm-monotonicity} (Monotonicity)}

To prove this proposition, we use \(h: \bbR \rightarrow \bbR\) and the core segment
\(\mathrm{spline}(r)\) as defined fully in \autoref{app-sec-trafo}. Note that the derivative of this transformation function is \small
\begin{equation}\protect\hypertarget{app-eq-transformation-deriv}{}{
        \frac{\partial}{\partial r}
        h(r) =
        \left\{
        \begin{array}{l @{\quad} l r l}
            1                                            & \text{if} \quad
            r \in (- \infty, a - \lambda )               & \text{(linear)},     \\[0.5em]
            \left(
            1 - \frac{a- r}{\lambda}
            \right)
            \cdot
            \frac{\partial}{\partial a} \text{spline}(a)
            +
            \frac{a- r}{\lambda}
                                                         & \text{if} \quad
            r \in [a - \lambda, a )                      & \text{(transition)}, \\[0.5em]
            \frac{\partial}{\partial r} \text{spline}(r) & \text{if} \quad
            r \in [a, b]                                 & \text{(core)},       \\[0.5em]
            \left(
            1 - \frac{r-b}{\lambda}
            \right)
            \cdot
            \frac{\partial}{\partial b} \text{spline}(b)
            +
            \frac{r-b}{\lambda}
                                                         & \text{if} \quad
            r \in(b, b + \lambda ]                       & \text{(transition)}, \\[0.5em]
            1                                            & \text{if} \quad
            r \in (b + \lambda,  \infty )                & \text{(linear)}.
        \end{array}
        \right.
    }\label{app-eq-transformation-deriv}\end{equation} \normalsize

\begin{proof}

    We prove the theorem by showing that
    \(\frac{\partial}{\partial r} h(r) > 0\) for all
    \(r \in \bbR\), which is equivalent to strict monotonicity. We
    proceed by cases.

    \begin{enumerate}

        \item
              \textit{Case 1: Linear segments.} Note that in the linear function segments, when $r \in (-\infty, a - \lambda) \cup (b+\lambda, \infty)$, we have $\frac{\partial}{\partial r} h(r) = 1$, which is strictly greater than zero. Thus, in the linear segments, $h$ is strictly monotonically increasing.

        \item
              \textit{Case 2: Core segment.} By Proposition~\ref{app-prop-trafo-derivative}, the derivative in the core segment when $r \in [a, b]$ is given by
              $$
                  \frac{\partial}{\partial r} \text{spline}(r) = \frac{1}{s(\bsdelta) \cdot d} \sum_{j=2}^{J} B_j^{(2)}(r) \exp(\delta_{j-1}).
              $$
              Note that, by definition, any basis function evaluation $B_j^{(2)}(r)$
              is nonnegative, and for any given $r \in [a, b]$, three basis function
              evaluations are positive. Note also that the distance
              between adjacent knots $d = k_{j+1} - k_j > 0$ by construction. Finally, observe that $\exp(\delta) > 0$ for all $\delta \in \bbR$, and consequently $s(\bsdelta) > 0$.
              As a result, $\frac{\partial}{\partial r} \text{spline}(r) > 0$
              for all $r \in [a, b]$.

        \item
              \textit{Case 3: Transition segments.} We start with the left transition segment. The derivative of this segment is given by
              $$
                  \frac{\partial}{\partial r}
                  \operatorname{left}(r)
                  =
                  \left(
                  1 - \frac{a- r}{\lambda}
                  \right)
                  \cdot
                  \frac{\partial}{\partial a} \text{spline}(a)
                  +
                  \frac{a- r}{\lambda}
              $$
              Observe that $a - \lambda \leq r < a$ and $\lambda > 0$, such that $\frac{a- r}{\lambda} \in (0, 1]$. Thus,
              $$
              \min \left\{\frac{\partial}{\partial r} \operatorname{left}(r): r \in [a- \lambda, a)\right\} =
              \min \left\{\frac{\partial}{\partial a} \text{spline}(a), 1\right\}.
              $$
              Since we know that $\frac{\partial}{\partial r} \text{spline}(r) > 0$
              for all $r \in [a, b]$ from Case 2, this means that
              $\frac{\partial}{\partial r} \operatorname{left}(r) > 0$ for all
              $r \in [a - \lambda, a]$. The proof for the right transition, when
              $r \in (b, b + \lambda]$, works analogously.
    \end{enumerate}

    We have shown that
    \(\frac{\partial}{\partial r} h(r) > 0\) if
    \(r \in (-\infty, a - \lambda) \cup (b+\lambda, \infty)\)
    (Case 1), if \(r \in [a - \lambda, a)\) or
        \(r \in (b, b + \lambda]\) (Case 3), and if
    \(r \in [a, b]\) (Case 2). Observe that \[
        \bbR = (-\infty, a - \lambda) \cup [a - \lambda, a) \cup [a, b] \cup (b, b + \lambda]  \cup (b+\lambda, \infty).
    \]

    Thus, we conclude that
    \(\frac{\partial}{\partial r} h(r) > 0\) for all
    \(r \in \bbR\), which is equivalent to saying that \(h\) is
    strictly monotonically increasing in \(r\) for all
    \(r \in \bbR\).

\end{proof}

\subsection{Proof of Theorem \mainref{thm-core-probability} (Core probability)}
\label{app-sec-proof-core-probability}

This proof has a longer setup. First, we derive the average slope of a monotonically increasing B-spline in Proposition~\ref{app-prp-avg-bspline-derivative} with the help of the Lemmas \ref{app-lem-bspline-derivative-segment} and \ref{app-lem-avg-bspline-derivative-segment}. Second, we use Proposition~\ref{app-prp-avg-bspline-derivative} to prove that the average slope of the transformation function $h$ over the subdomain $[a,b]$ is one in Lemma~\ref{app-lem-avg-slope-one}. Finally, we prove the theorem by showing first that $h(a) = a$, then using Lemma~\ref{app-lem-avg-slope-one} to show that $h(b)=b$.

\begin{lemma}[Derivative in a
        segment]\protect\hypertarget{app-lem-bspline-derivative-segment}{}\label{app-lem-bspline-derivative-segment}

    Let \(f(r) = \sum_{j=1}^J B_j(r) \omega_j\) be a monotonically increasing cubic B-spline (see Definition~\ref{app-def-inc-bspline}) with spline coefficients $\omega_j = \alpha + \sum_{\ell=2}^j \exp(\delta_{\ell-1})$, where $\alpha, \delta_1, \delta_2, \dots, \delta_{J-1} \in \bbR$. If
    \(r \in [k_{j}, k_{j+1}]\), where \(1 \leq j \leq J-3\), then the first derivative of \(f\) evaluated at \(r\)
    is
    \begin{equation}\protect\hypertarget{app-eq-mispline-derivative-segment}{}{
            \frac{\partial}{\partial r} f(r) = \sum_{j' = j}^{j+2} B_{j'}^{(2)}(r) \frac{\exp(\delta_{j'})}{d}.
        }\label{app-eq-mispline-derivative-segment}\end{equation}

\end{lemma}

\begin{proof}

    A well-known property of B-splines of order \(p\) is the fact that at
    any point \(R \in [k_1, k_{J-2}]\), we have \(p + 1\) positive
    basis functions, while the remaining basis functions evaluate to zero
    \parencite[see][p.~429]{Fahrmeir2013-RegressionModelsMethods}. Since the derivative of $f$ uses bases of order \(p=2\), that means at any point
    \(r \in [k_1, k_{J-2}]\), we have \(3\) positive basis
    functions. If we know that \(r \in [k_{j}, k_{j+1}]\), it
    follows from the definition of B-spline bases \eqref{app-eq-bspline-bases-equidist} that the positive basis
    functions are \(B_j\), \(B_{j+1}\), and \(B_{j+2}\), which directly lets the general form of the derivative simplify to
    \eqref{app-eq-mispline-derivative-segment}.

\end{proof}

\begin{lemma}[Average derivative in a
        segment]\protect\hypertarget{app-lem-avg-bspline-derivative-segment}{}\label{app-lem-avg-bspline-derivative-segment}

    Let \(f(r) = \sum_{j=1}^J B_j(r) \omega_j\) be a monotonically increasing cubic B-spline (see Definition~\ref{app-def-inc-bspline}) with spline coefficients $\omega_j = \alpha + \sum_{\ell=2}^j \exp(\delta_{\ell-1})$, where $\alpha, \delta_1, \delta_2, \dots, \delta_{J-1} \in \bbR$. If
    \(r \in [k_{j}, k_{j+1}]\), the average derivative of \(f\) in
    the corresponding function segment is given by

    \begin{equation}\protect\hypertarget{app-eq-mispline-avg-derivative-segment}{}{
            \frac{1}{d} \int\limits_{k_j}^{k_{j+1}} \left(\frac{\partial}{\partial r} f(r) \right) \mathrm d r =
            \frac{1}{d} \left(
            \frac{1}{6} \exp(\delta_{j}) + \frac{2}{3} \exp(\delta_{j+1}) + \frac{1}{6} \exp(\delta_{j+2})
            \right).
        }\label{app-eq-mispline-avg-derivative-segment}\end{equation}

\end{lemma}

\begin{proof}

    First, note that since \(r \in [k_{j}, k_{j+1}]\), we can
    apply Lemma~\ref{app-lem-bspline-derivative-segment} to write
    \footnotesize
    \[
        \begin{aligned}
            \frac{1}{d} \int\limits_{k_j}^{k_{j+1}} \left(\frac{\partial}{\partial r} f(r) \right) \text d r =
            \frac{1}{d}
            \left(
            \frac{\exp(\delta_{j})}{d} \int\limits_{k_j}^{k_{j+1}} B_j^{(2)}(r) \text d r \right.
            +
            \left.\frac{\exp(\delta_{j+1})}{d} \int\limits_{k_j}^{k_{j+1}} B_{j+1}^{(2)}(r) \text d r
            +
            \frac{\exp(\delta_{j+2})}{d} \int\limits_{k_j}^{k_{j+1}} B_{j+2}^{(2)}(r) \text d r
            \right).
        \end{aligned}
    \]
    \normalsize
    A sufficient condition for this expression to simplify to
    \eqref{app-eq-mispline-avg-derivative-segment} is given by the following
    three statements: \[
        \begin{aligned}
            \int\limits_{k_j}^{k_{j+1}} B_j^{(2)}(r) \text d r     & \overset{!}{=} \frac{d}{6},   &                  & \text{(s1)} \\
            \int\limits_{k_j}^{k_{j+1}} B_{j+1}^{(2)}(r) \text d r & \overset{!}{=} \frac{2 d}{3}, & \text{and} \quad & \text{(s2)} \\
            \int\limits_{k_j}^{k_{j+1}} B_{j+2}^{(2)}(r) \text d r & \overset{!}{=} \frac{d}{6}.   &                  & \text{(s3)}
        \end{aligned}
    \] In the following we show first that \(\text{s1}\) and \(\text{s3}\)
    hold, before showing that \(\text{s2}\) holds.

    \begin{enumerate}
        \item
              To show that $\text{s1}$ holds, we substitute the definition of a B-spline
              basis \eqref{app-eq-bspline-bases-equidist} and simplify, such that we obtain
              $$
                  \int\limits_{k_j}^{k_{j+1}} B_j^{(2)}(r) \text d r =
                  \frac{1}{2 d^2}\int\limits_{k_j}^{k_{j+1}} (k_{j+1} - r)^2 \text d r.
              $$
              The integral is readily solved using $u$-substitution with $u=k_{j+1} - r$.
              Inserting the integral limits, we get
              $$
                  \begin{aligned}
                      \frac{1}{2 d^2}\int\limits_{k_j}^{k_{j+1}} (k_{j+1} - r)^2 \text d r & = \frac{1}{2 d^2} \frac{(k_{j+1} - k_j)^3}{3} = \frac{d^3}{6 d^2}  = \frac{d}{6},
                  \end{aligned}
              $$
              confirming that $\text{s1}$ is true. The solution for $\text{s3}$ works in
              complete analogy.

        \item

              Substituting the definition of a B-spline basis into $\text{s2}$, we obtain
              \begin{equation}
                  \label{app-eq-bspline-segment-intermediate}
                  \int\limits_{k_j}^{k_{j+1}}B_{j+1}^{(2)}( r) \text d r = \frac{1}{2 d^2} \left(
                  - \int\limits_{k_j}^{k_{j+1}} ( r-k_{j-1})( r-k_{j+1}) \text d r
                  -
                  \int\limits_{k_j}^{k_{j+1}} ( r-k_{j})( r-k_{j+2}) \text d r
                  \right).
              \end{equation}
              The two integrals in \eqref{app-eq-bspline-segment-intermediate} can be solved
              analogously. Considering the left-hand term, we have
              $$
                  \begin{aligned}
                      - \int\limits_{k_j}^{k_{j+1}} (r-k_{j-1})(r-k_{j+1})\text dr
                       & = - \left[
                          \left( \frac{2k_{j+1}^3 - 6k_{j+1}^2d}{6} \right)
                          -
                          \left(
                          \frac{- 4k_j^3 + 6k_jk_{j-1}k_{j+1}}{6}
                          \right)
                          \right].
                  \end{aligned}
              $$
              Next, we substitute $k_j = k_{j+1} - d$ and $k_{j-1} = k_{j+1} - 2 d$ and simplify further, which results in
              $$
                  \begin{aligned}
                      - \int\limits_{k_j}^{k_{j+1}} (r -k_{j-1})(r -k_{j+1})\text dr & =
                      \frac{2d^3}{3}.
                  \end{aligned}
              $$
              Applying the analogous procedure to the right-hand term in \eqref{app-eq-bspline-segment-intermediate},
              and inserting both parts back in \eqref{app-eq-bspline-segment-intermediate}, we get
              $$
                  \int\limits_{k_j}^{k_{j+1}}B_{j+1}^{(2)}( r) \text d r
                  =
                  \frac{1}{2 d^2} \left( \frac{2d^3}{3} + \frac{2d^3}{3}\right) = \frac{2d}{3},
              $$
              concluding the proof of $\text{s2}$.

    \end{enumerate}

    Since we have shown that the sufficient conditions \(\text{s1}\),
    \(\text{s2}\), and \(\text{s3}\) are met, this concludes the proof.

\end{proof}

\begin{proposition}[Average slope of a monotonically increasing B-spline]\protect\hypertarget{app-prp-avg-bspline-derivative}{}\label{app-prp-avg-bspline-derivative}

    Let \(f(r) = \sum_{j=1}^J B_j(r) \omega_j\) be a monotonically increasing cubic B-spline (see Definition~\ref{app-def-inc-bspline}) with spline coefficients $\omega_j = \alpha + \sum_{\ell=2}^j \exp(\delta_{\ell-1})$, where $\alpha, \delta_1, \delta_2, \dots, \delta_{J-1} \in \bbR$. Let \(a = k_1\) and
    \(b = k_m\) denote the boundary knots. Then the average slope of
    \(f\) over the interval \([a, b]\) is given by
    \begin{equation}\protect\hypertarget{app-eq-avg_slope}{}{
            \begin{aligned}
                s(\bsdelta) & = \frac{1}{b-a} \int\limits_a^b \frac{\partial}{\partial r} f(r) \mathrm d r
                =
                \frac{1}{b-a}
                \sum_{j=1}^{J-3}
                \left(
                \frac{\exp(\delta_{j}) + \exp(\delta_{j+2})}{6} + \frac{2\exp(\delta_{j+1})}{3}  \right),
            \end{aligned}
        }\label{app-eq-avg_slope}\end{equation}

\end{proposition}

\begin{proof}

    The integral can be additively decomposed using the interior knots as
    the bounds of individual integrals, such that we can write \[
        \frac{1}{b-a} \int\limits_a^b \frac{\partial}{\partial r} f(r) \mathrm d r
        =
        \frac{1}{b-a}
        \sum_{j=1}^{J-3} \left(
        \int\limits_{k_j}^{k_{j+1}} \frac{\partial}{\partial r} f(r) \mathrm d r \right).
    \] By Lemma~\ref{app-lem-avg-bspline-derivative-segment}, the individual
    integrals in the right-hand-side sum can be written in terms of the
    parameters \(\delta_j, \delta_{j+1}\), and \(\delta_{j+2}\), such that
    we can write \[
        \frac{1}{b-a}
        \int\limits_a^b \frac{\partial}{\partial r} f(r) \mathrm d r
        =
        \frac{1}{b-a}
        \sum_{j=1}^{J-3}
        \left(
        \frac{1}{6} \exp(\delta_{j}) + \frac{2}{3} \exp(\delta_{j+1}) + \frac{1}{6} \exp(\delta_{j+2}) \right),
    \] which directly simplifies to \eqref{app-eq-avg_slope}.

\end{proof}

\begin{lemma}[Average slope over core segment]
    \label{app-lem-avg-slope-one} The average slope of the transformation function \(h: \bbR \rightarrow \bbR\) as defined in \autoref{app-sec-trafo} over the core segment $[a, b]$ is one.
\end{lemma}

\begin{proof}
    Observe that the derivative of $h$ with respect to $r$ for $r \in [a, b]$ is identified in Proposition~\ref{app-prop-trafo-derivative} as
    $$
        \frac{\partial}{\partial r} h(r) = \frac{1}{s(\bsdelta)\cdot d} \sum_{j=2}^J B_j^{(2)}(r) \exp(\delta_{j-1}).
    $$
    Note further that $h$ can be written as a B-spline $h(r) = s(\bsdelta)^{-1} \tilde{h}(r)$, where $\tilde{h}(r) = \sum_{j=1}^J B_j(r) \omega_j$ is a cubic B-spline with spline coefficients defined as $\omega_j = \alpha + \sum_{\ell = 2}^j \exp(\delta_{\ell-1})$.
    As a result, the derivative of $h$ can be written as
    $$
        \frac{\partial}{\partial r} h(r) = \frac{1}{s(\bsdelta)} \frac{\partial}{\partial r} \tilde h(r).
    $$
    Since by Proposition~\ref{app-prp-avg-bspline-derivative}, $s(\bsdelta)$ is the average derivative of $\tilde h$ over $[a,b]$, we can see that dividing by $s(\bsdelta)$ normalizes the average slope of $h$ over $[a,b]$ to one:
    $$
        \frac{1}{b-a} \int_a^b \frac{\partial}{\partial r} h(r) \mathrm d r = \frac{1}{s(\bsdelta)} \frac{1}{b-a} \int_a^b \frac{\partial}{\partial r} \tilde h(r) \mathrm d r = \frac{s(\bsdelta)}{s(\bsdelta)} = 1.
    $$
\end{proof}

\subsubsection*{Main proof of Theorem~\mainref{thm-core-probability}}

We use the transformation function \(h: \bbR \rightarrow \bbR\) as defined in \autoref{app-sec-trafo}.

\begin{proof}
    First, note that $h(a)=a$ by construction, since $\alpha = a - h(a | \alpha = 0)$, where $h(a | \alpha = 0)$ denotes a preliminary evaluation of $h$ where $\alpha$ is temporarily set to $0$.
    Second, note that by Lemma~\ref{app-lem-avg-slope-one}, the average slope of $h$ over $[a,b]$ is one. Since $h(a)=a$, this property directly implies $h(b)=b$. Next, recall that the cumulative distribution function of $R$ is defined as $\bbP(R \leq r) = F_Z(h(r))$. It follows that $\bbP(R \leq a) = F_Z(h(a))=F_Z(a)$ and $\bbP(R \leq b) = F_Z(h(b))=F_Z(b)$, concluding the proof.
\end{proof}

\subsection{Proof of Theorem \mainref{thm-identity} (Reduction to identity)}
\label{app-sec-proof-identity}

We use the transformation function \(h: \bbR \rightarrow \bbR\) as defined in \autoref{app-sec-trafo}.

\begin{proof}
    First, consider the core segment, where $r \in [a, b]$. If $\delta_j = c$ for all $j=1, \dots, J-1$, the derivative of $h$ in the core segment can be written as
    $$
        \begin{aligned}
            \frac{\partial}{\partial r} h(r) & = \frac{\exp(c)}{s(\bsdelta)\cdot d} \sum_{j=2}^J B_j^{(2)}(r) = \frac{\exp(c)}{s(\bsdelta)\cdot d},
        \end{aligned}
    $$
    where the second identity relies on the unity decomposition property of B-spline bases \parencite[see][p.~429]{Fahrmeir2013-RegressionModelsMethods}. Note that the property applies here even though the index starts at $j=2$, since we write the second-order B-spline in the derivative with indices based on the knots of the parent function, which causes $B_1^{(2)}(r)=0$. Inserting
    $\delta_j = c$ for all $j=1, \dots, J-1$, we further get
    $$
        \begin{aligned}
            s(\bsdelta) =
            \frac{1}{b-a}
            \sum_{j=1}^{J-3}
            \left(
            \frac{\exp(\delta_{j}) + \exp(\delta_{j+2})}{6} + \frac{2\exp(\delta_{j+1})}{3}  \right)
            =
            \frac{(J-3) \exp(c)}{b-a}.
        \end{aligned}
    $$
    Since there are $J-3$ function segments in $[a,b]$, defined by equidistant knots, we can write $b-a = (J-3)d$, where $d$ is the distance between adjacent knots, i.e. the width of each function segment. We can now write
    $$
        \frac{\partial}{\partial r} h(r) = \frac{\exp(c)}{(\exp(c) / d)\cdot d} = 1,
    $$
    which means that if all elements of $\bsdelta$ are equal, $h$ has a unit slope on $[a,b]$. Since we observed in the proof of Theorem~\mainref{thm-core-probability} that $h(a)=a$ and $h(b)=b$, we can conclude that, in this case, $h$ is the identity function on $[a, b]$.

    Second, consider the left transition segment as defined in \eqref{app-eq-trafo-left}.
    If $\delta_j = c$ for all $j=1, \dots, J-1$, we have seen that $\frac{\partial}{\partial a} \text{spline}(a) = 1$. Then the left transition segment becomes
    $$
        \text{left}(r) = r\left[
            \frac{a - r/2}{\lambda}
            \right]
        + r \left[
            1 - \frac{a - r/2}{\lambda}
            \right] = r.
    $$
    The shift $A = \operatorname{spline}(a) - \operatorname{left}(a)$ thus simplifies to $A=0$. The case of the right transition segment works analogously. Finally, the shift terms $\tilde A$ and $\tilde B$ in the linear extrapolation segments likewise simplify to zero, completing the identity core segment and identity transition segments with identity extrapolation segments.

\end{proof}

\subsection{Proposition~\ref{app-prop-tails} (Tail probabilities)}

\begin{proposition}[Tail probabilities]\label{app-prop-tails}
    Let $R$ be an absolutely continuous random variable with cumulative distribution
    function $F_R(r) = F_Z(h(r))$, where \(h: \bbR \rightarrow \bbR\) is given by
    \eqref{app-eq-transformation}, with \(\mathrm{spline}(r)\) as in
    \eqref{app-eq-transformation-core}, and where $F_Z$ is
    a fully specified continuous cumulative distribution function. Then
    $F_R(r) = F_Z(r + \tilde{A})$ for $r < a-\lambda$ and $F_R(r) = F_Z(r + \tilde{B})$ for $b+\lambda < r$.
\end{proposition}

\begin{proof}
    We consider first the case $r < a-\lambda$. In this case, $h(r) = r + \tilde A$, such that $F_R(r) = F_Z(h(r)) = F_Z(r + \tilde A)$ for $r < a-\lambda$. The second case works analogously.
\end{proof}

By this proposition, the tails of the distribution of $R$, where we take the tails to mean $R < (a-\lambda)$ or $(b+\lambda)<R$, are described by shifted versions of the reference CDF. The transition segments provide smooth transitions between the flexible core, in which the shape of $R$'s distribution is governed by $\text{spline}(r)$ and the shifted-reference tails. As $\lambda \rightarrow 0$, the transition segments vanish, and the tail probabilities become $F_R(r) \rightarrow F_Z(r)$ for both $a < r$ and $r > b$. As can be seen in \mainautoref{fig-trafo} (bottom left panel) of the main text, this comes at the price of a potentially discontinuous jump in $\frac{\partial}{\partial r}h(r)$ at $r=a$ and $r = b$, which causes problems for gradient-based estimation techniques. On the other hand, as $\lambda \rightarrow \infty$, the transformation function $h$ has simple linear extrapolation for $r < a$ and $b < r$, and the
tail probabilities become $F_R(r) \rightarrow F_Z(r / \text{spline}'(a))$ for $r < a$ and $F_R(r) \rightarrow F_Z(r / \text{spline}'(b))$ for $b < r$, where we write $\text{spline}'(\cdot)$ to denote the first derivative of $\text{spline}(\cdot)$ with respect to its input. Linear extrapolation
could thus lead to extreme tail behavior, if $\text{spline}'(a)$ or $\text{spline}'(b)$ happen to be estimated to be close to zero or considerably larger than one. The former case would induce heavy tails, while the latter case would induce light tails. We expect little information to be available for the estimation of $\text{spline}'(a)$ and $\text{spline}'(b)$ in many cases; and it is unclear
how well information about $\text{spline}'(a)$ and $\text{spline}'(b)$ can be directly
used to determine tail probabilities. So with a moderate $\lambda$, Proposition~\ref{app-prop-tails} can be thought of as a safeguard against potentially spurious extreme tail behavior while maintaining smoothness of $h$ and its derivative, compare the middle bottom panel of \mainautoref{fig-trafo} of the main text. Still, linear extrapolation is possible with $\lambda \rightarrow \infty$ if desired.

\clearpage
\section{Illustration of the random walk prior for \texorpdfstring{$\bsdelta$}{delta}}
\label{app-sec-rwprior-illustration}
In this section, we provide a brief illustration of how the random walk prior
for $\bsdelta$ works. For the sake of simplicity, we take a concrete case of $J=5$. The illustration generalizes easily. In this case, we have $J-1 = 4$ elements in $\bsdelta$. The first-difference matrix is
$$
    \bfD = \begin{bmatrix}
        -1 & 1  & 0  & 0 \\
        0  & -1 & 1  & 0 \\
        0  & 0  & -1 & 1
    \end{bmatrix},
$$
and the prior for $\bsdelta$ can be written in terms of $\bfK = \bfD^\sfT \bfD$:
$$
    \pi(\bsdelta | \tau^2_\delta) \propto \left(\frac{1}{\tau^2_\delta}\right)^{\operatorname{rk}(\bfK)/2}\exp \left( - \frac{1}{2 \tau^2_\delta} \bsdelta^T \bfD^T \bfD \bsdelta \right)
$$
The product $\bfD \bsdelta$ means:

$$
    \bfD \bsdelta = \begin{bmatrix}
        \delta_2 - \delta_1 \\
        \delta_3 - \delta_2 \\
        \delta_4 - \delta_3 \\
    \end{bmatrix}
    =
    \bsdelta_{[2:4]} - \bsdelta_{[1:3]}
    .
$$
The subscript in the expression $\bsdelta_{[i:j]}$, $i < j$, denotes a partition of $\bsdelta$ consisting of the $i^{th}$ to $j^{th}$ elements of $\bsdelta$, including the $i^{th}$ and $j^{th}$ elements, such that $\bsdelta_{[i:j]}$ has dimension $(3 \times 1)$.
We can write

$$
    \begin{aligned}
        \pi(\bsdelta | \tau^2_\delta) & \propto \left(\frac{1}{\tau^2_\delta}\right)^{\operatorname{rk}(\bfK)/2}\exp \left( - \frac{1}{2 \tau^2_\delta} (\bsdelta_{[2:4]} - \bsdelta_{[1:3]})^T (\bsdelta_{[2:4]} - \bsdelta_{[1:3]}) \right),
    \end{aligned}
$$

which is proportional to the density of a normal distribution with mean vector $\bsdelta_{[1:3]}$ and covariance matrix $\tau^2_\delta \bfI$, where $\bfI$ is the $(3 \times 3)$ identity matrix, for $\bsdelta_{[2:4]}$ and a constant prior for $\delta_1$:

$$
    \begin{aligned}
        \delta_1         & \sim \text{const}                                                \\[0.5em]
        \bsdelta_{[2:4]} & \sim \mathcal{N}\bigl(\bsdelta_{[1:3]}, \tau^2_\delta\bfI\bigr).
    \end{aligned}
$$

\clearpage

\section{Details on the scale-dependent prior}
\label{app-sec-sdprior-details}

\FloatBarrier

\begin{figure}[htb]
    \includegraphics[width=\textwidth]{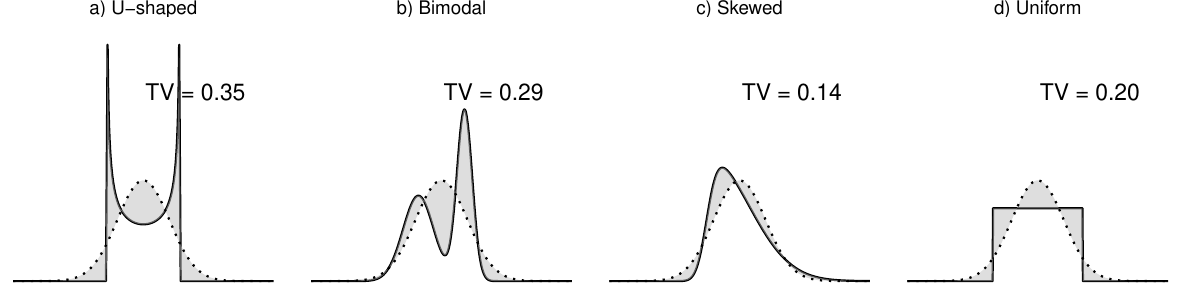}
    \caption[Illustration of the total variation distance (TV) for four example densities; densities used for data generation.]{\label{app-fig-sdprior-references}
        Illustration of the total variation distance (TV) for four example densities. The total variation distance is the shaded area.
    }
\end{figure}

\begin{figure}[hbt]
    \includegraphics[width=\textwidth]{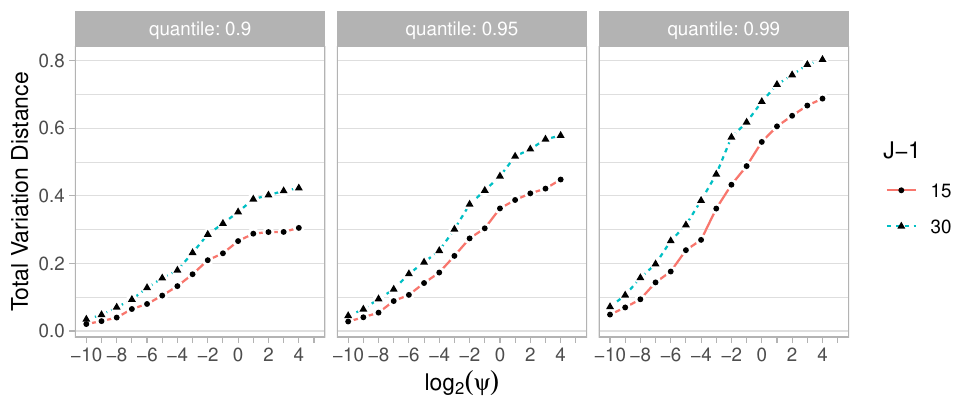}
    \caption[Quantiles of total variation distances.]{\label{app-fig-sdprior-15vs30}
        Quantiles of total variation distances observed using $J-1 = 15$ and $J-1 = 30$.
    }
\end{figure}

\clearpage

\hypertarget{app-sec-computational-aspects}{%
    \section{Details on the prior for \texorpdfstring{$\bsdelta$}{delta}}\label{app-sec-computational-aspects}}

\subsection{Cancellation of constants added to \texorpdfstring{$\bsdelta$}{delta} in \texorpdfstring{$h(r)$}{h(r)}}
\label{app-sec:cancellation-of-constants}

First, to provide intuition for the role of $\delta_1$ as an added constant, consider the simple setup where $p(a_1)\propto \text{const}$ and $a_2 \sim \mathcal{N}(0,\tau^2)$, collected in $\mathbf{a} = [a_1,a_2]^\top$. Define the shifted vector $\mathbf{c} = \mathbf{a} + b$. Then $c_1 = a_1 + b$ still has a flat prior, since $a_1$ did, while $c_2 = a_2 + b \sim \mathcal{N}(b,\tau^2)$. Equivalently, we may describe $\mathbf{c}$ in terms of $\mathbf{d} = [d_1,d_2]^\top$, where $d_1=b$ with $p(d_1)\propto \text{const}$ and $d_2 \sim \mathcal{N}(d_1,\tau^2)$. Thus, introducing the constant $b$ is equivalent to reparameterizing the model so that the first component acts as a free baseline, around which the second component is centered. This illustrates how $\delta_1$ plays the role of an added location constant.

Now consider adding a constant $\eta \in \bbR$ to $\bsdelta$. Then the core segment of the
transformation function for $r \in [a, b]$ is
\begin{align}
    h(r) & = \alpha + \frac{1}{s(\bsdelta + \eta)}
    \sum_{j=2}^J B_j(r) \sum_{\ell = 2}^j \exp\bigl(\delta_{\ell-1} + \eta\bigr) \\
         & = \alpha + \frac{\exp(\eta)}{s(\bsdelta + \eta)}
    \sum_{j=2}^J B_j(r) \sum_{\ell = 2}^j \exp\bigl(\delta_{\ell-1}\bigr).
\end{align}
and the normalization to slope one then is
\begin{align}
    s(\bsdelta + \eta) & = \frac{1}{b-a}
    \sum_{j=1}^{J-3}
    \left(
    \frac{\exp(\delta_{j} + \eta) + \exp(\delta_{j+2} + \eta)}{6} + \frac{2\exp(\delta_{j+1} + \eta)}{3}  \right) \\
                       & = \frac{\exp(\eta)}{b-a}
    \sum_{j=1}^{J-3}
    \left(
    \frac{\exp(\delta_{j}) + \exp(\delta_{j+2})}{6} + \frac{2\exp(\delta_{j+1})}{3}  \right)                      \\
                       & = \exp(\eta)s(\bsdelta),
\end{align}
such that we have
\begin{align}
    h(r)
     & = \alpha + \frac{\exp(\eta)}{\exp(\eta)s(\bsdelta)}
    \sum_{j=2}^J B_j(r) \sum_{\ell = 2}^j \exp\bigl(\delta_{\ell-1}\bigr),
\end{align}
where $\exp(\eta)$ cancels out in the fraction such that we have $h(r | \bsdelta) = h(r | \bsdelta + \eta)$ for any $\eta \in \bbR$.

\subsection{Reparameterization of \texorpdfstring{$\bsdelta$}{delta}}
\label{app-sec:sum-to-zero}

We implement the removal of the unpenalized constant from $\bsdelta$ via reparameterization using a mixed model decomposition. The description here is closely based on
the description in \textcite{Kneib2019-ModularRegressionLego}, Appendix A.
\begin{enumerate}
    \item Find the eigendecomposition of $\bfK = \bsGamma \bsOmega \bsGamma^\sfT$, where $\bsOmega = \operatorname{diag}(\omega_1, \dots, \omega_{J-1})$ with sorted eigenvalues $\omega_1 \leq \dots \leq \omega_{J-1}$, and $\bsGamma \bsGamma^\sfT = \bfI$ is the orthonormal matrix of corresponding eigenvectors.
    \item As the rank of $\bfK$ is $J-2$, we have $\omega_1 = 0$.
    \item We split $\bsGamma = [\bsGamma_1, \bsGamma_2]$ and $\bsOmega = \operatorname{blockdiag}(\bsOmega_1, \bsOmega_2)$, where $\bsOmega_1 = \omega_1$ and $\bsOmega_2 = \operatorname{diag}(\omega_2, \dots, \omega_{J-1})$. The matrices $\bsGamma_1$ and $\bsGamma_2$ are the corresponding matrices of eigenvectors of dimension $(J-1) \times 1$ and $(J-1) \times (J-2)$.
    \item We can now write
          $$
              \bsdelta = \bsGamma_1 \eta + \bsGamma_2 \bsOmega_2^{-1/2} \tilde \bsdelta,
          $$
          with priors
          $$
              \pi(\eta) \propto \text{const} \quad \text{and} \quad \tilde\bsdelta \sim \calN(\bfzero, \tau^2_\delta \bfI_{J-2}).
          $$
          The parameter $\eta$ represents the part of $\bsdelta$ that is unpenalized by the prior.
          In mixed model terminology, $\eta$ corresponds to the fixed effect, while
          $\tilde \bsdelta$ corresponds to the random effect.
    \item We can now define $\bsdelta \equiv \bsGamma_2 \bsOmega_2^{-1/2} \tilde\bsdelta$, with the reduced precision matrix $\bfK = \bsGamma_2 \bsOmega_2 \bsGamma_2^\sfT$. By this step, the unpenalized constant part of the prior is fixed to zero.
    \item We conduct inference on the level of $\tilde\bsdelta$.
\end{enumerate}

\clearpage
\section{Details on posterior inference}
\label{app-sec:inference}

\subsection{Basis matrix approximation}
\label{app-sec:basis-grid}
To avoid having to compute
the basis function evaluations of the transformation function's spline
segment in every MCMC iteration, we use a grid approximation. That is,
we obtain one matrix of basis function evaluations on a tight
equidistant grid
\(\tilde r_1, \dots, \tilde r_{1000}\), with
\(\tilde r_1 = a\) and \(\tilde r_{1000} = b\) set
to the lower and upper boundary knot, respectively. For a basis function
evaluation of a specific value \(r\), we then find the
greatest \(\tilde r_{\text{lo}}\) and the smallest
\(\tilde r_{\text{hi}}\) such that
\(r \in [\tilde r_{\text{lo}}, \tilde r_{\text{hi}}]\)
and interpolate between the two corresponding elements
\(B(\tilde r_{\text{lo}})\) and
\(B(\tilde r_{\text{hi}})\) of the pre-computed basis matrix:
\[
    \begin{aligned}
        B_j(r) & \approx \left(1-\frac{r - \tilde r_{\text{lo}}}{(b-a)/1000}\right)\cdot B(\tilde r_{\text{lo}}) + \frac{r - \tilde r_{\text{lo}}}{(b-a)/1000} \cdot B(\tilde r_{\text{hi}}).
    \end{aligned}
\]

\subsection{Initialization}
\label{app-sec:init}

Algorithm~\ref{app-alg-init} details the initialization process. The temporary swap of the
variance parameters' priors in Step 2 is a protection against non-concavity induced by
potential scale-dependent Weibull priors on the variance parameters. Note that the
initialization need not be perfect, it mainly serves to provide sensible starting
values for subsequent MCMC inference. Therefore, the temporary alteration of the model
in Step 2 is negligible for overall inference. Similar justifications apply for two
other features of the initialization algorithm:
\begin{enumerate}
    \item We jointly optimize the parameters $\bsbeta$, $\bsgamma$ and
          $\tilde \bsdelta$ with their respective hyperparameters instead of optimizing for the
          hyperparameters using a marginalized likelihood with $\bsbeta$, $\bsgamma$ and
          $\tilde \bsdelta$ integrated out; a practice that can induce over-smoothing.
          This point is partly alleviated by our use of a lower bound $>0$ for the variance
          parameters during optimization.
    \item We optimize in stages: We first optimize for the parameters in the location and
          scale model parts while keeping $\tilde \bsdelta = \bfzero$ fixed. We then optimize
          for $\tilde \bsdelta$ and $\tau^2_\delta$, keeping the other parameters fixed.
          Essentially, $\bsbeta$, $\bsgamma$ and their hyperparameters are found under
          an auxiliary assumption of Gaussianity.
          This point is partly alleviated by the fact that, due to the model's
          assumptions, the transformation function is independent of the location and scale
          model parts.
\end{enumerate}

\paragraph{Inverse softclip link function.}
The link function $g: [\underline{x}, \overline{x}] \to \bbR$ is specified based on the
softplus function $\operatorname{sp}: \bbR \to (0, \infty) $.
The softplus function can be modified to work on inputs with
arbitrary lower bounds $\underline{x}$ and upper bounds $\overline{x}$, yielding
\begin{align}
    \operatorname{sp}(x) & = \ln(1 + \exp(x))                                 & \bbR \to (0, \infty)             \\
    \operatorname{sl}(x) & = \ln(1 + \exp(x - \underline{x})) + \underline{x} & \bbR \to (\underline{x}, \infty) \\
    \operatorname{su}(x) & = -\ln(1 + \exp(x - \overline{x})) + \overline{x}  & \bbR \to (0, \overline{x}).
\end{align}
The respective inverses are
\begin{align}
    \operatorname{sp}^{-1}(x) & = \ln(\exp(x) - 1)                                 & (0, \infty) \to \bbR             \\
    \operatorname{sl}^{-1}(x) & = \ln(\exp(x - \underline{x}) - 1) + \underline{x} & (\underline{x}, \infty) \to \bbR \\
    \operatorname{su}^{-1}(x) & = \ln(\exp(\overline{x} - x)) + \overline{x}       & (0, \overline{x}) \to \bbR.
\end{align}
We define the link function $g$ and its inverse $g^{-1}$ by composing
$\operatorname{sl}$ and $\operatorname{su}$ and their respective inverses, i.e.,
\begin{align}
    \label{app-eq:invsoftclip}
    g(x)      & = (\operatorname{sl}^{-1} \circ \operatorname{su}^{-1})(x) & (\underline{x}, \overline{x}) \to \bbR  \\
    g^{-1}(x) & = (\operatorname{su} \circ \operatorname{sl})(x)           & \bbR \to (\underline{x}, \overline{x}).
\end{align}
The function $g^{-1}$ is also known under the name \textit{softclip}, since it provides
a continuous map from the real line to a bounded interval.

\begin{algorithm}[bth]
    \DontPrintSemicolon
    \caption{Initialization of Bayesian penalized transformation models.\label{app-alg-init}}
    \vspace{0.5em}
    \begin{enumerate}[leftmargin=*,itemsep=0pt]
        \small
        \item Initialize the parameters in $\bsbeta$, $\bsgamma$ and $\tilde \bsdelta$ to zero. Initialize the hyperparameters in $\bstau^2_\ell$ and  $\bstau^2_s$ to ten and $\tau^2_\delta$ to one.
        \item Replace the prior distributions for all parameters in $(\bstau^2_\ell, \bstau^2_s)$ and $\tau^2_\delta$ with a uniform prior $\calU(0.025, 10\,000)$.
        \item Transform the variance parameters in $(\bstau^2_\ell, \bstau^2_s)$ and $\tau^2_\delta$ to the real line using the inverse softclip link function $g: [\underline{x}, \overline{x}] \to \bbR$ using $\underline{x} = 0.025$ and $\overline{x} = 10\,000$. The temporary uniform priors are adjusted accordingly using the change of variables theorem. The function $g$ is given in~\eqref{app-eq:invsoftclip}.
        \item Find the posterior modes for $(\bsbeta, \bsgamma, g(\bstau^2_\ell), g(\bstau^2_s))$
              via gradient ascent while holding the remaining parameters fixed at their initial values, i.e., $$(\bsbeta^{[0]}, \bsgamma^{[0]}, g(\bstau^{2}_\ell)^{[0]}, g(\bstau^{2}_s)^{[0]}) = \argmax_{(\bsbeta, \bsgamma, g(\bstau^2_\ell), g(\bstau^2_s))} \ln p(\bsbeta, \bsgamma, g(\bstau^2_\ell), g(\bstau^2_s) | \tilde \bsdelta, \tau^2_\delta, \bsy).$$
        \item Find the posterior modes for $\tilde \bsdelta, g(\tau^2_\delta)$ via gradient ascent while holding the remaining parameters fixed at their values found in the previous step, i.e.,
              $$\tilde \bsdelta^{[0]}, g(\tau^{2}_\delta)^{[0]} = \argmax_{\tilde \bsdelta, g(\tau^2_\delta)} \ln p( \tilde \bsdelta, g(\tau^2_\delta), \bsy | \bsbeta^{[0]}, \bsgamma^{[0]}, g(\bstau^{2}_\ell)^{[0]}, g(\bstau^{2}_s)^{[0]}).$$
        \item Transform the variance parameters in $(\bstau^2_\ell, \bstau^2_s)$ and $\tau^2_\delta$ to their original scale using the inverse link function $g^{-1}: \bbR \to [\underline{x}, \overline{x}]$.
        \item Reset the prior distributions for the parameters in $(\bstau^2_\ell, \bstau^2_s)$ and $\tau^2_\delta$ to their original specifications.
    \end{enumerate}
\end{algorithm}

\subsection{Constant intercepts}
\label{app-seq:intercepts}
We treat the intercepts
\(\beta_0\), \(\gamma_0\) in the covariate models as constants that are identified by the model
assumptions when the remaining parameters are held fixed. We assume
\(\bbE(R) = 0\) and \(\bbVar(R) = 1\).
To identify \(\beta_0\) and \(\gamma_0\), we first write our covariate
models for the location and scale without the respective intercept
terms, i.e. \(\tilde{\mu}(\bsx) = \mu(\bsx) - \beta_0\) and
\(\tilde{\sigma}(\bsx) = \sigma(\bsx) / \exp(\gamma_0)\). Now, if we
treat \(\beta_0\) and \(\gamma_0\) as constants, we obtain the
expressions \begin{equation}\protect\hypertarget{app-eq-intercepts}{}{
        \begin{aligned}
            \beta_0 = \bbE \left( \frac{1}{\tilde{\sigma}(\bsx)} \right)^{-1} \bbE \left(\frac{y - \tilde{\mu}(\bsx)}{\tilde{\sigma}(\bsx)}\right)
            \quad \text{and} \quad
            \gamma_0 & = \ln \left(\sqrt{\bbVar\left(
                \frac{y - \beta_0 - \tilde{\mu}(\bsx)}{\tilde{\sigma}(\bsx)}
                \right)}\right).
        \end{aligned}
    }\label{app-eq-intercepts}\end{equation} During MCMC sampling, we replace
the expected value and the variance with their respective sample
estimators, so that \(\beta_0\) and \(\gamma_0\) are always
available as deterministic functions of the data and the remaining model
parameters.
Consequently, \(\beta_0\) and \(\gamma_0\) can be updated via~\eqref{app-eq-intercepts}
in each iteration of the sampling algorithm.

\subsection{IWLS proposals for \texorpdfstring{$\bsbeta_\ell$ and $\bsgamma_s$}{beta ell and gamma s}}

\textcite{Klein2015-BayesianStructuredAdditiveb} found that using the expectation of
the negative second derivative $\bbE(\bfF(\bstheta^{[t-1]}))$
in $\bfF(\bstheta^{[t-1]})^{-1}$ improved sampling efficiency
in distributional regression models.
For $\bstheta = \bsbeta_\ell$ and $\bstheta = \bsgamma_s$, we can approximate
$\bbE(\bfF(\cdot))$ by the expected negative second derivatives of the log full
conditionals derived under the assumption of a Gaussian likelihood. They are then given by
$\bbE(\bfF(\bstheta)) \approx \bfB_\theta^\sfT \bfW_\theta \bfB_\theta + \frac{1}{\tau^2_\theta}\bfK_\theta$.
The matrix $\bfB_\theta$ is the respective $N \times D$ basis matrix of each term,
$\bfK_\theta$ is the $D \times D$ penalty matrix and $\tau^2_\theta$ is
the variance parameter corresponding to each term, with $N$
denoting the number of response observations.
The matrices $\bfW_\theta$ are $N \times N$ diagonal weight matrices with
elements $w^\beta_{ii} = 1 / \sigma(\bsx_i)^2$ and $w^\gamma_{ii} = 2$.
Different from the approach described in the main text, this alternative does not rely
on initial values $\bstheta^{[0]}$.

\subsection{Clipping of \texorpdfstring{$\tau^2_\delta$}{tau delta squared}}

We encountered numerical issues in some cases, when the model estimates
$\tau^2_\delta$ to be very close to zero: Sampling for $\tilde \bsdelta$ stopped
completely, with no proposals getting accepted. The problem can be solved by putting
a lower bound on $\tau^2_\delta$ for estimation. Specifically, we bound
$\ln \tau^2_\delta$ to be $\geq -11$ during sampling, such that $\tau^2_\delta \geq \exp(-11) \approx 0.000017$.

\clearpage

\section{Data-generating mechanisms for simulation studies}
\label{app-sec:data}

The data-generating mechanisms used in our simulations are as follows:
\begin{itemize}
    \item Gaussian: Draw $r_i \sim \calN(0, 1)$ for $i = 1, \dots, N$.
    \item PTM: Draw $z_i \sim \calN(0, 1)$ for $i = 1, \dots, N$. Next, draw $\bsdelta$ from its prior using $\tau_\delta = 0.2$. This is done by drawing $\tilde \bsdelta \sim \calN(\bfzero, \tau^2 \bfI_{J-2})$ and computing $\bsdelta = \bsGamma_2 \bsOmega_2^{-1/2} \tilde \bsdelta$. See \autoref{app-sec:sum-to-zero} for details on $\bsGamma_2$ and $\bsOmega_2$.
          Next, compute the expectation and variance of the implied distribution with CDF $\tilde F_R(r) = \Phi(h(r | \bsdelta))$ by numerical integration; denote them as $m$ and $s^2$. Compute $\tilde z_i = s z_i + m$.
          Then, numerically invert the transformation function to obtain $r_i = h^{-1}(\tilde z_i | \bsdelta)$, which then has expectation zero and variance one. We use $J-1 = 15$ parameters and $-a = b = 3$ and set $\lambda = 0.1(b-a) = 0.6$.

    \item Skewnorm: Draw $z_i$ for $i = 1, \dots, N$ from a standardized skew-normal distribution with density $f(z_i) = 2\phi(z_i)\Phi(\alpha z_i)$, where $\phi(\cdot)$ and $\Phi(\cdot)$ denote the standard normal density and cumulative distribution function, respectively and $\alpha$ is the slant parameter. We use $\alpha = 5$ for a notable right-skew. Compute standardized samples $r_i = (z_i - \mathrm{E}) / \sqrt{\mathrm{V}}$, where the expectation is $\mathrm{E} = (\alpha / \sqrt{1 + \alpha^2}) \cdot \sqrt{2 / \pi} \approx 0.78$ and the variance is $\mathrm{V} = 1 - (2 / \pi) \cdot (\alpha^2 / (1 + \alpha^2)) \approx 0.39$.

    \item Mixture: Draw $z_i$ for $i = 1, \dots, N$ from a two-component mixture of Gaussian distributions with means $\bsmu = (-2, 1)$, standard deviations $\bssigma = (1, 0.5)$ and mixing probabilities $p_1 = p_2 = 0.5$. This creates a bimodal distribution. Compute standardized samples $r_i = (z_i - \mathrm{E}) / \sqrt{\mathrm{V}}$, where the expectation is $\mathrm{E} = p_1 \mu_1 + p_2 \mu_2 = 0.75$ and the variance is $\mathrm{V} = p_1 \sigma_1^2 + p_2 \sigma_2^2 + p_1 p_2 (\mu_1 - \mu_2)^2 = 2.875$.

    \item U-shaped: Draw $z_i \sim \mathrm{Beta}(a, b)$ for $i = 1, \dots, N$ with $a = b = 0.5$, which is a U-shaped distribution. Compute standardized samples $r_i = (z_i - \mathrm{E}) / \sqrt{\mathrm{V}}$, where the expectation is $\mathrm{E} = a / (a + b) = 0.5$ and the variance is $\mathrm{V} = (ab) / ((a+b)^2(a+b+1)) = 0.125$.

    \item Uniform: Draw $z_i \sim \mathrm{Unif}(0, 0.1)$ for $i = 1, \dots, N$. Compute standardized samples $r_i = (z_i - \mathrm{E}) / \sqrt{\mathrm{V}}$, where the expectation is $\mathrm{E} = 0.5$ and the variance is $\mathrm{V} = 1 / 12$.
\end{itemize}

\autoref{app-fig-sdprior-references} displays the U-shaped, bimodal, skewed, and uniform densities used here.

\section{Simulation study 1: Unconditional model}
\label{app-sec:sim1}

Our first simulation study is designed to explore the influence of hyperparameters and the effect of prior specifications without considering covariates. We conduct the simulation in two parts. First, we focus on hyperparameter settings for the PTM, comparing model performance on Gaussian data and data generated from varying PTMs. Second, we apply the model to four additional, hand-crafted datasets and compare PTM performance to a Gaussian model with constant priors and to kernel density estimations.

We use sample sizes $N_{\text{train}} \in \{25, 50, 100, 500\}$ to cover small-sample situations. We generate data from the Gaussian, PTM, skewnorm, mixture, u-shaped and uniform scenarios described in \autoref{app-sec:data}.

To demonstrate independence from starting values, we initialize each element of $\bsdelta$ with a draw from a $\mathrm{Unif}(-1, 1)$ distribution. We set the initial $\tau_\delta = 1$.

Note that our model is misspecified for the U-shaped and uniform data-generating mechanisms, since these produce sharply bounded data, while our model assumes that the response can take values on the full real line. This is intentional; it provides a challenging situation for our model. For each data-generating mechanism, we draw $100$ datasets of size $N=2\,000$. We apply the PTM, the Gaussian model, and the kernel density estimation to a subset $r_1, \dots, r_{N_{\text{train}}}$ observations, where ${N_{\text{train}}} \in \{25, 50, 100, 500\}$ as mentioned above. We use $1\,500$ observations as testing data when evaluating out-of-sample predictive performance. Note that, since we randomly draw $\bsdelta$, this procedure actually creates a different underlying distributional shape for each dataset of the PTM data-generating mechanism. This is intentional, since it creates a wide variety of shapes and thus allows for a broad exploration of model behavior.

\subsection{Performance criteria}

We consider the following diagnostic criteria:

\begin{itemize}
    \item Minimum effective samples drawn per minute for any parameter in $\bsdelta$ and for $\tau^2_\delta$. We compute the \enquote{bulk} and \enquote{tail} effective sample sizes separately.
          This quantifies sampling efficiency and helps identify potential issues in sampling difficult parameters. Separate evaluation of bulk and tail ESS provides insight into information in central vs. extreme parts of the posterior. The quantities are computed using the Python library Arviz \parencite{Kumar2019-ArviZUnifiedLibrary}, version 0.21.0.

    \item Maximum $\hat{R}$ for any parameter in $\bsdelta$ and for $\tau^2_\delta$.
          The $\hat{R}$ diagnostic assesses convergence across chains; values close to 1 indicate good convergence. Taking the maximum highlights the worst-performing parameter.

    \item Acceptance probability for proposed draws.
          A well-tuned sampler typically achieves acceptance rates within a target range (e.g., 0.6–0.9 in HMC), helping to avoid overly small or overly large step sizes.

    \item Relative frequency of divergent transitions.
          Divergences indicate regions of high curvature or problematic geometry and are critical warnings in HMC-based methods.
\end{itemize}

We consider the Kullback-Leibler divergence (KLD) as a metric for predictive performance. Specifically, we compute the divergence from the true log density to the estimated predictive log density on the test observations with indices $i = 501, \dots, N$ as
$$
    \text{KLD} = \frac{1}{T(N-500)} \sum_{t=1}^T \sum_{i=501}^N (\ln f(r_i) - \ln \hat{f}^{(t)}(r_i)).
$$
The index $t = 1, \dots, T$ identifies the individual MCMC samples. Smaller KLD values indicate a better approximation of the true distribution, with a minimum of zero.

\subsection{Hyperparameter configurations}

\begin{figure}
    \includegraphics[width=\textwidth]{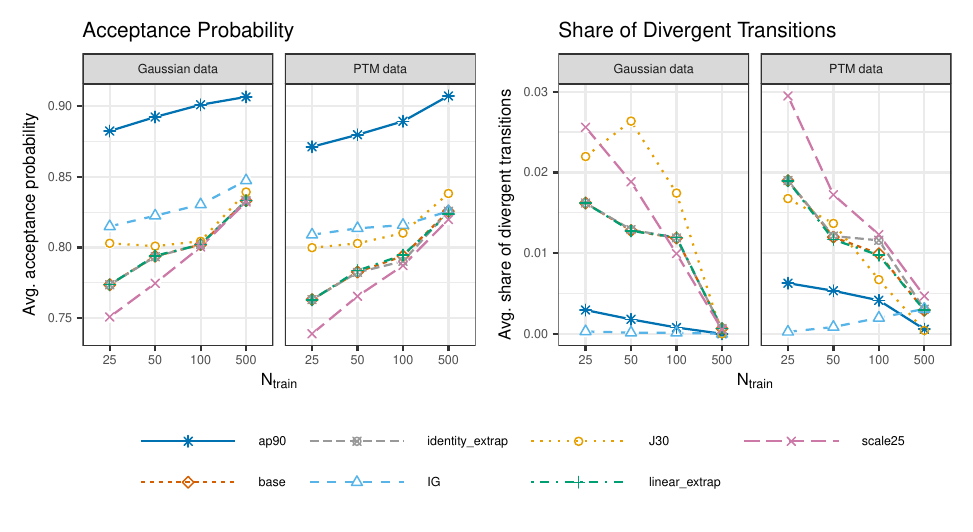}
    \caption[Acceptance probabilities and divergent transitions, Simulation 1, Part 1.]{Acceptance probabilities and relative frequencies of divergent transitions from the first part of the unconditional simulation study, averaged over the results from $100$ simulation runs for each sample size, data type, and hyperparameter configuration.\label{app-fig:sim1-aprob-errors}}
\end{figure}

We compare seven hyperparameter configurations:

\begin{itemize}
    \tightlist
    \item \texttt{base}: A penalized transformation model without covariates, such that $F(r_i) = \Phi(h( (r_i - \mu) / \sigma | \bsdelta))$. The prior for $\bsdelta$ is the random walk prior as described in the main text, with random walk variance $\tau^2_\delta$. The prior for $\tau^2_\delta$ is the scale-dependent prior $\text{Weibull}(0.5, \theta)$ with $\theta = 0.5$. We set $-a = b = 4$ and use $J-1 = 15$ parameters in $\bsdelta$. We set $\lambda = 0.1(b-a)$.
          We sample $\bsdelta$ and $\log(\tau^2)$ jointly using the No-U-Turn Sampler (NUTS) and use a target acceptance probability of $.8$ in the warmup for NUTS. We use $5\,000$ warmup samples and $10\,000$ posterior samples, each in four independent chains, resulting in a total of $40\,000$ posterior samples.
          We consider this configuration as the base configuration. The other configurations are created by varying individual settings.
    \item \texttt{IG}: Like \texttt{base}, but using an inverse gamma hyperprior for $\tau^2$, with concentration $a = 0.01$ and scale $b = 0.01$, which are common default settings for producing an uninformative prior. This hyperprior allows us to sample $\tau^2_\delta$ using a Gibbs update. The full conditional is  $\tau^2 | \bsdelta \sim \calI \calG(\tilde a, \tilde b)$ with $\tilde a = a + \frac{1}{2} \mathrm{rank}(\bfK)$ and $\tilde b = b + \frac{1}{2} \bsdelta^\sfT \bfK \bsdelta$, where $\bfK$ is the penalty matrix used in the prior for $\bsdelta$. In this case, we sample $\bsdelta$ and $\tau^2$ in separate blocks: 1) Draw a sample of $\bsdelta | \tau^2$ using NUTS, 2) draw a sample of $\tau^2$ from its full conditional. The inverse gamma prior is an important comparison, because it is widely used for P-splines and computationally convenient.
    \item \texttt{ap90}: Like \texttt{base}, but using a target acceptance probability of $.9$ in the warmup for NUTS. This leads to smaller step sizes in proposal generation and may make it easier for the model to explore complex posterior landscapes.
    \item \texttt{J30}: Like \texttt{base}, but using $J-1 = 30$ parameters in $\bsdelta$. This equips the model with additional flexibility.
    \item \texttt{scale25}: Like \texttt{base}, but using a high scale $\theta = 25$ in the Weibull hyperprior for $\tau^2_\delta$.
    \item \texttt{identity\_extrap}: Like \texttt{base}, but using $\lambda \rightarrow 0$ for extrapolating the transformation function with the identity function outside $[a, b]$.
    \item \texttt{linear\_extrap}: Like \texttt{base}, but using $\lambda \rightarrow \infty$ for linear extrapolation of the transformation function.
\end{itemize}
\begin{figure}
    \includegraphics[width=\textwidth]{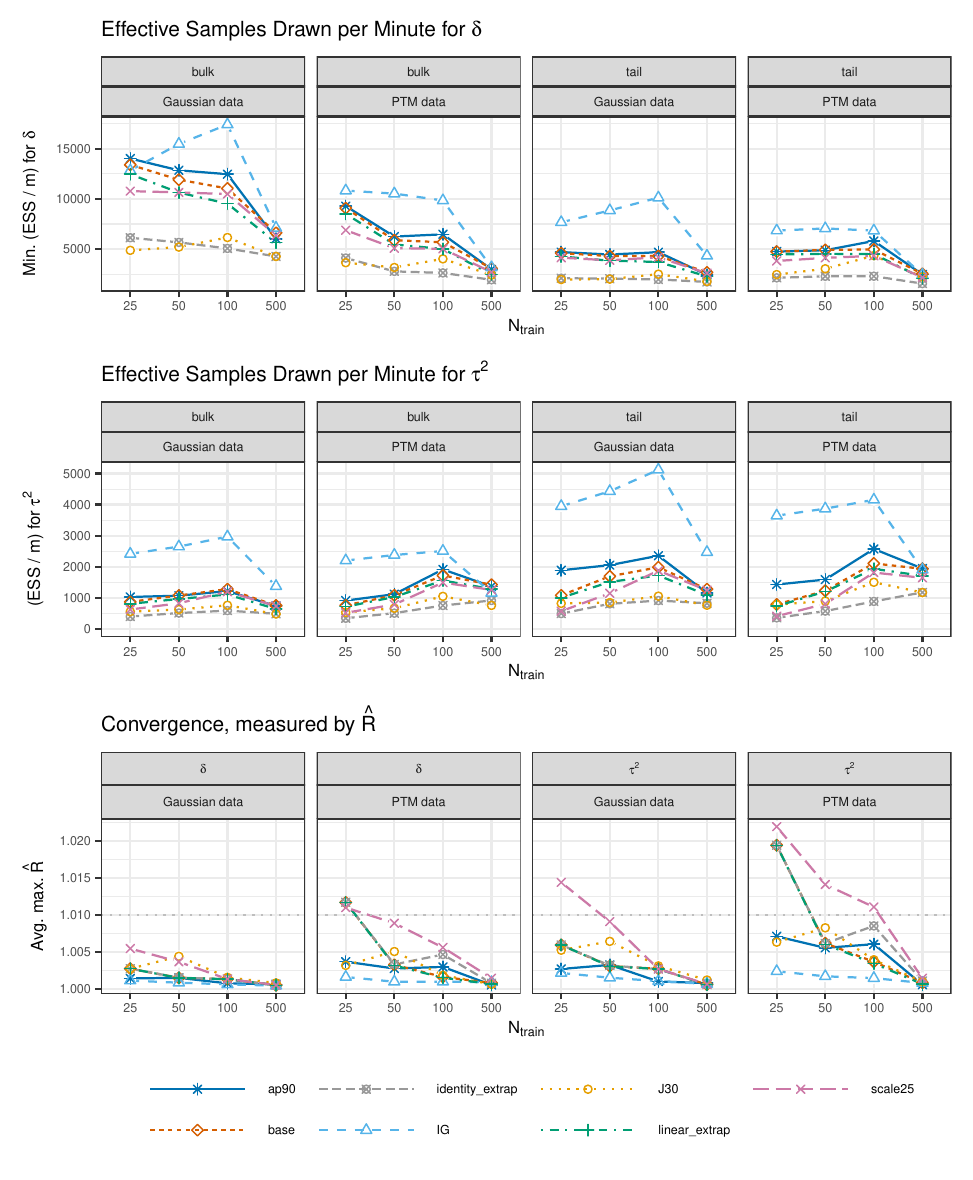}
    \caption[Diagnostics, Simulation 1, Part 1.]{Effective samples drawn per minute and convergence measure $\hat{R}$ from the first part of the unconditional simulation study, averaged over the results from $100$ simulation runs for each sample size, data type, and hyperparameter configuration. For $\bsdelta$, we display the average over the minimum effective sample size for any of the $J-1$ parameters in $\bsdelta$ to provide a conservative estimate.\label{app-fig:sim1-ess-rhat}}
\end{figure}
\autoref{app-fig:sim1-ess-rhat} summarizes this first part of the simulation study with respect to sampling efficiency and convergence. We note the following:
\begin{itemize}
    \tightlist
    \item The inverse gamma hyperprior (\texttt{IG}) leads to superior sampling efficiency and convergence, providing consistently higher numbers of effective samples and lower $\hat{R}$ than the other configurations. The advantage of the inverse gamma prior however seems to vanish with a moderately high sample size of $N_{\text{train}} = 500$.
    \item With effective samples per minute consistently close to or larger than $2\,500$, sampling efficiency appears to be generally feasible for moderately-sized datasets.
    \item Generally, sampling efficiency is higher for $\bsdelta$ than for $\tau^2$ and drops considerably with increasing sample size, indicating that scaling to higher sample sizes is challenging for the PTM.
    \item Of the configurations with a scale-dependent prior for $\tau^2_\delta$, \texttt{ap90} appears to be most efficient and show the best convergence diagnostics, with an average $\hat{R} < 1.01$ in all conditions and for all data types.
    \item The configuration \texttt{scale25} appears to converge worse and be less efficient than other configurations.
    \item Differences between \texttt{base}, and \texttt{linear\_extrap} are small, but \texttt{identity\_extrap} appears to be notably less efficient with small sample sizes.
\end{itemize}
\autoref{app-fig:sim1-aprob-errors} additionally summarizes the achieved acceptance probabilities and divergent transitions that occurred during MCMC sampling. We note the following:

\begin{itemize}
    \tightlist
    \item The acceptance probability is close to the target level in all cases. Generally, acceptance probability rises slightly as sample size increases.
    \item All configurations show sharply reduced relative frequencies of divergent transitions as sample size increases to $N_{\text{train}} = 500$. For small sample sizes, all configurations except for \texttt{ap90} and \texttt{IG} show considerable relative frequencies of divergent transitions between one and three percent. \texttt{IG} yields the lowest numbers, staying close to 0\% divergent transitions at all sample sizes. Of the other configurations, \texttt{ap90} shows the lowest relative frequency, staying well below 1\% even for the most challenging case of PTM data and $N_{\text{train}} = 25$.
\end{itemize}

\begin{figure}
    \includegraphics[width=\textwidth]{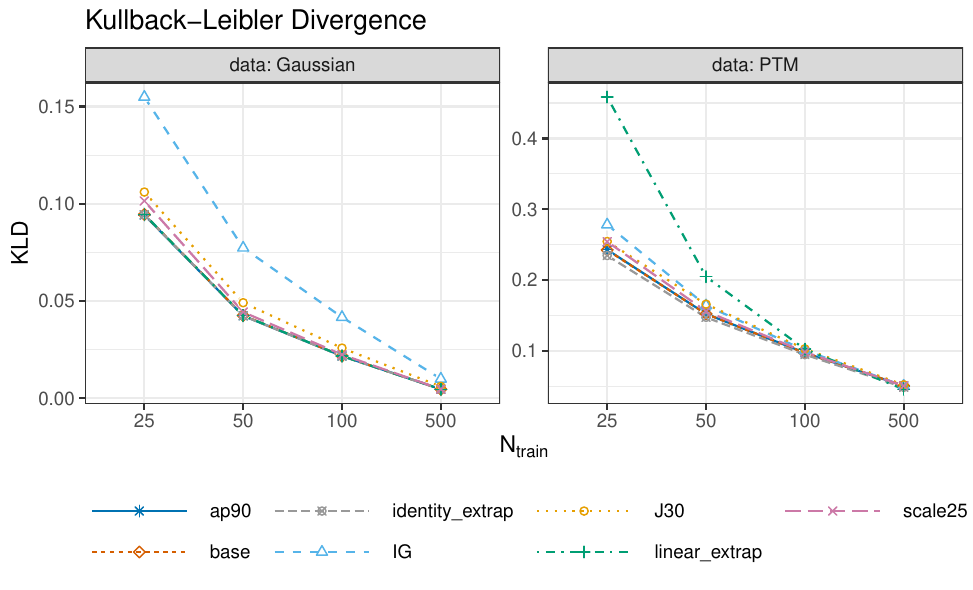}
    \caption[KL divergence, Simulation 1, Part 1.]{Kullback-Leibler divergence for evaluating predictive distributional performance in the first part of the unconditional simulation study, averaged over the results from $100$ simulation runs for each sample size, data type, and hyperparameter configuration. Smaller values indicate a better performance.\label{app-fig:sim1-performance-hyper}}
\end{figure}

\autoref{app-fig:sim1-performance-hyper} shows our measures for distributional performance in the first part of the unconditional simulation study. We note the following:
\begin{itemize}
    \tightlist
    \item Performance converges as sample size increases, implying that hyperparameter configurations matter most for small sample sizes.
    \item \texttt{IG} stands out as a configuration that consistently performs slightly worse than the other configurations, particularly for Gaussian data and small sample sizes. This result suggests that the scale-dependent prior for $\tau^2_\delta$ better allows the PTM to fall back to the Gaussian base model, if the data do not provide strong evidence for deviations from Gaussianity. Thus, the observation is consistent with the reasoning for our choice of the scale-dependent prior.
    \item \texttt{linear\_extrap} performs notably worse than other configurations for PTM data with small sample sizes. Recall that the PTM data were generated using a transition from linear to identity extrapolation, such that the linear extrapolation constitutes a model misspecification in the tails here. It appears reasonable that this pattern would become visible in the Kullback-Leibler divergence. The result suggests that theoretical considerations about the model's desired tail behavior can be relevant if tail behavior is of interest.
    \item Except for \texttt{linear\_extrap} and \texttt{IG}, all configurations show practically indistinguishable performance.
\end{itemize}

\paragraph{Conclusion.} In sum, we observe that the scale-dependent prior yields the expected benefit for Gaussian data with small sample sizes. Among the configurations with scale-dependent prior, \texttt{ap90} shows superior diagnostic measures in terms of convergence and the relative frequency of divergent transitions, and, to a lesser extent, sampling efficiency, while there is no discernible difference in terms of predictive distributional performance. As a result, we use the configuration \texttt{ap90} as the default configuration in the following.

\subsection{Different data scenarios}

In the second part of the unconditional simulation study, we compare the following models:

\begin{itemize}
    \item \texttt{ptm-15}: A penalized transformation model, using the same configuration as \texttt{base} in the first part of the unconditional simulation, but with target acceptance probability $.9$ for hyperparameter tuning in the warmup period of the No-U-Turn Sampler.
    \item \texttt{ptm-30}: Like \texttt{ptm-15}, but using $J-1 = 30$ instead of $J-1 = 15$. We include this model, since a higher number of parameters may be useful for the misspecified models.
    \item \texttt{gaussian}: A Gaussian model $r_i \sim \calN(\mu, \sigma^2)$, where $\mu$ and
          $\log(\sigma)$ are sampled jointly with the No-U-Turn Sampler using constant priors. Provides a baseline comparison.
    \item \texttt{kde}: Kernel density estimation using a Gaussian kernel with bandwidth selected via the Sheather–Jones plug-in method \parencite{Sheather1991-ReliableDatabasedBandwidth}. Provides a baseline comparison for data-driven nonparametric density estimation without uncertainty quantification.
\end{itemize}

\begin{figure}
    \includegraphics{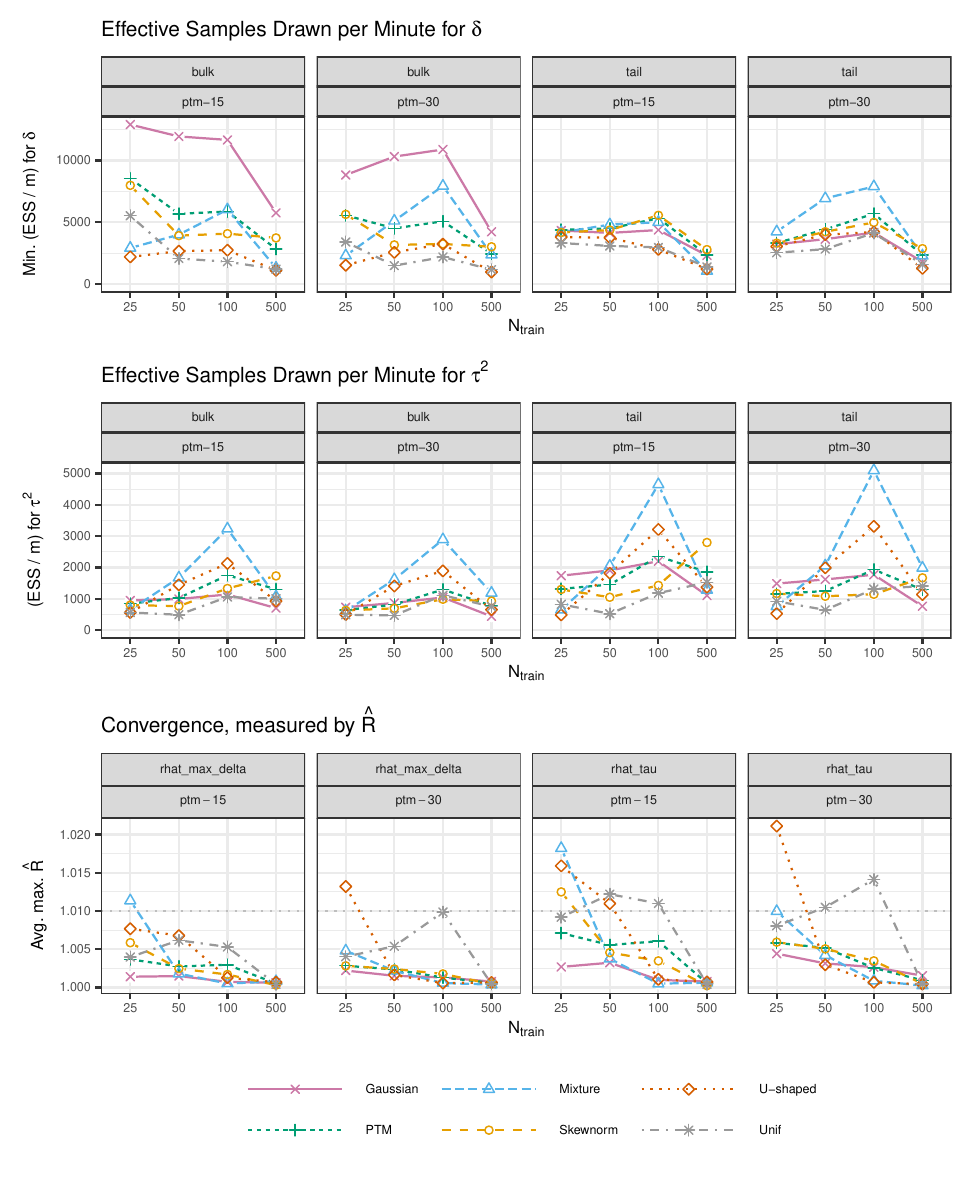}
    \caption[Diagnostics, Simulation 1, Part 2.]{Effective samples drawn per minute and convergence measure $\hat{R}$ from the second part of the unconditional simulation study, averaged over the results from $100$ simulation runs for each sample size, data type, and model. For $\bsdelta$, we display the average over the minimum effective sample size for any of the $J-1$ parameters in $\bsdelta$ to provide a conservative estimate.\label{app-fig:sim1-ess-rhat-all_data}}
\end{figure}

\begin{figure}
    \includegraphics{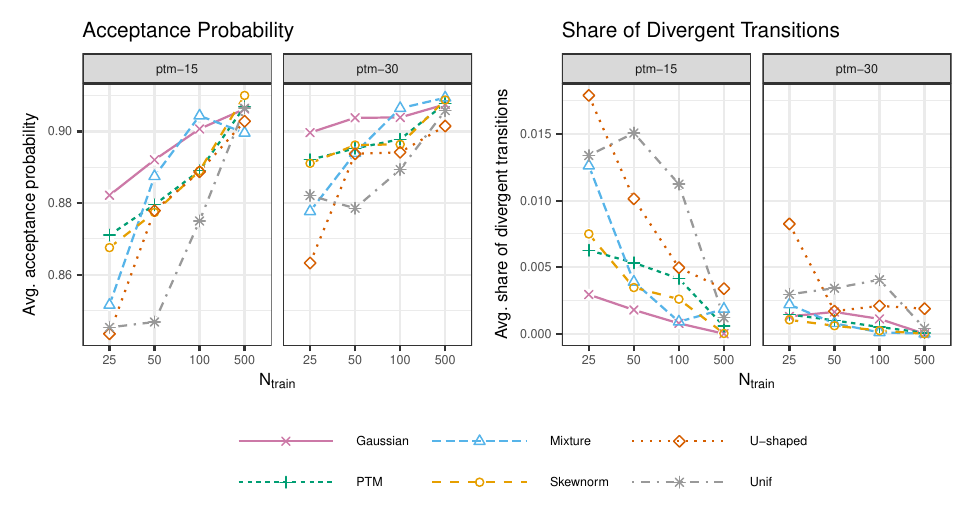}
    \caption[Acceptance probabilities and divergent transitions, Simulation 1, Part 2.]{Acceptance probabilities and relative frequencies of divergent transitions from the second part of the unconditional simulation study, averaged over the results from $100$ simulation runs for each sample size, data type, and model.\label{app-fig:sim1-aprob-errors-all_data}}
\end{figure}

Diagnostic information for the models \texttt{ptm-15} and \texttt{ptm-30} is displayed in \autoref{app-fig:sim1-ess-rhat-all_data}, showing the number of effective samples drawn per minute and convergence measured by $\hat{R}$, and \autoref{app-fig:sim1-aprob-errors-all_data}, showing the acceptance probability for MCMC proposals and the relative frequency of divergent transitions. The challenging nature of the investigated scenarios, particularly \texttt{Mixture} and the misspecified scenarios \texttt{Unif} and \texttt{U-shaped} surfaces in a mild tendency for lower sampling efficiency and larger $\hat{R}$, which particularly with small sample sizes crosses the rule-of-thumb threshold of $1.01$ in some cases. Nonetheless, $\hat{R}$ stays below $1.02$ in all but one case and only minimally above $1.02$ in the exception, indicating that convergence issues do not tend to be severe. The acceptance probability of MCMC proposals stays well above $80\%$ in all cases. The relative frequency of divergent transitions remains below $0.02$ in all cases and below $0.01$ for \texttt{ptm-30}. We interpret this as a slight indication in favor of the more complex model \texttt{ptm-30}.

\begin{figure}
    \includegraphics[width=\textwidth]{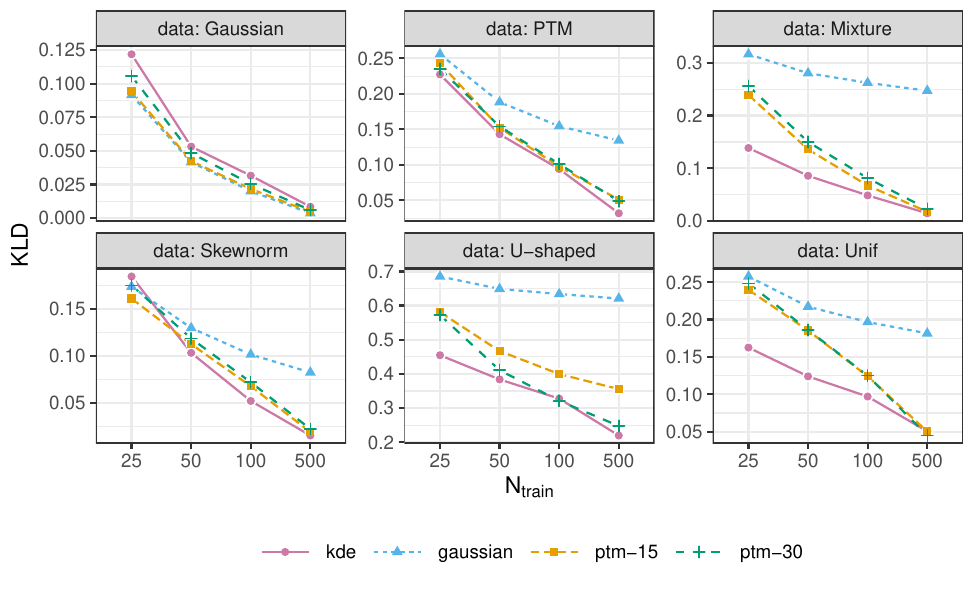}
    \caption[KL divergence, Simulation 1, Part 2.]{Kullback-Leibler divergence for evaluating predictive distributional performance in the second part of the unconditional simulation study, averaged over the results from $100$ simulation runs for each sample size, data type, and model. Smaller values indicate a better performance.\label{app-fig:sim1-kld-all_data}}
\end{figure}

\begin{figure}
    \includegraphics[width=\textwidth]{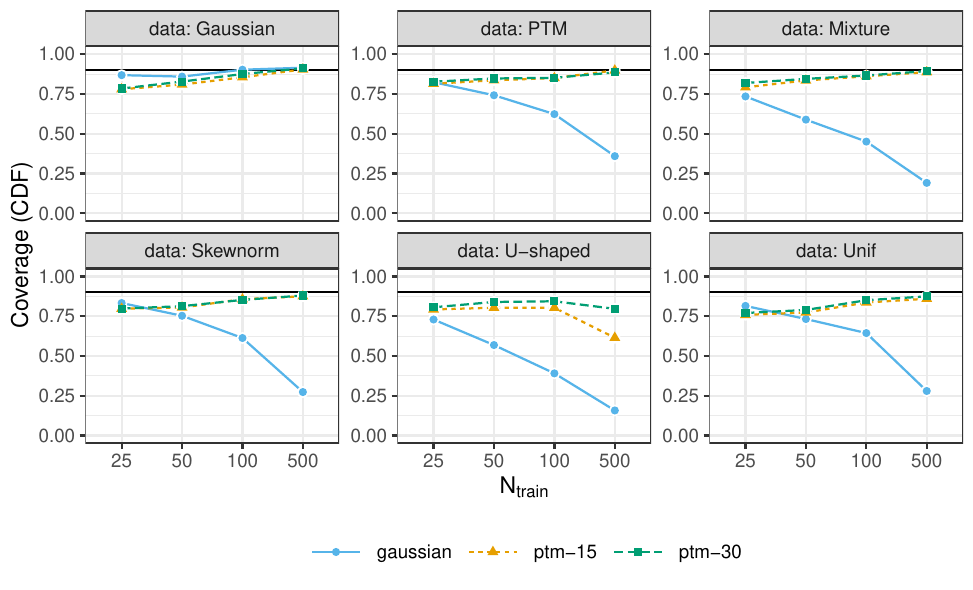}
    \caption[Coverage, Simulation 1, Part 2.]{Average pointwise coverage probability for $90\%$ highest posterior density intervals. The target level $.90$ is marked by a solid black horizontal line.\label{app-fig:sim1-coverage}}
\end{figure}

\autoref{app-fig:sim1-kld-all_data} shows the Kullback-Leibler divergence to the true distribution, our metric for evaluating distributional fit. We note the following:
\begin{itemize}
    \tightlist
    \item For Gaussian data, the performance of all models is very similar. Particularly, \texttt{ptm-15} and \texttt{gaussian} appear to be indistinguishable, while \texttt{ptm-30} shows a minimally higher divergence with small sample sizes. The kernel density estimate again shows a minimally higher divergence than \texttt{ptm-30}.
    \item In all scenarios, the performance of the PTMs converges to the performance of the kernel density estimation with growing sample size. With $N_{\text{train}}=100$, the difference to kernel density estimation can be considered small, and with the next higher sample size in our setup, $N_{\text{train}}=500$, PTM performance is almost indistinguishable from kernel density estimation in all scenarios.
    \item For \texttt{PTM} data and \texttt{Skewnorm} data, the patterns look very similar. All models perform roughly on the same level with the smallest sample size $N_{\text{train}}=25$, and differences start to show and grow larger as sample size increases. Specifically, the divergence measured for \texttt{ptm-15}, \texttt{ptm-30} and \texttt{kde} drops notably faster than it does for \texttt{gaussian}; and the three former models appear to get closer to the true distribution at a similar pace.
    \item For \texttt{Mixture} data and \texttt{Unif} data, the divergence measured for \texttt{ptm-15}, \texttt{ptm-30} and \texttt{kde} again drops notably faster than it does for \texttt{gaussian}. However, \texttt{kde} gets closer to the true distribution than the PTM models with small sample sizes, the performance only converges with $N_{\text{train}}=500$.
    \item For \texttt{U-shaped} data, \texttt{kde} likewise gets closer to the true distribution than the PTMs with small sample sizes. However, \texttt{ptm-30} performs better than \texttt{ptm-15}, getting close to \texttt{kde} with $N_{\text{train}}=50$ and becoming indistinguishable from it at $N_{\text{train}}=100$.
\end{itemize}
In this part of the simulation, we also considered an aggregated estimate of pointwise calibration of highest posterior density intervals for the CDF, computed as
$$
    \text{Coverage}(\hat{F}) = \frac{1}{N-500} \sum_{i=501}^N \bbI(F_R(r_i) \in [\hat{F}_{\text{low}}(r_i), \hat{F}_{\text{high}}(r_i)]),
$$
where $\hat{F}_{\text{low}}(r_i)$ and $\hat{F}_{\text{high}}(r_i)$ are the lower and upper boundaries of a $0.90$ highest posterior density interval computed on the full sample of posterior CDF evaluations $\hat{F}^{(t)}(r_i), t = 1, \dots, T$ and $\bbI(\cdot)$ denotes the indicator function for the enclosed condition.
\autoref{app-fig:sim1-coverage} shows the results. We can observe that the PTMs yield highest posterior density intervals with close to nominal coverage, and that coverage converges to the nominal level as sample size increases. Unsurprisingly, the same is not true for the Gaussian baseline comparison model, which yields similar coverage levels as the PTMs only for Gaussian data and the smallest sample size $N_{\text{train}=25}$. In the \texttt{U-shaped} data scenario, the \texttt{ptm-15} model shows a notable drop in coverage as sample size increases from $N_{\text{train}=100}$ to $N_{\text{train}=500}$, arriving at $.615$ for $N_{\text{train}=500}$. While this is still drastically higher than the Gaussian comparison at $.158$, it may reflect a limitation of the less flexible model with $J-1 = 15$ parameters in the transformation function to accurately capture the features of the observed data; thus growing more certain about a suboptimal fit. While the more flexible \texttt{ptm-30} at $.797$ also does not reach the nominal level, it stays much closer, indicating that the added flexibility helps.

\paragraph{Conclusion.}
As the PTMs perform better than the Gaussian distribution in all non-Gaussian scenarios and converge to be on par with kernel density estimation as sample size grows, we interpret these results as demonstrating the ability of PTMs to effectively capture distributional information from observed data. Notably, the convergence of PTM and kernel density estimation performance even holds for challenging scenarios like the multi-modal \texttt{Mixture} case and, due to their strict boundedness, drastically misspecified scenarios like \texttt{U-shaped} and \texttt{Unif}. The discrepancy to kernel density estimation at smaller sample sizes in these three scenarios can be attributed to our prior setup in the PTMs, which penalizes the model towards the Gaussian reference distribution. For PTMs, strong deviations from this reference are only warranted with strong empirical evidence, so the observed pattern shows us that the model is behaving as expected and intended. We observe some indications that a higher number of parameters in the transformation function provides benefits without doing harm: First, we observe a smaller relative frequency of divergent transitions with \texttt{ptm-30} than with \texttt{ptm-15}; second, we observe superior performance of \texttt{ptm-30} compared to \texttt{ptm-15} in terms of Kullback-Leibler divergence and coverage; and third, we do not observe substantial disadvantages of \texttt{ptm-30} compared to \texttt{ptm-15} in terms of any other diagnostic or performance measure.

\clearpage

\section{Simulation study 2: Conditional model}
\label{app-sec:sim2}

\subsection{Covariate functions}
\label{app-sec:covfun}
\FloatBarrier

\begin{figure}[ht]
    \includegraphics[width=\textwidth]{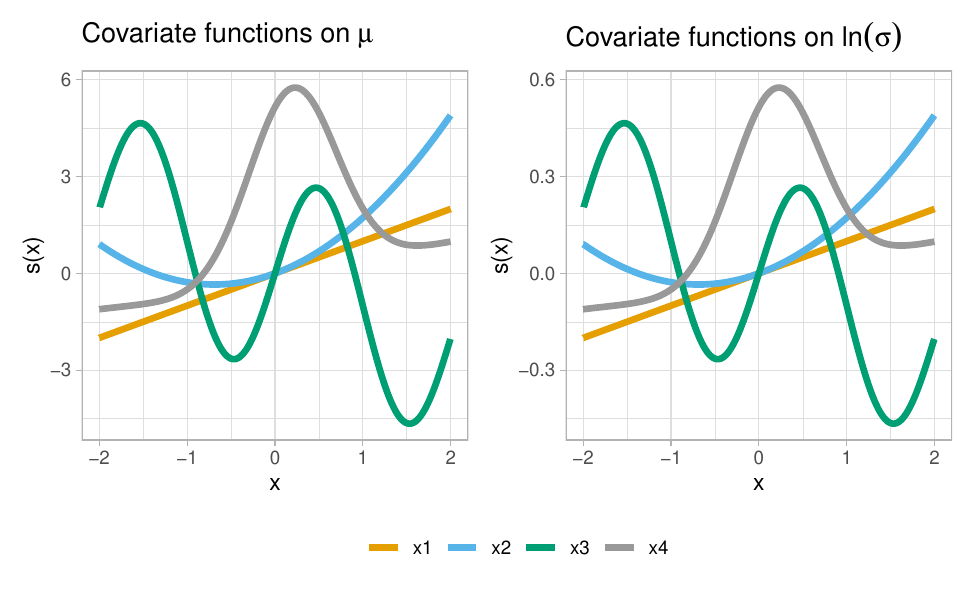}
    \caption[Covariate functions in Simulation 2.]{Covariate functions used in the simulation study. Note the different scales of the y-axes. The lower scale for the scale functions is due to them operating on log-level, leading to multiplicative effects on the level of $\sigma$. \label{app-fig:sim2-covariate_functions}}
\end{figure}

\FloatBarrier

\subsection{Model specifications}
\label{app-sec:sim2-models}

\begin{itemize}
    \tightlist
    \item PTM:
          Bayesian penalized transformation models (PTMs) with a standard Gaussian
          reference distribution and \(J-1 = 30\) transformation function parameters in \(\bsdelta\).
          The prior for \(\bsdelta\) is the random walk prior described in \mainautoref{sec-ptm-prior},
          with random walk variance $\tau^2_\delta \sim \text{Weibull}(0.5, 0.5)$.
          We set up the knot base with $-a = b = 4$, and use $\lambda = 0.1(b-a)$ in the transformation function.
          The models can be written as
          \[\bbP(Y \leq y_i) = \Phi\left( h \left(\frac{y_i - \mu(\bsx_i)}{\sigma(\bsx_i)} \right) \right),\]
          where \(\bsx_i = [x_{i1}, x_{i2}, x_{i3}, x_{i4}]^\sfT\). The location
          model part is specified as
          \[\mu(\bsx_i) = \beta_0 + s_1(x_{i1}) + s_2(x_{i2}) + s_3(x_{i3}) + s_4(x_{i4}),\]
          where each \(s_1, \dots, s_4\) is a 20-parameter Bayesian P-spline.
          The inverse smoothing parameters of these P-splines receive inverse
          gamma hyperpriors with concentration \(1\) and scale \(0.001\). The
          scale model part is specified analogously, additionally using a
          logarithmic link function. For estimation, we apply MCMC as described in the main
          text. We draw
          $25\,000$ samples each in four chains after a warmup of $5\,000$ iterations and use a thinning factor of $5$, yielding a total of $20\,000$ samples.
    \item Gaussian: Gaussian location-scale models, which can be written in analogy to PTMs
          as
          \[\bbP(Y \leq y_i) = \Phi\left(\frac{y_i - \mu(\bsx_i)}{\sigma(\bsx_i)} \right)\]
          by letting the transformation function \(h\) equal the identity
          function. The covariate models \(\mu(\bsx_i)\) and \(\sigma(\bsx_i)\)
          are specified exactly as in the PTMs. For estimation, we apply MCMC with iteratively
          weighted least squares proposals \parencite{Klein2015-BayesianStructuredAdditiveb} in
          Liesel \parencite{Riebl2023-LieselProbabilisticProgramming}. We draw
          $50\,000$ samples each in four chains after a warmup of $5\,000$ iterations and use a thinning factor of $10$, yielding a total of $20\,000$ samples.
    \item TAMLS: A transformation additive model \parencite{Siegfried2023-DistributionfreeLocationscaleRegression}
          of the form
          $$
              \bbP(Y \leq y_i) = \Phi\left( \sigma(\bsx_i)^{-1}h(y_i) - \mu(\bsx_i) \right),
          $$
          with
          \begin{align}
              \mu(\bsx_i) & = \beta_0 + s_1(x_{i1}) + s_2(x_{i2}) + s_3(x_{i3}) + s_4(x_{i4}).
          \end{align}
          where each \(s_1, \dots, s_4\) is a 20-parameter P-spline. The
          scale model part is specified analogously, but using a
          logarithmic link function and omitting the intercept for identifiability.
          Implementation was done by adapting the example code
          included in the R package \texttt{tram}, available on GitHub: \url{https://github.com/cran/tram/blob/master/demo/stram.R}.
    \item SBGP: A semiparametric Bayesian Gaussian process model as proposed by \textcite{Kowal2024-MonteCarloInference}. The model is
          $g(y_i) = f_\theta(\bsx_i) + \sigma \epsilon_i$
          where $\bsx_i = (x_{i1}, x_{i2}, x_{i3}, x_{i4})^\sfT$, $\epsilon_i \sim \calN(0, 1)$ and $f_\theta \sim \calG \calP(m_\theta, K_\theta)$. In the Gaussian process,
          $m_\theta$ is an unknown constant mean and $K_\theta$ is an isotropic Matérn covariance
          function with unknown variance, range, and smoothness parameters.
          We tested using an anisotropic Matérn covariance function instead, but encountered
          overflow errors in this model variant.
          We also tested
          a version of the model that includes a simple linear term
          $g(y_i) = \bsx_i^\sfT \bsbeta + f_\theta(\bsx_i) + \sigma \epsilon_i$.
          We found that the results
          improved when including this term and report the results for this model version.
          Computations are
          carried out using the code of the R package \texttt{SeBR} as provided by the authors
          in the supplementary materials of \textcite{Kowal2024-MonteCarloInference}.
          We save $5\,000$ posterior predictive samples for each row of the test dataset.
    \item DDPstar: A density regression via Dirichlet process mixtures of normal structured additive regression models \parencite{Rodriguez-Alvarez2024-DensityRegressionDirichleta}. The model sets up
          the conditional density of $y_i$ given $\bsx_i = (x_{i1}, x_{i2}, x_{i3}, x_{i4})^\sfT$ as
          $$
              p(y_i | \bsx_i) = \int \phi(y_i | \mu(\bsx_i), \sigma^2) \mathrm{d} G_{\bsx_i}(\mu(\bsx_i), \sigma^2),
          $$
          with $\phi(\cdot | \mu, \sigma^2)$ denoting the Gaussian density with mean $\mu$ and variance $\sigma^2$.
          The means of the mixture components are modeled as
          \begin{align}
              \mu_l(\bsx_i) & = \beta_{0,l} + s_{1,l}(x_{i1}) + s_{2,l}(x_{i2}) + s_{3,l}(x_{i3}) + s_{4,l}(x_{i4}),
          \end{align}
          where each \(s_1, \dots, s_4\) is a 20-parameter P-spline using the default concentration $1$ and scale $0.005$.
          As \textcite{Rodriguez-Alvarez2024-DensityRegressionDirichleta} do in their example code, we use $20$ mixture components, and otherwise stick to the defaults of their R library \texttt{DDPstar}, which we use to carry out the computations. We draw $15\,000$ samples after a warmup of $5\,000$ iterations and use a thinning factor of $10$, yielding a total of $1\,500$ samples. The aggressive thinning was necessary to keep the computation of performance criteria feasible.
    \item BCTM-LS: Bayesian conditional transformation models \parencite[BCTM,][]{Carlan2023-BayesianConditionalTransformation} with standard normal reference distribution. These BCTMs
          use the form
          \[\bbP(Y \leq y_i) = \Phi\bigl(\gamma_0 + h_0(y) + h_1(x_{i1}) + h_2(x_{i2}) + h_3(x_{i3}) + h_4(x_{i4}) \bigr),\]
          where $h_1, \dots, h_4$ are 20-parameter Bayesian P-splines and $h_0$ is a 20-parameter monotonically
          increasing Bayesian P-spline. Overall intercepts are removed from the partial transformation functions, and a global intercept is added back into the model as $\gamma_0$.
          We use inverse gamma hyperpriors with
          concentration \(1\) and scale \(0.001\) for all inverse smoothing
          parameters of the P-spline terms. We draw
          $5\,000$ samples each in four chains after a warmup of $5\,000$ iterations, yielding a total of $20\,000$ samples. Since sampling is done exclusively with NUTS, we use no thinning.
    \item BCTM-TE: Bayesian conditional transformation models \parencite[BCTM,][]{Carlan2023-BayesianConditionalTransformation} with standard normal reference distribution. These BCTMs
          use the form
          \[\bbP(Y \leq y_i) = \Phi\bigl(h_1(y_i|x_{i1}) + h_2(y_i|x_{i2}) + h_3(y_i|x_{i3}) + h_4(y_i|x_{i4}) \bigr),\]
          where
          \[h_j(y_i | x_{ij}) = \bigl(\gamma_0 + \bsa(y_i)^\sfT \otimes \bsb_j(x_{ij})^\sfT\bigr)^\sfT \bsgamma_j, \qquad j = 1, \dots, 4,\]
          are the BCTM's partial transformation functions, with \(\otimes\)
          denoting the Kronecker product. The bases \(\bsa\) and \(\bsb\) are
          chosen as 8-parameter B-spline bases. Overall intercepts are removed from the partial transformation functions, and a global intercept is added back into the model as $\gamma_0$. The BCTM uses inverse gamma hyperpriors with
          concentration \(1\) and scale \(0.001\) for all inverse smoothing
          parameters of the tensor product terms. We apply MCMC via the No-U-Turn Sampler as available in
          the Python library \texttt{liesel-bctm} \parencite{Brachem2023-LieselbctmBayesianConditional}. We draw
          $5\,000$ samples each in four chains after a warmup of $5\,000$ iterations, yielding a total of $20\,000$ samples. Since sampling is done exclusively with NUTS, we use no thinning.
    \item QGAM: A fully specified additive quantile regression model with P-splines for all four covariates, as available in the R package \texttt{qgam} \parencite{Fasiolo2021-FastCalibratedAdditive,Fasiolo2021-QgamBayesianNonparametric}. We fit QGAM for a grid of 25 evenly spaced probability levels in $\{0.005, \dots, 0.995\}$. Each quantile model is of the form
          \[Q_Y(\alpha | \bsx_i) = \beta_0 + s_{\alpha, 1}(x_{i1}) + s_{\alpha, 2}(x_{i2}) + s_{\alpha, 3}(x_{i3}) + s_{\alpha, 4}(x_{i4}),\]
          where each \(s_{\alpha, 1}, \dots, s_{\alpha, 4}\) is a 20-parameter
          P-spline and \(Q_Y(\alpha | \bsx_i)\) is the conditional
          \(\alpha\)-quantile of the response \(Y\). We include a model $\kappa(\bsx)$ for the variance of the preliminary location-scale GAM fit used in
          the qgam inference algorithm. The term $\kappa(\bsx)$ includes 7-parameter P-splines for all covariates.
\end{itemize}

\subsection{Performance criteria}
\label{app-sec:sim2-performance}

\begin{enumerate}
    \def\labelenumi{\arabic{enumi}.}
    \tightlist
    \item
          The widely-applicable information criterion (WAIC) can be viewed as an
          estimator for the expected log pointwise predictive density \parencite{Vehtari2017-PracticalBayesianModel, Watanabe2010-AsymptoticEquivalenceBayes}. It thus measures predictive performance. On the deviance
          scale, the WAIC is given by
          \(\text{WAIC} = -2(\text{lppd} - p_{\text{WAIC}}),\) where the log
          pointwise predictive density (lppd) is
          \[\text{lppd} = \sum_{i=1}^{N} \ln \left( \frac{1}{T} \sum_{t=1}^T \hat{f}_Y^{[t]}(y_i | \bsx_i) \right)\]
          and the effective number of parameters is given by
          \[p_{\text{WAIC}} = \sum_{i=1}^{N} \widehat{\text{Var}}_{t=1}^T\bigl( \ln \hat{f}_Y^{[t]}(y_i | \bsx_i) \bigr).\]
          The WAIC is a fully Bayesian model choice criterion and can be applied
          without knowledge of the true distribution. On the deviance scale, a
          smaller WAIC indicates a better predictive performance. Here, $N$ refers to the size of the training data set. 
    \item
          The mean absolute difference (MAD) between the true cumulative density
          function (CDF) evaluations of the test data and the estimated CDF for
          the test data, given by
          \[\text{MAD}^{[t]} = \frac{1}{N_{\text{test}}} \sum_{i=1}^{N_{\text{test}}}\ \left| F_{\text{true}}(y_i | \bsx_i) - \hat{F}^{[t]}(y_i | \bsx_i) \right|.\]
    \item
          The Kullback-Leibler divergence (KLD) is a common measure to assess
          the statistical distance between two distributions. We estimate the
          KLD as
          \[\widehat{\text{KLD}}^{[t]} = \frac{1}{N_{\text{test}}} \sum_{i=1}^{N_{\text{test}}} \bigl( \ln f_{\text{true}}(y_i | \bsx_i) - \ln \hat{f}^{[t]}(y_i | \bsx_i) \bigr).\]
    \item
          The continuous ranked probability score (CRPS) is given by
          \begin{equation}\protect\hypertarget{app-eq-crps}{}{\text{CRPS} = \int\limits_0^1 \text{QS}_\alpha (\hat{F}, y) \ \text d \alpha,}\label{app-eq-crps}\end{equation}
          using a generic observations $y$ here,
          see \textcite{Gneiting2007-StrictlyProperScoring} and
          \textcite{Gneiting2011-QuantilesOptimalPoint}. We use different methods to compute the CRPS,
          based on availability of computations:
          \begin{itemize}
              \item Predictive samples. The CRPS can be estimated by comparing a set of
                    predictive samples to a set of test observations. Implementations are available in
                    the R package \texttt{scoringRules}
                    \parencite{Jordan2019-EvaluatingProbabilisticForecasts} and the
                    Python library \texttt{properscoring} \parencite{2015-ProperscoringProperScoring}.
                    We use this method for SBGP, Gaussian, and PTM.
              \item Integrating over the quantile score. We
                    compute the quantile score \(\text{QS}_\alpha\) for a probability level
                    \(\alpha\) for a test observation \(y_i\) as
                    \[\text{QS}_\alpha (\hat{F}, y_i) = \frac{2}{T} \sum_{t=1}^T \biggl( \bigl(\mathbb{I}_{y_i < \hat F^{-1,[t]}(\alpha)} - \alpha) \cdot \bigl(\hat F^{-1,[t]}(\alpha) - y_i\bigr) \biggr),\]
                    where \(\mathbb{I}_{(\cdot)}\) is the indicator function for the
                    condition \((\cdot)\). We evaluate the quantile score for a grid of 25
                    probability levels \(\alpha\), evenly spaced between \(0.005\) and
                    \(0.995\), and evaluate the integral in \eqref{app-eq-crps} numerically
                    using the trapezoid rule. We use this method for TAMLS, DDPstar, and QGAM.
              \item For the BCTM models, we do not record CRPS computations in the
                    simulation study, since we do not have an efficient method for computation available
                    for these models; and they can be compared by KLD and WAIC instead.
          \end{itemize}

    \item CDF credible and confidence intervals: For the Bayesian models, we compute
          credible intervals based on the empirical quantiles of the empirical posterior samples
          $\{\hat{F}^{[t]}_Y(y_i)\}_{t=1}^T$ for each test observation
          $i = 1, \dots, N_{\text{test}}$. We then note whether the true CDF $F_Y(y_i)$ lies
          inside the so-constructed interval and record a value of $1$ if that is the case and
          $0$ otherwise. We compute the coverage by averaging these indicators over all
          test observations. We compute the width by averaging over the difference of the
          upper and lower quantile for all test observations.
          This method is applicable for DDPstar, Gaussian, PTM, BCTM-LS, and BCTM-TE. For TAMLS,
          we used bootstrapping, then applied the same approach to the bootstrapped samples.
    \item MSE of covariate effect: We measure the performance of the covariate effect estimators using the
          mean squared error, $\text{MSE}\bigl(\hat{s}^{[t]}\bigr) = \frac{1}{N}\sum_{i=1}^N \bigl(s_{\text{true}}(x_i) - \hat{s}^{[t]}(x_i)\bigr)^2,$
          averaged over all \(T\) samples.
    \item Credible and confidence intervals for covariate effects: To avoid side-effects
          from sum-to-zero constraints applied to P-splines, we center all true and estimated
          covariate functions within each sample using the test observations. We then apply the
          same procedure as for the CDF credible and confidence intervals.
\end{enumerate}

\subsubsection{Evaluating the conditional CDF and density for SBGP}

The semiparametric Bayesian Gaussian Process model (SBGP) implementation provided by 
\textcite{Kowal2024-MonteCarloInference} currently does not include functionality for 
directly evaluating
the conditional CDF and density. We compute them based on the following
considerations using the quantities provided by the implementation. The SBGP model is
$$
g(y_i) = f_\theta(\bsx_i) + \sigma \epsilon_i, \quad \epsilon_i \sim \calN(0, 1)
$$
where $i = 1, \dots, N$ identifies observations. This implies
$$
F_Y(y_i \mid \bsx_i, \sigma) = \Phi
\left(
\frac{g(y_i) - f_\theta(\bsx_i)}{\sigma}
\right)
$$
where $\Phi$ denotes the standard normal CDF. 
The corresponding conditional density follows by change of variables:
\[
p_Y(y_i \mid \bsx_i, \sigma)
=
\phi\!\left(
\frac{g(y_i)-\mu_i}{\sigma}
\right) \frac{1}{\sigma} g'(y_i),
\]
where $\phi$ denotes the standard normal density and $g'$ the derivative of $g$ with
respect to $y$.

Since the SBGP implementation allows access to posterior draws
$g^{[t]}$, of the transformation and the scale $\sigma^{[t]}$, $t=1,\dots,T$,
we get posterior draws of the 
conditional CDF and PDF, $\hat{F}^{[t]}_Y(y \mid \bsx, \sigma^{[t]})$ and 
$\hat{p}^{[t]}_Y(y \mid \bsx, \sigma^{[t]})$.
These draws use a fixed predictive mean of the latent Gaussian process at $\bsx_i$, as
provided by SBGP. For training sample sizes $>1000$, SBGP uses a fixed $\sigma$ at its
point estimate instead of samples. We use evaluations of the CDF and PDF on the test
data for the Kullback-Leibler divergence and the mean absolute deviation (MAD) of the
CDF, and evaluations on the training data for the WAIC.

\subsection{Diagnostics}
\label{app-sec:sim2-diagnostics}

\FloatBarrier

\begin{table}[ht]
    \small
    \caption[Diagnostics for PTMs with and without jittering of start values in Simulation 2.]{Diagnostic information for PTMs in the simulation study. The model PTM-jitter denotes a PTM with jittered initial values drawn from $\calN(0, 1)$ for all parameters. Results are grouped by sample size and MCMC kernel and aggregated over all seeds and data types. The column $\alpha$ displays the average acceptance probability for proposals in the respective kernel. Proposals from Gibbs kernels always get accepted. Proposals from IWLS and NUTS kernels are consistent with the target acceptance probabilities of $.5$ and $.9$, respectively. The column Max $\hat{R}$ shows the average maximum $\hat{R}$ within each model run; the maximum is taken over all sampled parameters. The effective sample sizes are average minimums over all parameters' ESS within each model run. The two left ESS columns show ESS per minute.\label{app-tab:sim2-diagnostics}}
    
\begin{tabu} to \linewidth {>{\raggedright\arraybackslash}p{7em}>{\raggedright}X>{\raggedleft}X>{\raggedright}X>{\raggedright}X>{\raggedright}X>{\raggedright}X>{\raggedright}X>{\raggedright}X}
\toprule
\multicolumn{5}{c}{ } & \multicolumn{4}{c}{Minimum Effective Sample Size} \\
\cmidrule(l{3pt}r{3pt}){6-9}
Model & MCMC & $N_{\text{train}}$ & $\alpha$ & Max. $\hat{R}$ & Bulk / Min. & Tail / Min. & Bulk & Tail\\
\midrule
PTM & GibbsKernel & 250 & 1.00 & 1.00 & 419.9 & 685.9 & 4209.0 & 6896.8\\
PTM & GibbsKernel & 500 & 1.00 & 1.00 & 384.2 & 599.9 & 5707.0 & 8948.9\\
PTM & GibbsKernel & 1000 & 1.00 & 1.00 & 276.6 & 409.4 & 7258.9 & 10811.2\\
\addlinespace
PTM & IWLSKernel & 250 & 0.51 & 1.00 & 489.9 & 655.6 & 4874.0 & 6524.9\\
PTM & IWLSKernel & 500 & 0.51 & 1.00 & 435.5 & 567.0 & 6353.8 & 8318.5\\
PTM & IWLSKernel & 1000 & 0.51 & 1.00 & 297.9 & 376.9 & 7514.6 & 9673.1\\
\addlinespace
PTM & NUTSKernel & 250 & 0.88 & 1.01 & 417.6 & 421.2 & 4143.6 & 4273.6\\
PTM & NUTSKernel & 500 & 0.89 & 1.00 & 448.1 & 491.4 & 6764.4 & 7566.7\\
PTM & NUTSKernel & 1000 & 0.90 & 1.00 & 329.2 & 369.8 & 8991.7 & 10374.9\\
\addlinespace
PTM-jitter & GibbsKernel & 250 & 1.00 & 2.71 & 0.4 & 1.4 & 5.4 & 20.9\\
PTM-jitter & GibbsKernel & 500 & 1.00 & 2.64 & 0.4 & 1.5 & 5.8 & 21.8\\
PTM-jitter & GibbsKernel & 1000 & 1.00 & 2.55 & 0.3 & 1.3 & 6.6 & 37.7\\
\addlinespace
PTM-jitter & IWLSKernel & 250 & 0.27 & Inf & 0.3 & 0.3 & 4.3 & 4.4\\
PTM-jitter & IWLSKernel & 500 & 0.26 & 4.84 & 0.3 & 0.3 & 4.5 & 4.6\\
PTM-jitter & IWLSKernel & 1000 & 0.31 & 4.43 & 0.2 & 0.2 & 4.8 & 5.2\\
\addlinespace
PTM-jitter & NUTSKernel & 250 & 0.83 & Inf & 0.3 & 0.5 & 4.2 & 7.9\\
PTM-jitter & NUTSKernel & 500 & 0.80 & Inf & 0.3 & 0.4 & 4.2 & 7.2\\
PTM-jitter & NUTSKernel & 1000 & 0.76 & Inf & 0.2 & 0.2 & 4.2 & 6.5\\
\bottomrule
\end{tabu}
\end{table}

\begin{table}[ht]
    \small
    \caption[Divergent transitions and maximum tree depths for PTMs with and without jittering of start values in Simulation 2.]{Notes captured during MCMC sampling for the PTMs in the simulation study. The values are averaged over all seeds, data types, and sample sizes.  \label{app-tab:sim2-errors}}
    
\begin{tabu} to \linewidth {>{\raggedright\arraybackslash}p{10em}>{\raggedleft\arraybackslash}p{15em}>{\raggedright}X}
\toprule
Model & Note & Rel. freq.\\
\midrule
PTM & divergent transition & 0.004\\
PTM & maximum tree depth & 0.000\\
PTM & nan acceptance prob & 0.000\\
\addlinespace
PTM-jitter & divergent transition & 0.939\\
PTM-jitter & maximum tree depth & 0.889\\
PTM-jitter & nan acceptance prob & 0.363\\
\bottomrule
\end{tabu}
\end{table}

\FloatBarrier

\subsection{Additional results}
\label{app-sec:sim2-results}

\FloatBarrier

\begin{table}[ht]
    \small
    \caption{Comparison of performance of PTMs and the true Gaussian model on Gaussian data. \label{app-tab:sim2-gauss}}
    
\begin{tabu} to \linewidth {>{\raggedleft}X>{\centering}X>{\centering}X>{\centering}X>{\centering}X>{\centering}X>{\centering}X>{\raggedleft}X}
\toprule
\multicolumn{6}{c}{ } & \multicolumn{2}{c}{CDF CI} \\
\cmidrule(l{3pt}r{3pt}){7-8}
Model & $N_{\text{train}}$ & KLD $\downarrow$ & CRPS $\downarrow$ & WAIC $\downarrow$ & MAD $\downarrow$ & Coverage & Width\\
\midrule
Gaussian & 1000 & 0.028 & 0.817 & 3417 & 0.061 & 0.923 & 0.184\\
PTM & 1000 & 0.028 & 0.817 & 3415 & 0.061 & 0.923 & 0.184\\
\addlinespace
Gaussian & 500 & 0.052 & 0.829 & 1726 & 0.084 & 0.917 & 0.248\\
PTM & 500 & 0.052 & 0.829 & 1724 & 0.084 & 0.916 & 0.247\\
\addlinespace
Gaussian & 250 & 0.097 & 0.853 & 876 & 0.114 & 0.910 & 0.332\\
PTM & 250 & 0.097 & 0.853 & 873 & 0.114 & 0.908 & 0.330\\
\bottomrule
\end{tabu}
\end{table}

\begin{figure}[ht]
    \includegraphics[width=\textwidth]{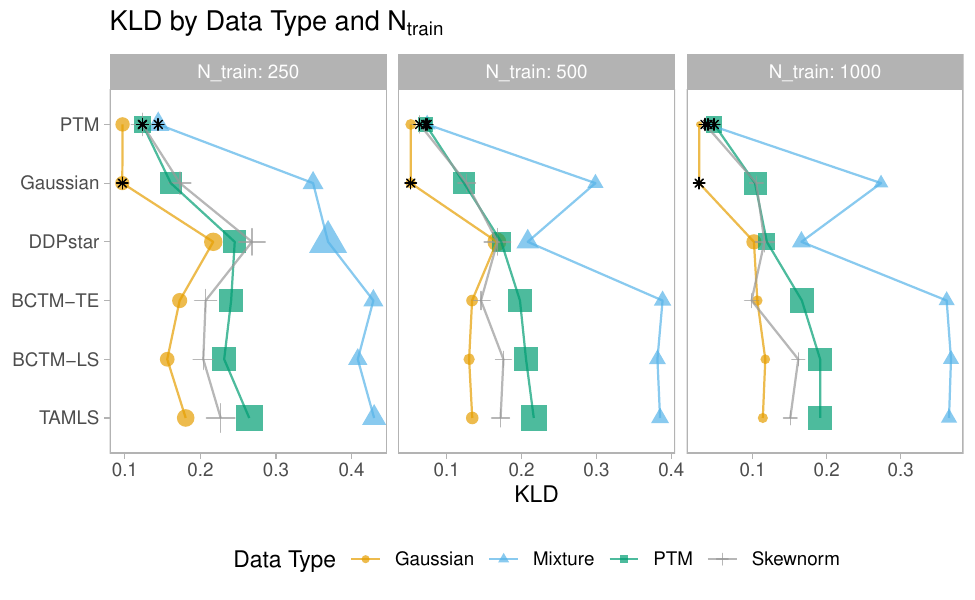}
    \includegraphics[width=\textwidth]{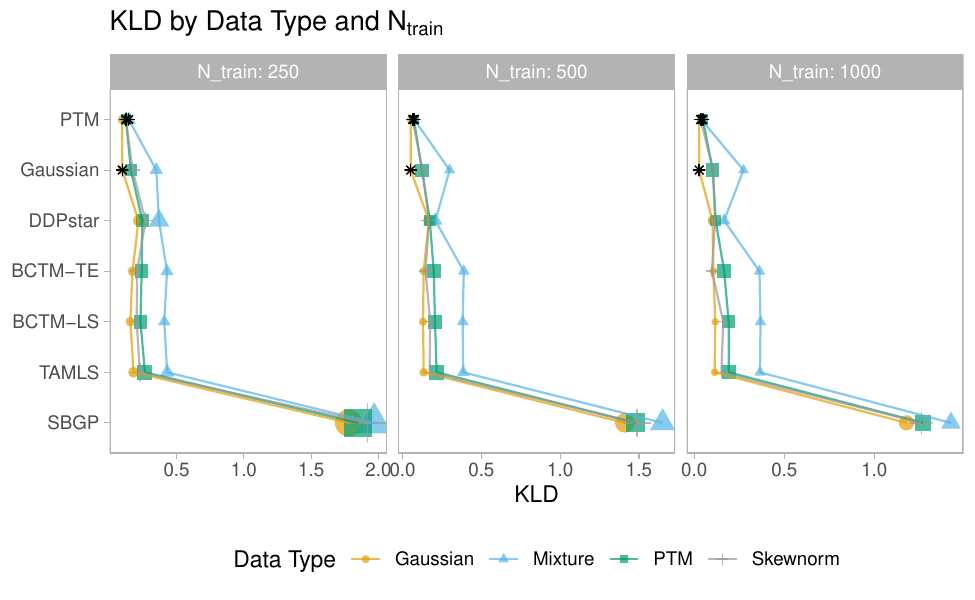}
    \caption[KL divergence in Simulation 2.]{Kullback-Leibler divergence. The plot shows the average of 100 replications with different seeds for pseudo-random number generation. The size of the point shapes is proportional to the standard deviation over these 100 replications. Within each panel and data type, a black star marks the best-performing observation.
    The top panel shows the plot without SBGP, since the inclusion of SBGP requires a wider scale.
    \label{app-fig:sim2-kld}}
\end{figure}

\begin{figure}[ht]
    \includegraphics[width=\textwidth]{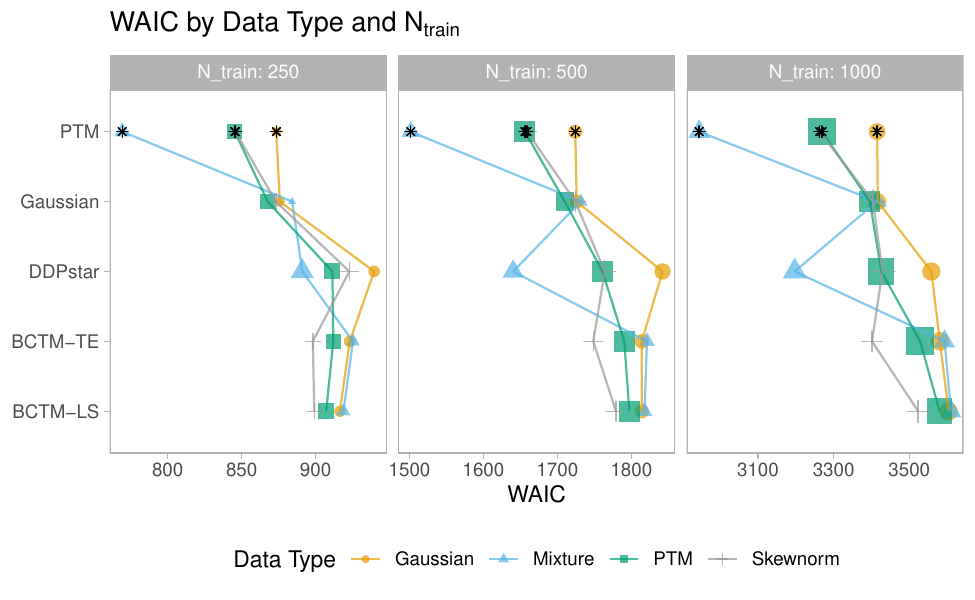}
    \includegraphics[width=\textwidth]{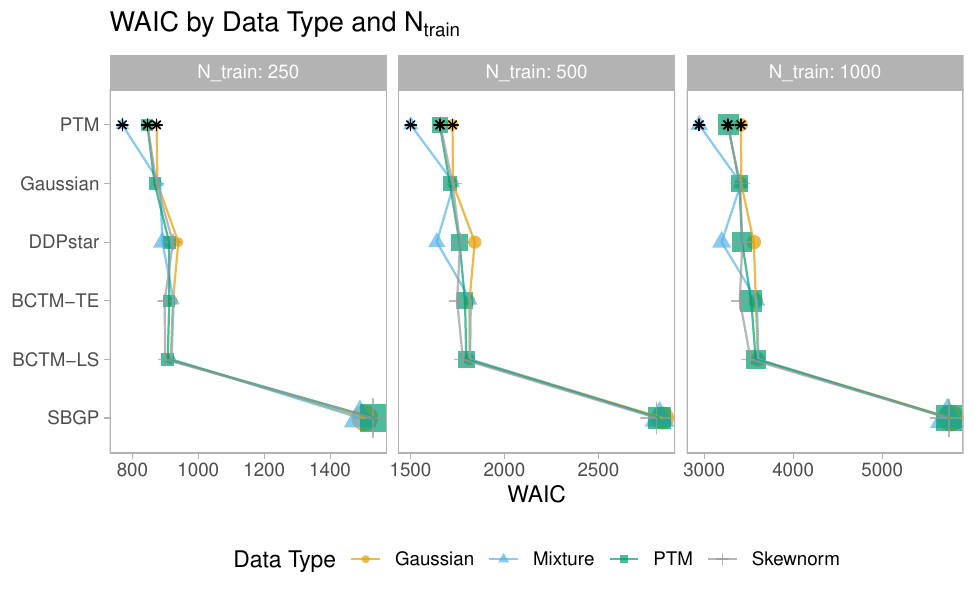}
    \caption[WAIC in Simulation 2.]{WAIC. The plot shows the average of 100 replications with different seeds for pseudo-random number generation. The size of the point shapes is proportional to the standard deviation over these 100 replications. Within each panel and data type, a black star marks the best-performing observation.
    The top panel shows the plot without SBGP, since the inclusion of SBGP requires a wider scale.\label{app-fig:sim2-waic}}
\end{figure}

\begin{figure}[ht]
    \includegraphics[width=\textwidth]{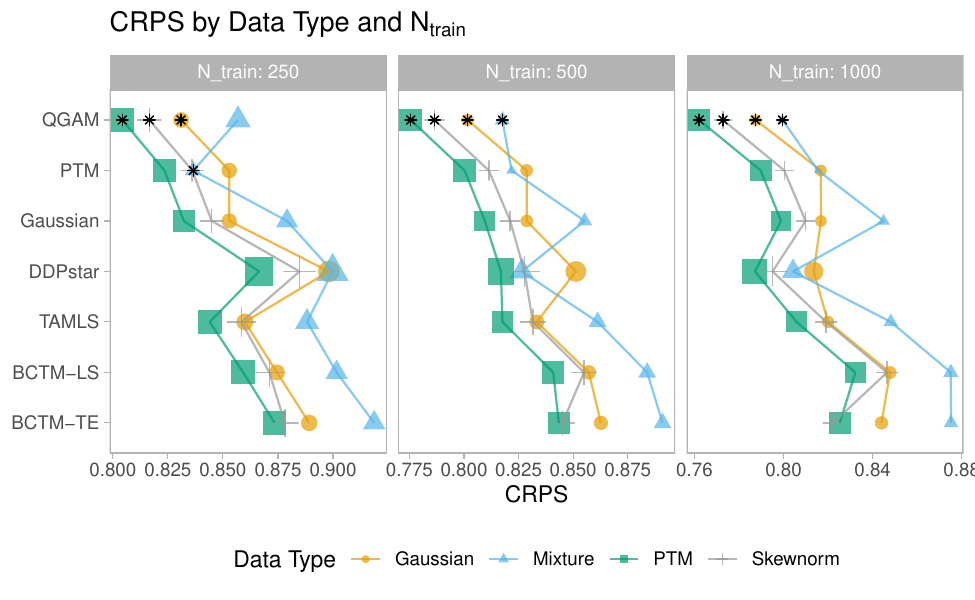}
    \includegraphics[width=\textwidth]{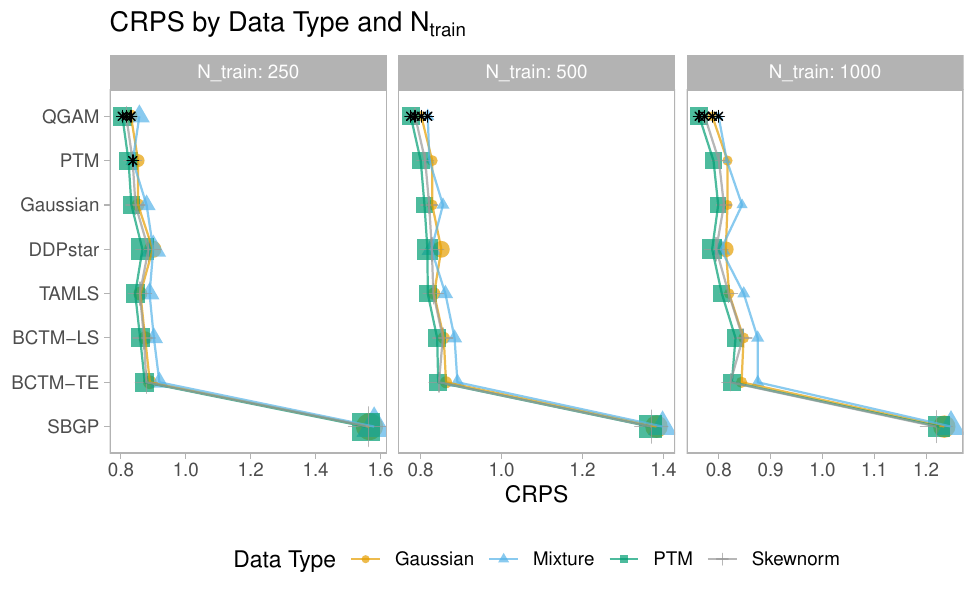}
    \caption[CRPS in Simulation 2.]{Continuous ranked probability score. The plot shows the average of 100 replications with different seeds for pseudo-random number generation. The size of the point shapes is proportional to the standard deviation over these 100 replications. Within each panel and data type, a black star marks the best-performing observation. The top panel shows the plot without SBGP, since the inclusion of SBGP requires a wider scale.\label{app-fig:sim2-crps}}
\end{figure}

\begin{figure}[ht]
    \includegraphics[width=\textwidth]{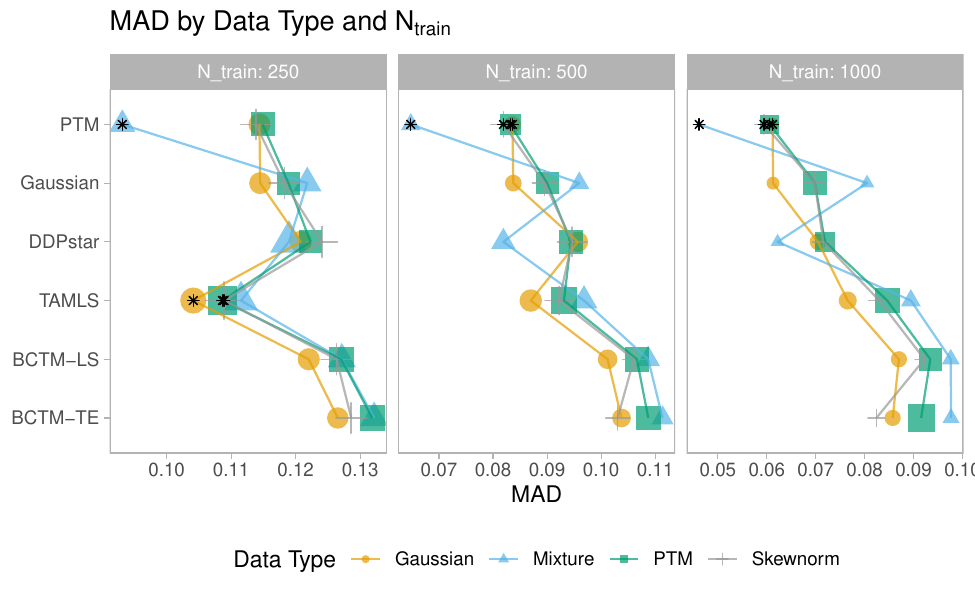}
    \includegraphics[width=\textwidth]{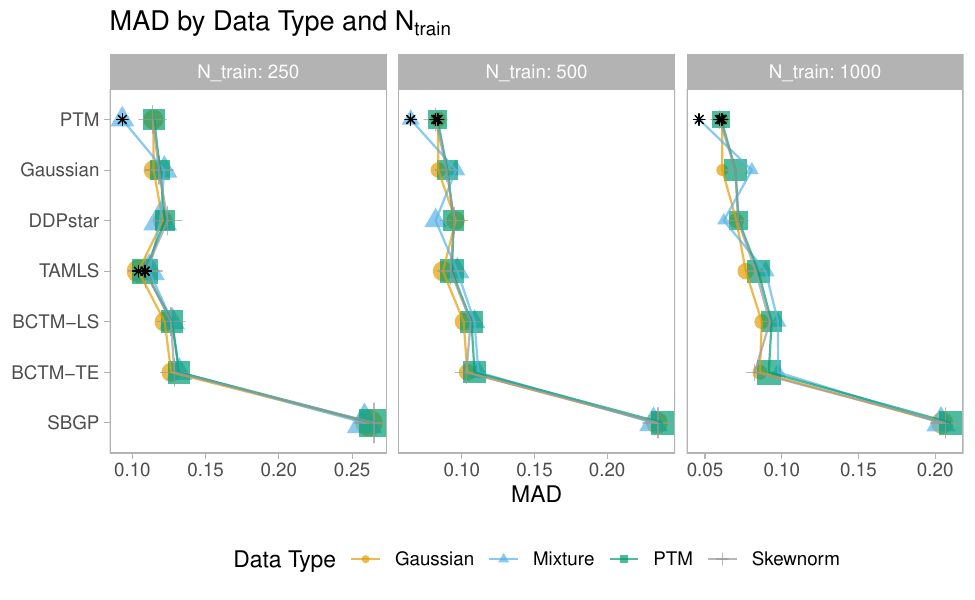}
    \caption[MAD in Simulation 2.]{Mean absolute difference in the CDF. The plot shows the average of 100 replications with different seeds for pseudo-random number generation. The size of the point shapes is proportional to the standard deviation over these 100 replications. Within each panel and data type, a black star marks the best-performing observation. The top panel shows the plot without SBGP, since the inclusion of SBGP requires a wider scale.\label{app-fig:sim2-mad}}
\end{figure}

\begin{figure}[ht]
    \includegraphics[width=\textwidth]{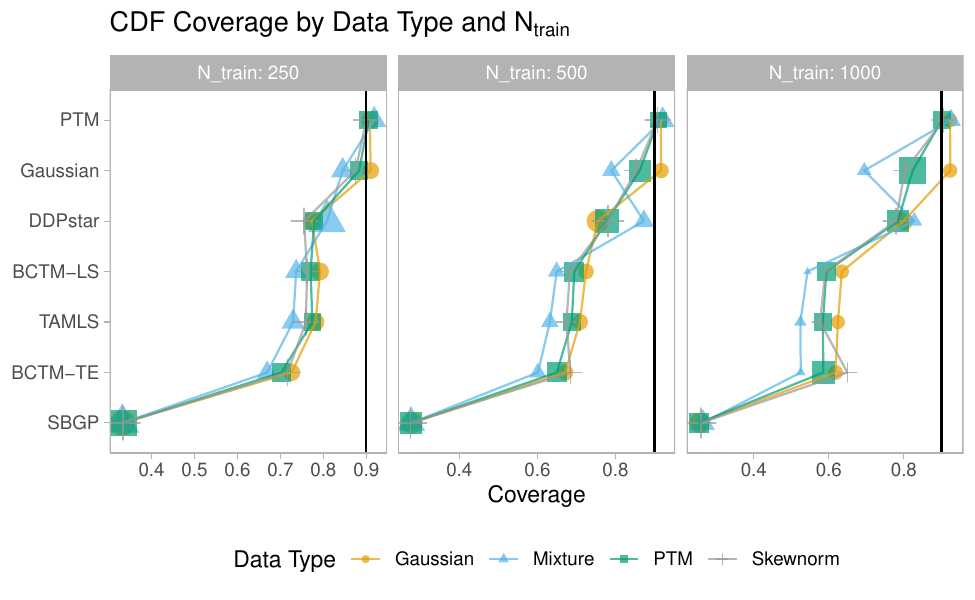}
    \caption[Coverage in Simulation 2.]{Average pointwise coverage for $90\%$ credible and confidence intervals of the CDF. Coverage within each replication is computed based on $5\,000$ random test observations from the target distribution. The plot shows the average of 100 replications with different seeds for pseudo-random number generation. The size of the point shapes is proportional to the standard deviation over these 100 replications. The vertical black line marks the nominal target level of $.9$.\label{app-fig:sim2-cdf_coverage}}
\end{figure}

\begin{figure}[ht]
    \includegraphics[width=\textwidth]{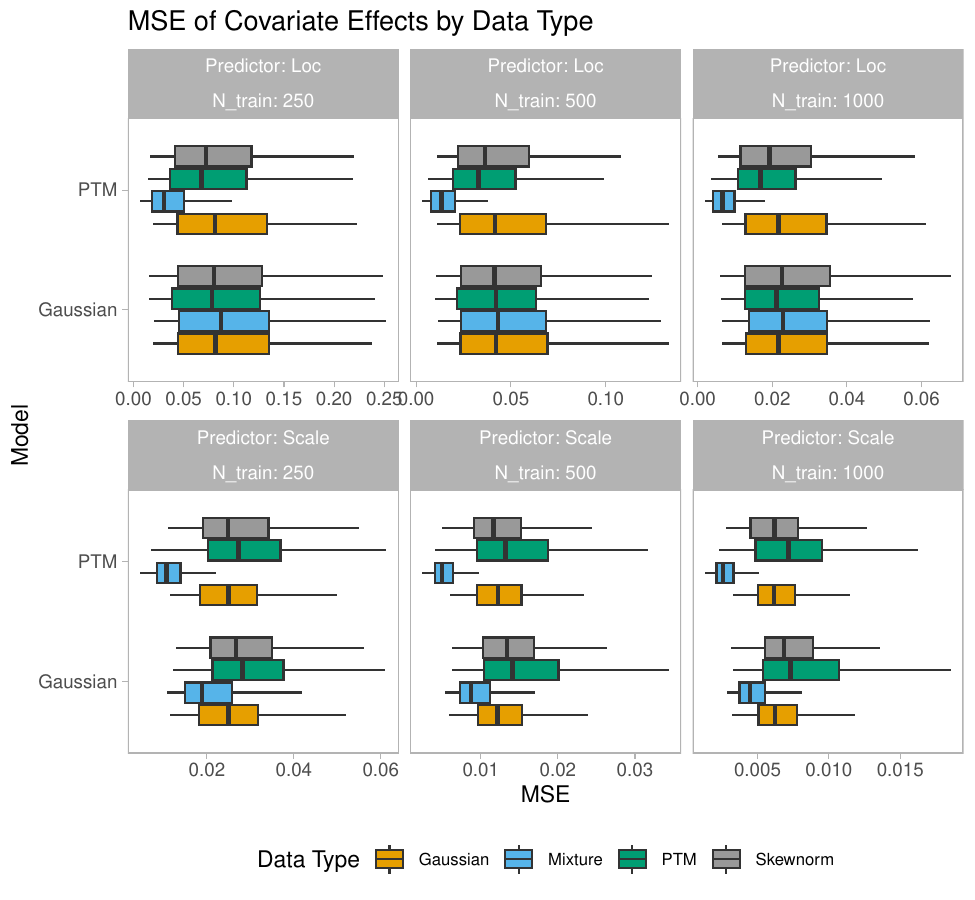}
    \caption[MSE of covariate effects in Simulation 2 by data type.]{Mean squared error for covariate effects. Each boxplot includes values from 100 replications and all four covariate functions, totaling 400 observations.\label{app-fig:sim2-mse_by_data}}
\end{figure}

\begin{figure}[ht]
    \includegraphics[width=\textwidth]{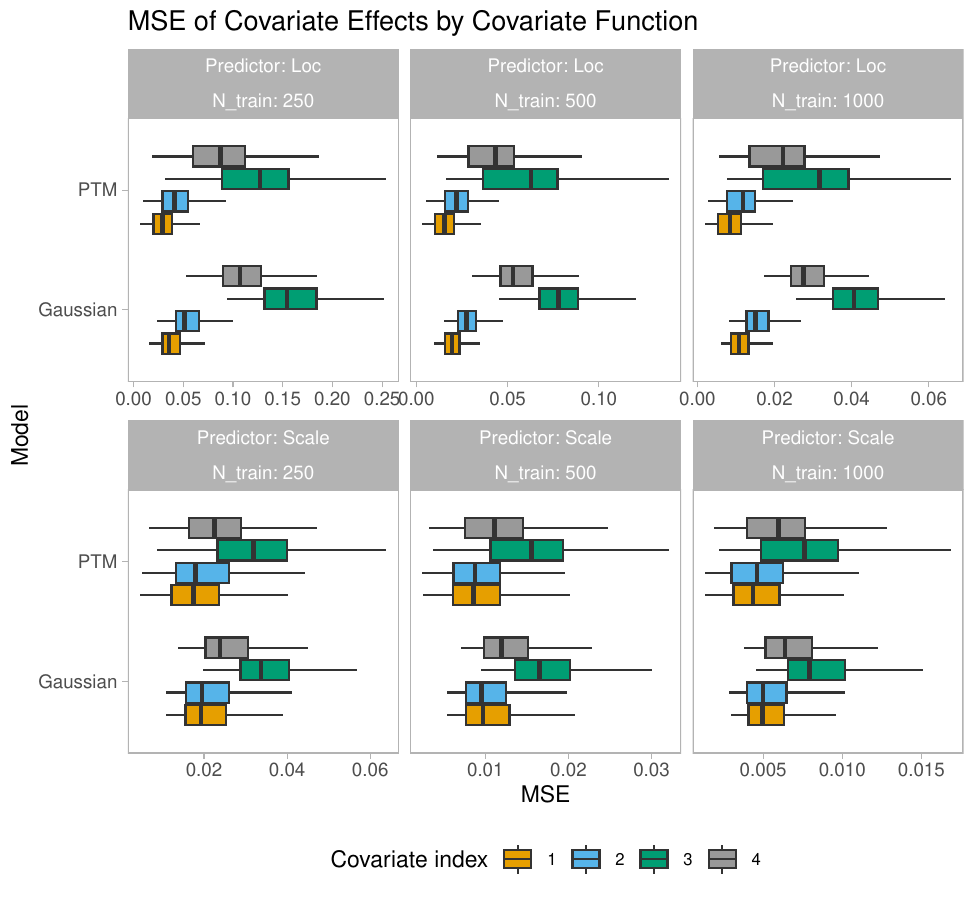}
    \caption[MSE of covariate effects in Simulation 2 by covariate function.]{Mean squared error for covariate effects. Each boxplot includes values from 100 replications and all four data generating mechanisms, totaling 400 observations.\label{app-fig:sim2-mse_by_cov}}
\end{figure}

\begin{figure}[ht]
    \includegraphics[width=\textwidth]{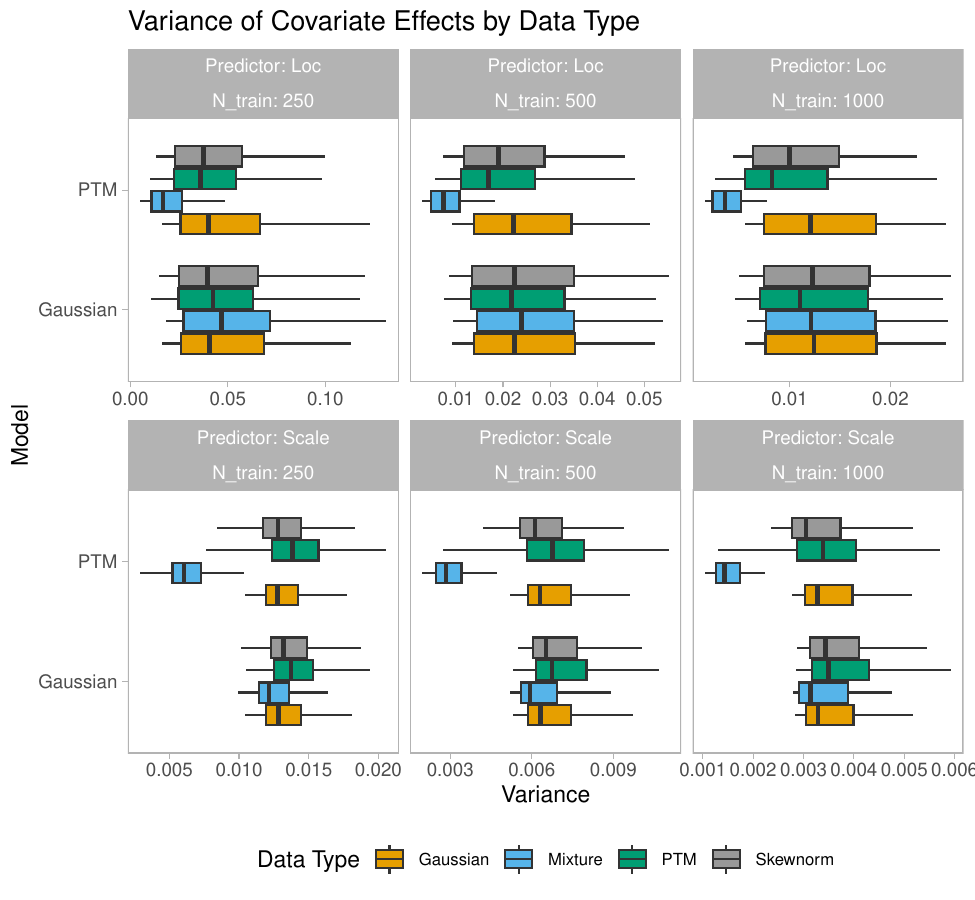}
    \caption[Variance of covariate effects in Simulation 2 by data type.]{Variance of covariate effects. Each boxplot includes values from 100 replications and all four covariate functions, totaling 400 observations.\label{app-fig:sim2-var_by_data}}
\end{figure}

\begin{figure}[ht]
    \includegraphics[width=\textwidth]{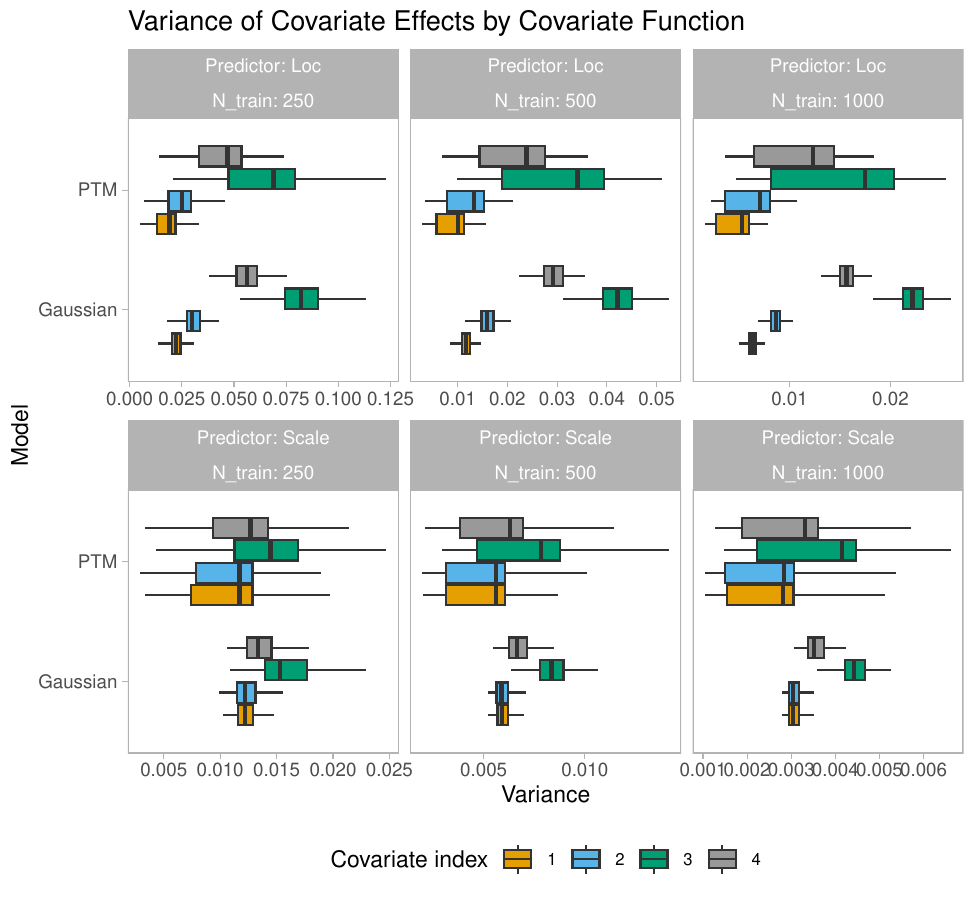}
    \caption[Variance of covariate effects in Simulation 2 by covariate function.]{Variance of covariate effects. Each boxplot includes values from 100 replications and all four data generating mechanisms, totaling 400 observations.\label{app-fig:sim2-var_by_cov}}
\end{figure}

\begin{figure}[ht]
    \includegraphics[width=\textwidth]{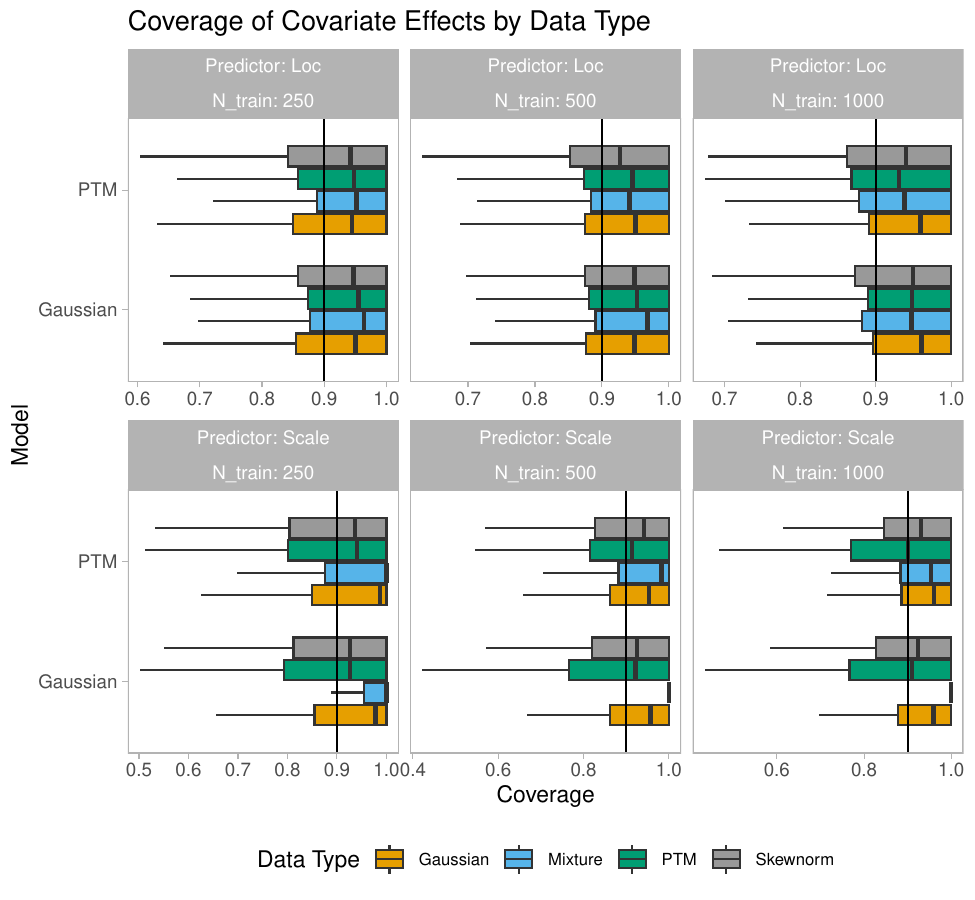}
    \caption[Coverage of covariate effects in Simulation 2 by data type.]{
        Pointwise coverage for $90\%$ credible and confidence intervals for covariate effects. Coverage within each replication is computed based on $5\,000$ random test observations from the target distribution. Each boxplot includes values from 100 replications and all four covariate functions, totaling 400 observations.\label{app-fig:sim2-cov_coverage_by_data}
    }
\end{figure}

\begin{figure}[ht]
    \includegraphics[width=\textwidth]{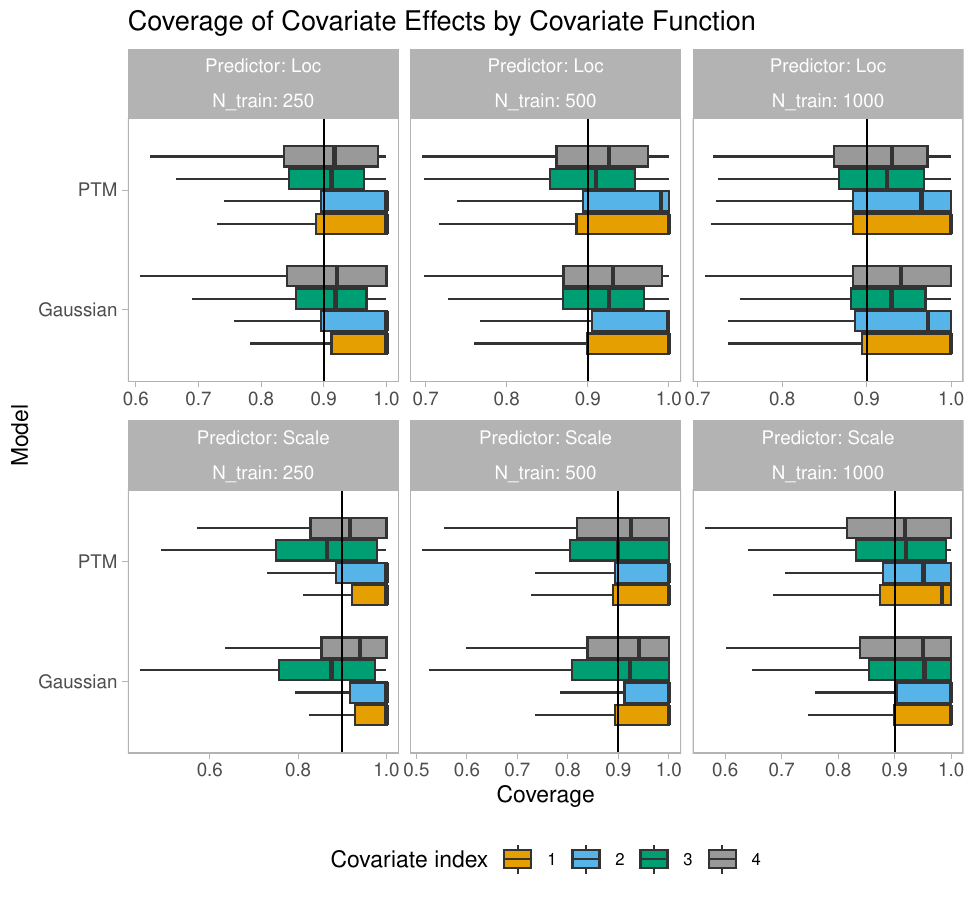}
    \caption[Coverage of covariate effects in Simulation 2 by covariate function.]{Pointwise coverage for $90\%$ credible and confidence intervals for covariate effects. Coverage within each replication is computed based on $5\,000$ random test observations from the target distribution. Each boxplot includes values from 100 replications and all four data generating mechanisms, totaling 400 observations.\label{app-fig:sim2-cov_coverage_by_cov}}
\end{figure}

\clearpage

\subsection{Comparison of different inverse Gamma prior specifications}

\FloatBarrier
When considering the variance parameters $\tau^2$ in the regularization priors used
in our covariate effects, the inverse Gamma prior $\tau^2 \sim \mathcal{IG}(1, 0.001)$
used by us has a peak at small values and a heavy tail. When used as a prior for P-splines, this reflects skepticism about highly variable (\enquote{wiggly}) functions, while still allowing the model to arrive at such functions if warranted by the data. When considering the precision $1/\tau^2$, an inverse Gamma prior with $a=1$ and $b \rightarrow 0$ approaches a flat prior for the precision $1/\tau^2$. We ran the simulation study for the PTM in the two small-data scenarios ($N_{\text{obs}} \in \{250, 500\}$) using two other common settings for the inverse Gamma hyperpriors in the covariate effects:
\begin{itemize}
    \tightlist
    \item $a = b = 0.01$, which is a common prior that approaches a flat prior on $\ln \tau^2$ as $a,b \rightarrow 0$.
    \item $a = 1.0, b = 0.005$, which is another common variant of the "$a=1$, $b$ small" variety of inverse Gamma priors.
\end{itemize}
The two following tables show the results alongside the data underlying the results reported in the paper. Note that to keep the results comparable, we used only data from the $N_{\text{obs}} \in \{250, 500\}$ conditions for all models for these tables. While a small effect of the hyperprior is discernible, it does not lead to any change in conclusions.

\begin{table}[!ht]
    \caption{Results for covariate effects. The results are averaged over all seeds, data types, sample sizes ($N_{\text{obs}} \in \{250, 500\}$),
        and over all four functions for the location and scale model parts, respectively. Coverages are reported
        for 90\% credible intervals.\label{app-tab:sim2-igprior-covariates}}
    \footnotesize
    \begin{tabu} to \linewidth {>{\raggedleft\hsize=2\hsize}X>{\centering}X>{\centering}X>{\centering}X>{\centering}X>{\centering}X>{\centering}X}
        \toprule
        \multicolumn{1}{c}{ } & \multicolumn{3}{c}{Location Terms} & \multicolumn{3}{c}{Scale Terms}                                               \\
        \cmidrule(l{3pt}r{3pt}){2-4} \cmidrule(l{3pt}r{3pt}){5-7}
        Model                 & MSE $\downarrow$                   & Coverage                        & Width & MSE $\downarrow$ & Coverage & Width \\
        \midrule
        PTM (IG 0.01, 0.01)   & 0.065                              & 0.914                           & 0.553 & 0.027            & 0.929    & 0.374 \\
        PTM (IG 1, 0.005)     & 0.059                              & 0.916                           & 0.526 & 0.022            & 0.922    & 0.336 \\
        PTM (IG 1, 0.001)     & 0.054                              & 0.916                           & 0.496 & 0.018            & 0.905    & 0.291 \\
        \addlinespace
        Gaussian              & 0.070                              & 0.922                           & 0.579 & 0.020            & 0.912    & 0.318 \\
        DDPstar               & 0.082                              & 0.907                           & 0.606 & 0.035            & 0.451    & 0.173 \\
        \bottomrule
    \end{tabu}
\end{table}

\normalsize

\begin{table}[!ht]
    \caption{Predictive performance and CDF coverage for the tested models. The results are averaged over all seeds, data types, and sample sizes ($N_{\text{obs}} \in \{250, 500\}$). For the KLD, MAD, and CDF, smaller values indicate a better
        predictive performance. Coverages are reported for 90\% credible intervals.\label{app-tab:sim2-igprior-dist}}
    \footnotesize
    \begin{tabu} to \linewidth {>{\raggedleft\hsize=2\hsize}X>{\centering}X>{\centering}X>{\centering}X>{\centering}X>{\centering}X>{\raggedleft}X}
        \toprule
        \multicolumn{5}{c}{ } & \multicolumn{2}{c}{CDF CI}                                                                               \\
        \cmidrule(l{3pt}r{3pt}){6-7}
        Model                 & KLD $\downarrow$           & CRPS $\downarrow$ & WAIC $\downarrow$ & MAD $\downarrow$ & Coverage & Width \\
        \midrule
        PTM (IG 0.01, 0.01)   & 0.113                      & 0.826             & 1231              & 0.104            & 0.918    & 0.303 \\
        PTM (IG 1, 0.005)     & 0.101                      & 0.821             & 1246              & 0.099            & 0.914    & 0.287 \\
        PTM (IG 1, 0.001)     & 0.094                      & 0.826             & 1234              & 0.094            & 0.911    & 0.272 \\
        \addlinespace
        Gaussian              & 0.172                      & 0.840             & 1299              & 0.104            & 0.867    & 0.290 \\
        DDPstar               & 0.227                      & 0.859             & 1334              & 0.107            & 0.788    & 0.275 \\
        BCTM-LS               & 0.237                      & -                 & 1356              & 0.116            & 0.727    & 0.268 \\
        BCTM-TE               & 0.246                      & -                 & 1355              & 0.118            & 0.678    & 0.252 \\
        TAMLS                 & 0.251                      & 0.849             & -                 & 0.100            & 0.719    & 0.288 \\
        SBGP                  & -                          & 1.492             & -                 & 0.253            & 0.898    & 0.898 \\
        QGAM                  & -                          & 0.811             & -                 & -                & -        & -     \\
        \bottomrule
    \end{tabu}
\end{table}
\footnotesize

\normalsize

\clearpage

\hypertarget{app-application-dutch-boys}{%
    \section{Application: Fourth Dutch Growth Study}\label{app-application-dutch-boys}}

\FloatBarrier

This section includes additional information about the
Fourth Dutch Growth Study application example.

\subsection{Model specifications}

\begin{itemize}
    \tightlist
    \item PTM: As described in the main text.
          The inverse smoothing parameters of the P-splines receive inverse
          gamma hyperpriors with concentration \(1\) and scale \(0.001\).
          We draw
          $10\,000$ samples each in four chains after a warmup of $5\,000$ iterations and use a thinning factor of $5$, yielding a total of $8\,000$ samples.
    \item Gaussian: Gaussian location-scale model analogous to the PTM, changing the transformation function $h$ to the identity function. We draw
          $20\,000$ samples each in four chains after a warmup of $5\,000$ iterations and use a thinning factor of $10$, yielding a total of $8\,000$ samples.
    \item TAMLS: A transformation additive model \parencite{Siegfried2023-DistributionfreeLocationscaleRegression}
          \begin{align}
              \bbP(BMI \leq \text{bmi}_{i}) & = \Phi(\sigma_{i}^{-1}h(\text{bmi}_{i}) - \mu_{i}) \\
              \mu_{i}                       & = \beta_0 + s(\text{age}_i)                        \\
              \ln \sigma_{i}                & = g(\text{age}_i)
          \end{align}
          with $\mu(\text{age}_i)$ and $\sigma(\text{age}_i)$
          given by 20-parameter P-splines; $\sigma(\text{age}_i)$ does not include an intercept.
          The transformation function $h$ is parameterized as an increasing Bernstein polynomial of order 10.
          Implementation was done by adapting the example code
          included in the R package \texttt{tram}, available on GitHub: \url{https://github.com/cran/tram/blob/master/demo/stram.R}
    \item SBGP: A semiparametric Bayesian Gaussian process model as proposed by \textcite{Kowal2024-MonteCarloInference}. The model is
          $g(\text{bmi}_i) = f_\theta(\text{age}_i) + \sigma \epsilon_i$
          where $\epsilon_i \sim \calN(0, 1)$ and $f_\theta \sim \calG \calP(m_\theta, K_\theta)$. In the Gaussian process,
          $m_\theta$ is an unknown constant mean and $K_\theta$ is an isotropic Matérn covariance
          function with unknown variance, range, and smoothness parameters. We also tested
          a version of the model that includes a simple linear term
          $g(y_i) = \text{age}_i\beta + f_\theta(\text{age}_i) + \sigma \epsilon_i$.
          We found that the results
          improved when including this term and report the results for this model version.
          Computations are
          carried out using the code of the R package \texttt{SeBR} as provided by the authors
          in the supplementary materials of \textcite{Kowal2024-MonteCarloInference}.
          We save $5000$ posterior predictive samples for each row of the test dataset.
    \item DDPstar: A density regression via Dirichlet process mixtures of normal structured additive regression models \parencite{Rodriguez-Alvarez2024-DensityRegressionDirichleta}.
          The means of the $l = 1, \dots, L$ mixture components are modeled as
          20-parameter P-splines $\mu(\text{age}_{i,l})$ using the default concentration $1$ and scale $0.005$.
          As \textcite{Rodriguez-Alvarez2024-DensityRegressionDirichleta} do in their example code, we use $L=20$ mixture components, and otherwise stick to the defaults of their R library \texttt{DDPstar}, which we use to carry out the computations. We draw $15\,000$ samples after a warmup of $5\,000$ iterations and use a thinning factor of $10$, yielding a total of $1\,500$ samples.
    \item BCTM: Bayesian conditional transformation model \parencite[BCTM,][]{Carlan2023-BayesianConditionalTransformation} with standard normal reference distribution of the form
          \[\bbP(BMI \leq \text{bmi}_i) = \Phi\bigl(h(\text{bmi}_i|\text{age}_{i}) \bigr),\]
          where
          \[h(\text{bmi}_i|\text{age}_{i}) = \bigl(\bsa(\text{bmi}_i)^\sfT \otimes \bsb(\text{age}_{i})^\sfT\bigr)^\sfT \bsgamma\]
          is the BCTM's transformation function, with \(\otimes\)
          denoting the Kronecker product. The bases \(\bsa\) and \(\bsb\) are
          chosen as 9-parameter B-spline bases, yielding a total of $81$ parameters (including an intercept). The BCTM uses inverse gamma hyperpriors with
          concentration \(1\) and scale \(0.001\) for all inverse smoothing
          parameters of the tensor product terms. We apply MCMC via the No-U-Turn Sampler as available in
          the Python library \texttt{liesel-bctm} \parencite{Brachem2023-LieselbctmBayesianConditional}. We draw
          $5\,000$ samples each in four chains after a warmup of $5\,000$ iterations, yielding a total of $20\,000$ samples. Since sampling is done exclusively with NUTS, we use no thinning.
    \item QGAM: A fully specified additive quantile regression model with P-splines for all four covariates, as available in the R package \texttt{qgam} \parencite{Fasiolo2021-FastCalibratedAdditive,Fasiolo2021-QgamBayesianNonparametric}. We fit QGAM for a grid of 25 evenly spaced probability levels in $\{0.005, \dots, 0.995\}$. Each quantile model is of the form
          \[Q_{\text{bmi}}(\alpha | \text{age}_{i}) = s_{\alpha}(\text{age}_{i}) \]
          where \(s_{\alpha}(\text{age}_{i})\) is a 20-parameter
          P-spline (including an intercept) and \(Q_{\text{bmi}}(\alpha | \text{age}_{i})\) is the conditional
          \(\alpha\)-quantile of the response.
\end{itemize}

\subsection{Additional results}
\begin{figure}[ht]
    \includegraphics[width=\textwidth]{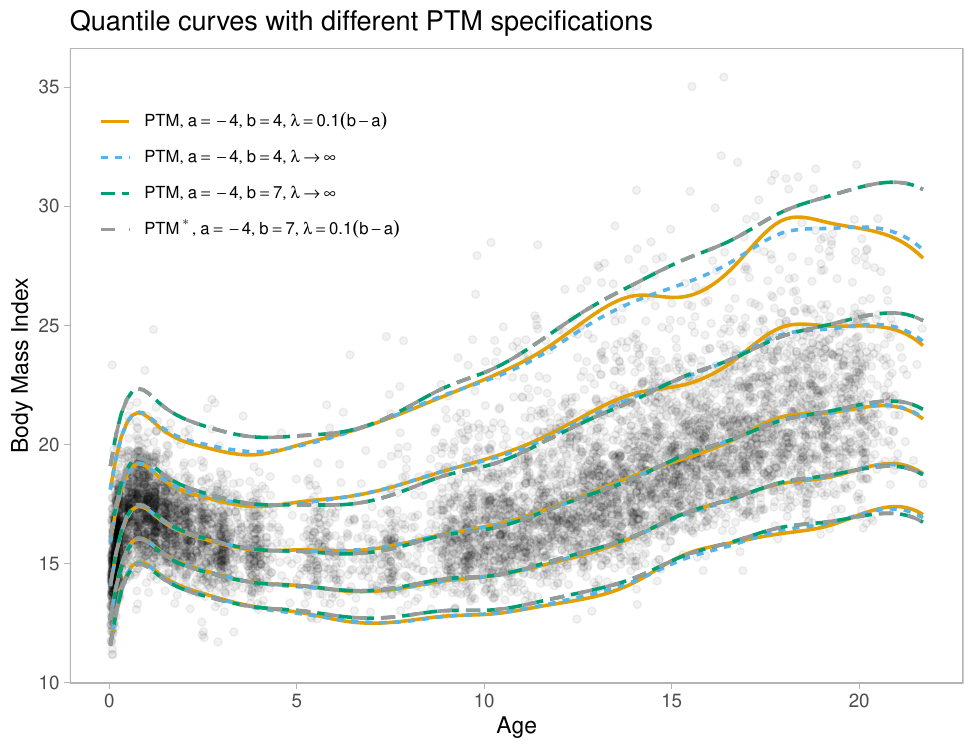}
    \caption[Quantile curves for PTM variants in 4th Dutch Growth Study.]{Quantile curves at probability levels $0.01, 0.1, 0.5, 0.9, 0.99$ for PTM variants. The curves were produced with full-sample model runs. \label{app-fig:db-qcurves-ptm}}
\end{figure}

\autoref{app-tab:db-dist_ptm} shows predictive performance for all PTM variants that
were run. \autoref{app-fig:db-qcurves-ptm} shows the corresponding
quantile curves for the PTM variants. From both the plot and the table it becomes obvious that all models with $b=7$ are practically equivalent, while the models with $b=4$ struggle with accurately capturing the highest quantile of the data, and the model with $a=-4, b=4$ and $\lambda=0.1(b-a)$ yields notably more wiggly quantile curves than the other models.

\begin{figure}[ht]
    \includegraphics[width=\textwidth]{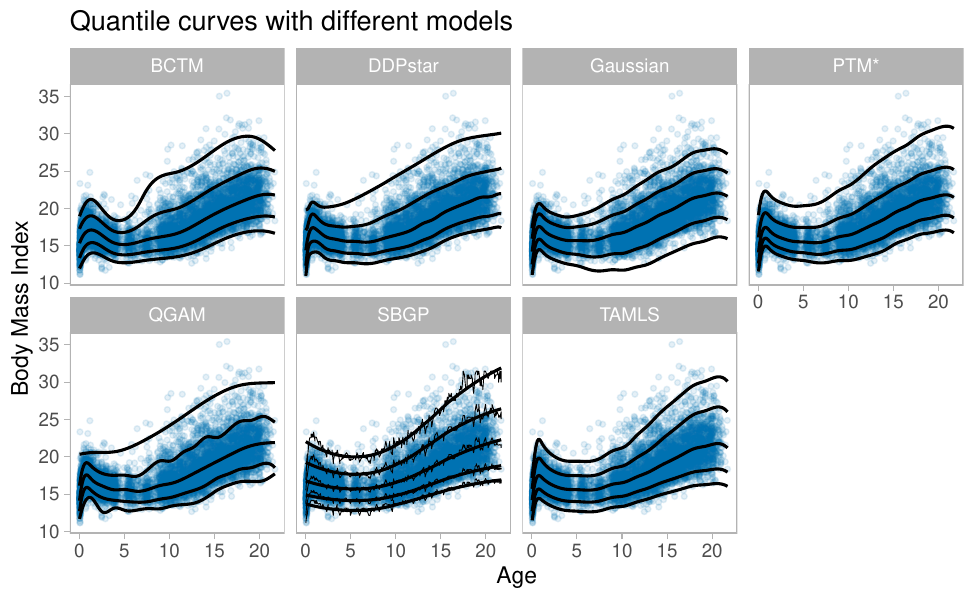}
    \caption[Quantile curves for all model variants in 4th Dutch Growth Study.]{
        Quantile curves at probability levels $0.01, 0.1, 0.5, 0.9, 0.99$ for different models. The curves were produced with full-sample model runs. 
    For SBGP, the available implementation returns posterior predictive samples at the specified age values rather than an analytic conditional quantile function. The SBGP curves shown here were therefore computed pointwise as empirical quantiles of 5000 posterior predictive draws each for a grid of 150 values for age in $[\min(age), \max(age)]$. Their local wiggliness reflects this sample-based reconstruction, including finite Monte Carlo variation and numerical inversion of the estimated transformation, and should not be interpreted as evidence for an intrinsic lack of smoothness. The overlaid smooth lines for SBGP are cubic regression splines applied only for visualization (using the \texttt{ggplot2} default in \texttt{geom\_smooth}).
    \label{app-fig:db-qcurves-all}
        }
\end{figure}

\autoref{app-fig:db-qcurves-all}
shows quantile curves for all models, including the PTM with
$a = -4$, $b = 7$ and $\lambda = 0.1(b-a)$. We note a highly volatile quantile curve for the highest quantile in the BCTM and a visible underestimation of the lowest quantile in the Gaussian model.

\autoref{app-tab:db-diagnostics} contains diagnostic information about posterior MCMC samples for the Gaussian and PTM models. The quantities are computed using the Python library Arviz \parencite{Kumar2019-ArviZUnifiedLibrary}, version 0.21.0.
\begin{itemize}
    \item ESS (Bulk): Effective sample size for the bulk of the distribution, indicating how many independent draws the Markov chain is equivalent to in regions of high posterior mass.
    \item ESS (Tail): Effective sample size for the tail of the distribution, quantifying how well the sampler explores the extremes of the posterior.
    \item $\hat{R}$: The rank-normalized split $\hat{R}$ statistic, which compares within-chain to between-chain variance; values close to 1 indicate convergence, while values larger than 1.01 suggest problematic mixing \parencite{Gelman1992-InferenceIterativeSimulation}.
\end{itemize}
The maximum $\hat{R}$ does not exceed $1.01$ in any case, suggesting unproblematic mixing between the four sampled chains. The minimum effective sample sizes far exceed $400$, suggesting sufficient information available in the posterior samples for all sampled quantities.

\autoref{app-tab:db-dist_ptm} contains notes captured during MCMC sampling for the PTM and Gaussian models. The notes have the following meaning:
\begin{itemize}
    \item \texttt{nan acceptance prob}: A numerical issue during sampling, leading to an invalid acceptance probability in a Metropolis-Hastings acceptance step. For Gaussian proposal distributions like the one used in iteratively weighted least squares proposals, large numbers here can indicate problems with the covariance matrix of the proposal distribution.
    \item \texttt{divergent transition}: Divergent transitions in NUTS sampling are numerical warnings that the Hamiltonian simulation could not accurately follow the posterior’s geometry, often occurring in regions of high curvature or when parameters are poorly scaled. A few isolated divergences may not matter, but a persistent nonzero fraction (e.g. >1–2\% of all transitions) indicates a serious problem with the sampler’s validity \parencite{Betancourt2018-ConceptualIntroductionHamiltonian}.
    \item \texttt{maximum tree depth}: Maximum tree depth in NUTS sampling is a safeguard that limits how far the algorithm can explore a trajectory during each iteration. This note is emitted when the maximum tree depth is reached. It means that no U-turn was encountered in the simulated trajectory that generates the proposal, i.e. the leapfrog algorithm was not stopped early.
\end{itemize}

We observe minimal shares of divergent transitions and maximum tree depths, and virtually no NaN acceptance probabilities, showing no serious cause for concern.

\begin{table}[ht]
    \caption[Predictive performance for PTM model variants in the Fourth Dutch Growth Study.]{Predictive performance for PTM model variants, based on 10-fold cross-validation. Smaller values for the Log Score and CRPS indicate better performance. \label{app-tab:db-dist_ptm}}
    \small
    
\begin{tabu} to \linewidth {>{\raggedright\arraybackslash}p{3em}>{\raggedleft\arraybackslash}p{3em}>{\raggedleft\arraybackslash}p{3em}>{\raggedright}X>{\raggedleft}X>{\raggedleft}X}
\toprule
Model & a & b & $\lambda$ & Log Score $\downarrow$ & CRPS $\downarrow$\\
\midrule
PTM & -4 & 7 & $\lambda \to \infty$ & 625.69 & 0.365\\
PTM & -4 & 7 & $\lambda = 0.1(b-a)$ & 625.69 & 0.365\\
PTM & -7 & 7 & $\lambda \to \infty$ & 625.72 & 0.365\\
PTM & -7 & 7 & $\lambda = 0.1(b-a)$ & 625.72 & 0.365\\
PTM & -4 & 4 & $\lambda \to \infty$ & 638.65 & 0.364\\
PTM & -4 & 4 & $\lambda = 0.1(b-a)$ & 640.25 & 0.364\\
\bottomrule
\end{tabu}
\end{table}

\begin{table}[ht]
    \caption[Diagnostics for PTM model variants in the Fourth Dutch Growth Study.]{Diagnostic information for the PTM and Gaussian models in the Fourth Dutch Growth Study application example. The values are averaged over all
        folds of the cross-validation. $\hat{R}$ values are based on four parallel chains. \label{app-tab:db-diagnostics}}
    \small
    
\begin{tabu} to \linewidth {>{\raggedright}X>{\centering\arraybackslash}p{3em}>{\centering\arraybackslash}p{3em}>{\centering\arraybackslash}p{5em}>{\raggedright}X>{\centering}X>{\centering}X}
\toprule
Model & a & b & $\lambda$ & Min. ESS (Bulk) & Min. ESS (Tail) & Max. $\hat{R}$\\
\midrule
Gaussian & - & - & - & 2 372 & 4 632 & 1.002\\
PTM & -4 & 4 & $\to \infty$ & 1 510 & 3 407 & 1.003\\
PTM & -4 & 4 & 0.1(b-a) & 2 685 & 5 362 & 1.002\\
PTM & -4 & 7 & $\to \infty$ & 3 979 & 7 831 & 1.001\\
PTM & -4 & 7 & 0.1(b-a) & 4 000 & 7 786 & 1.001\\
\bottomrule
\end{tabu}
\end{table}

\begin{table}[ht]
    \caption[Divergent transitions and maximum tree depths for PTM model variants in the Fourth Dutch Growth Study.]{Notes captured during MCMC sampling for the PTM and Gaussian models in the Fourth Dutch Growth Study application example. The values are averaged over all
        folds of the cross-validation.
        \label{app-tab:db-errors}}
    \small
    
\begin{tabu} to \linewidth {>{\raggedright\arraybackslash}p{4em}>{\centering\arraybackslash}p{3em}>{\centering\arraybackslash}p{3em}>{\centering\arraybackslash}p{5em}>{\raggedleft\arraybackslash}p{13em}>{\centering}X}
\toprule
Model & a & b & $\lambda$ & Note & Rel. freq.\\
\midrule
Gaussian & - & - & - & nan acceptance prob & 0\\
\addlinespace
PTM & -4 & 4 & $\to \infty$ & divergent transition & 0\\
PTM & -4 & 4 & $\to \infty$ & maximum tree depth & 0\\
PTM & -4 & 4 & 0.1(b-a) & divergent transition & 0\\
PTM & -4 & 4 & 0.1(b-a) & maximum tree depth & 0\\
\addlinespace
PTM & -4 & 7 & $\to \infty$ & divergent transition & 0\\
PTM & -4 & 7 & $\to \infty$ & maximum tree depth & 0\\
PTM & -4 & 7 & 0.1(b-a) & divergent transition & 0\\
PTM & -4 & 7 & 0.1(b-a) & maximum tree depth & 0\\
\bottomrule
\end{tabu}
\end{table}

\clearpage

\hypertarget{app-application-framingham-heart-study}{%
    \section{Application: Framingham heart
      study}\label{app-application-framingham-heart-study}}

\subsection{Model specifications}

\begin{itemize}
    \tightlist
    \item PTM: As described in the main text.
          The inverse smoothing parameters of the P-splines receive inverse
          gamma hyperpriors with concentration \(1\) and scale \(0.001\).
          We draw
          $25\,000$ samples each in four chains after a warmup of $5\,000$ iterations and use a thinning factor of $5$, yielding a total of $20\,000$ samples.
          For the random intercept, we use proposals from a No-U-Turn sampler instead of IWLS proposals, since we found that this improved sampling efficiency.
    \item Gaussian: Gaussian location-scale model analogous to the PTM, changing the transformation function $h$ to the identity function. We draw
          $25\,000$ samples each in four chains after a warmup of $5\,000$ iterations and use a thinning factor of $5$, yielding a total of $20\,000$ samples. For the random intercept, we use proposals from a No-U-Turn sampler instead of IWLS proposals, since we found that this improved sampling efficiency.
    \item TAMLS: A transformation additive model \parencite{Siegfried2023-DistributionfreeLocationscaleRegression}
          \begin{align}
              \bbP(Cholst \leq \text{cholst}_{ij}) & = \Phi(\sigma_{ij}^{-1}h(\text{cholst}_{ij}) - \mu_{ij})       \\
              \mu_{ij}                             & = \beta_0 + s(\text{age}_i) + \text{sex}_i \beta_1 + \zeta_{i} \\
              \ln \sigma_{ij}                      & = g(\text{age}_i) + \text{sex}_i \gamma_1
          \end{align}
          with $s$ and $g$ given by 20-parameter P-splines; $\sigma_{ij}$ does not include an intercept. On the location, we use an i.i.d. random intercept $\zeta_i \sim \calN(0, \psi^2)$. The transformation function $h$ is parameterized as an increasing Bernstein polynomial of order 10. We also include a model variant without random intercept.
          Implementation was done by adapting the example code
          included in the R package \texttt{tram}, available on GitHub: \url{https://github.com/cran/tram/blob/master/demo/stram.R}
    \item SBGP: A semiparametric Bayesian Gaussian process model as proposed by \textcite{Kowal2024-MonteCarloInference}. The model is
          $g(\text{cholst}_i) = \text{sex}_i \beta_1 + f_\theta(\text{age}_i) \beta_1 + \sigma \epsilon_i$
          where $\epsilon_i \sim \calN(0, 1)$ and $f_\theta \sim \calG \calP(m_\theta, K_\theta)$. In the Gaussian process,
          $m_\theta$ is an unknown constant mean and $K_\theta$ is an isotropic Matérn covariance
          function with unknown variance, range, and smoothness parameters. We also tested
          a version of the model that includes a simple linear term
          $g(\text{cholst}_i) = \text{sex}_i \beta_1 +\text{age}_i\beta + f_\theta(\text{age}_i) + \sigma \epsilon_i$.
          We found that the results
          improved when including this term and report the results for this model version.
          Computations are
          carried out using the code of the R package \texttt{SeBR} as provided by the authors
          in the supplementary materials of \textcite{Kowal2024-MonteCarloInference}.
          We save $5\,000$ posterior predictive samples for each row of the test dataset.
    \item DDPstar: A density regression via Dirichlet process mixtures of normal structured additive regression models \parencite{Rodriguez-Alvarez2024-DensityRegressionDirichleta}.
          The means of the $l = 1, \dots, L$ mixture components are modeled as
          $\mu_{i,l} = \beta_{0,l} + \text{sex}_i \beta_{1,l} + s(\text{age}_{i,l})$, where $s$ are 20-parameter P-splines using the default concentration $1$ and scale $0.005$.
          As \textcite{Rodriguez-Alvarez2024-DensityRegressionDirichleta} do in their example code, we use $L=20$ mixture components, and otherwise stick to the defaults of their R library \texttt{DDPstar}, which we use to carry out the computations. We draw $15000$ samples after a warmup of $5000$ iterations and use a thinning factor of $10$, yielding a total of $1500$ samples. While \texttt{DDPstar} allows inclusion of
          random intercepts, it does not allow for prediction for unknown clusters in random
          intercept models, so we do not include a random intercept version.
    \item BCTM: Bayesian conditional transformation model \parencite[BCTM,][]{Carlan2023-BayesianConditionalTransformation} with standard normal reference distribution of the form
          \[\bbP(Cholst \leq \text{cholst}_{ij}) = \Phi\bigl(h(\text{cholst}_{ij}|\text{age}_{ij}) + \text{sex}_i \gamma_2 + \zeta_i\bigr),\]
          where
          \[h(\text{cholst}_{ij}|\text{age}_{ij}) = \bigl(\bsa(\text{cholst}_{ij})^\sfT \otimes \bsb(\text{age}_{ij})^\sfT\bigr)^\sfT \bsgamma_1,\] with \(\otimes\)
          denoting the Kronecker product. The bases \(\bsa\) and \(\bsb\) are
          chosen as 9-parameter B-spline bases, yielding a total of $81$ parameters (including an intercept). The BCTM uses inverse gamma hyperpriors with
          concentration \(1\) and scale \(0.001\) for all inverse smoothing
          parameters of the tensor product terms.
          The term $\zeta_i \sim \calN(0, \psi^2)$ is an i.i.d. random intercept. We also include a model variant without random intercept.
          We apply MCMC via the No-U-Turn Sampler as available in
          the Python library \texttt{liesel-bctm} \parencite{Brachem2023-LieselbctmBayesianConditional}. We draw
          $5\,000$ samples each in four chains after a warmup of $5\,000$ iterations, yielding a total of $20\,000$ samples. Since sampling is done exclusively with NUTS, we use no thinning.
    \item QGAM: A fully specified additive quantile regression model with P-splines for all four covariates, as available in the R package \texttt{qgam} \parencite{Fasiolo2021-FastCalibratedAdditive,Fasiolo2021-QgamBayesianNonparametric}. We fit QGAM for a grid of 25 evenly spaced probability levels in $\{0.005, \dots, 0.995\}$. Each quantile model is of the form
          \[Q_{\text{cholst}}(\alpha | \text{age}_{ij}) = s_{\alpha}(\text{age}_{ij}) + \text{sex}_i \beta_1 \]
          where \(s_{\alpha}(\text{age}_{i})\) is a 20-parameter
          P-spline (including an intercept) and \(Q_{\text{cholst}}(\alpha | \text{age}_{ij})\) is the conditional
          \(\alpha\)-quantile of the response. While \texttt{qgam} allows inclusion of
          random intercepts, we found model fitting to be computationally infeasible, so we do not include a random intercept version.
\end{itemize}

\subsection{Additional results}

\begin{figure}[ht]
    \includegraphics[width=\textwidth]{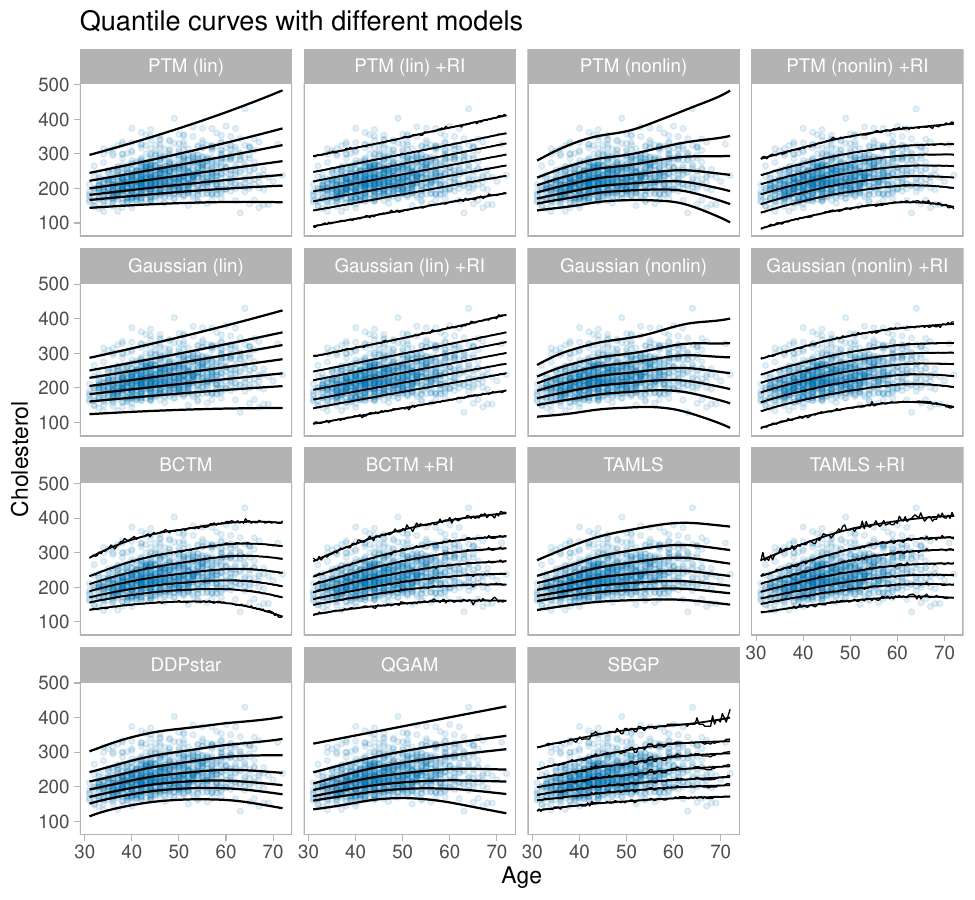}
    \caption[Quantile curves for PTM variants in Framingham Heart Study.]{
        Quantile curves at probability levels $0.01, 0.1, 0.5, 0.9, 0.99$ for different models. The curves were produced with full-sample model runs.
        The models labeled with +RI include an i.i.d. random intercept term for each patient.
        For SBGP, the available implementation returns posterior predictive samples at the specified age values rather than an analytic conditional quantile function. The SBGP curves shown here were therefore computed pointwise as empirical quantiles of 5000 posterior predictive draws each for a grid of 150 values for age in $[\min(age), \max(age)]$.
        For BCTM and the random intercept models, we similarly reconstructed the quantile curves from predictive samples. For each grid point, we drew samples from the posterior predictive distribution, drawing a random intercept term from $\calN(0, \psi^2)$ for the random intercept models. For the random intercept variance $\psi^2$, we use the variance point estimate (TAMLS) or draws from the variance parameter's posterior distribution (Bayesian models).
        The local wiggliness of the reconstructed quantile curves reflects this sample-based reconstruction and should not be interpreted as evidence for an intrinsic lack of smoothness.
        The overlaid smooth lines are cubic regression splines applied only for visualization (using the \texttt{ggplot2} default in \texttt{geom\_smooth}).
        \label{app-fig:fh-qcurves-all}
        }
\end{figure}

\autoref{app-fig:fh-qcurves-all}
shows quantile curves for all models. \autoref{app-tab:fh-diagnostics} contains diagnostic information about posterior MCMC samples for the Gaussian and PTM models. The quantities are computed using the Python library Arviz \parencite{Kumar2019-ArviZUnifiedLibrary}, version 0.21.0.
The maximum $\hat{R}$ exceeds $1.01$ for the PTMs with random intercept,
suggesting a problem with convergence. When we inspected these models more closely, we
found that the posterior distribution of the random intercept term closely
resembles the posterior estimate for $f_R(r)$ in the corresponding model without random
intercept. This could indicate that the transformation function estimate and the
random intercept capture similar information, leading to challenging identification
and hence problematic convergence.

The minimum effective sample sizes exceed $400$ for all models, suggesting sufficient information available in the posterior samples for all sampled quantities.

\autoref{app-tab:fh-errors} contains notes captured during MCMC sampling for the PTM and Gaussian models.
We observe minimal shares of divergent transitions and maximum tree depths, and virtually no NaN acceptance probabilities, showing no serious cause for concern.

\begin{table}[ht]
    \caption[Diagnostic information for the PTM in the Framingham Heart Study.]{Diagnostic information for the PTM and Gaussian models in the Framingham Heart Study application example. The values are averaged over all
        folds of the cross-validation. $\hat{R}$ values are based on four parallel chains. \label{app-tab:fh-diagnostics}}
    \small
    
\begin{tabu} to \linewidth {>{\raggedright}X>{\raggedleft}X>{\raggedleft}X>{\centering}X}
\toprule
Model & Minimum ESS (Bulk) & Minimum ESS (Tail) & Maximum $\hat{R}$\\
\midrule
PTM (lin) & 4 569 & 8 698 & 1.001\\
PTM (lin) +RI & 1 501 & 1 006 & 1.004\\
PTM (nonlin) & 2 247 & 4 380 & 1.002\\
PTM (nonlin) +RI & 1 456 & 2 005 & 1.002\\
\addlinespace
Gaussian (lin) & 13 977 & 16 360 & 1.000\\
Gaussian (lin) +RI & 2 207 & 4 327 & 1.002\\
Gaussian (nonlin) & 4 975 & 8 633 & 1.001\\
Gaussian (nonlin) +RI & 2 264 & 4 316 & 1.002\\
\bottomrule
\end{tabu}
\end{table}

\begin{table}[ht]
    \caption[Divergent transitions and maximum tree depths for the PTM in the Framingham Heart Study]{Notes captured during MCMC sampling for the PTM and Gaussian models in the Framingham Heart Study application example. The values are averaged over all
        folds of the cross-validation.
        \label{app-tab:fh-errors}}
    \small
    
\begin{tabu} to \linewidth {>{\raggedright\arraybackslash}p{10em}>{\raggedleft\arraybackslash}p{15em}>{\raggedright}X}
\toprule
Model & Note & Rel. freq.\\
\midrule
PTM (lin) & divergent transition & 0\\
PTM (lin) & maximum tree depth & 0\\
\addlinespace
PTM (lin) +RI & divergent transition & 0\\
PTM (lin) +RI & maximum tree depth & 0\\
\addlinespace
PTM (nonlin) & divergent transition & 0\\
PTM (nonlin) & maximum tree depth & 0\\
\addlinespace
PTM (nonlin) +RI & divergent transition & 0\\
PTM (nonlin) +RI & maximum tree depth & 0\\
\addlinespace
PTM (nonlin) +RI & nan acceptance prob & 0\\
Gaussian (lin) & nan acceptance prob & 0\\
Gaussian (lin) +RI & divergent transition & 0\\
Gaussian (lin) +RI & nan acceptance prob & 0\\
Gaussian (nonlin) & nan acceptance prob & 0\\
Gaussian (nonlin) +RI & divergent transition & 0\\
Gaussian (nonlin) +RI & nan acceptance prob & 0\\
\bottomrule
\end{tabu}
\end{table}

\clearpage

\stopcontents[appendix]
\stoplist[appendix]{lof}
\stoplist[appendix]{lot}
\printbibliography
\end{refsection}

\end{document}